\documentstyle[12pt]{article}
\newcommand{\be}{\begin{equation}}
\newcommand{\ee}{\end{equation}}
\newcommand{\bn}{\begin{eqnarray}}
\newcommand{\en}{\end{eqnarray}}
\newcommand{\p}{\partial}
\newcommand{\dslash}{\partial\!\!\!/}
\newcommand{\Dslash}{D\!\!\!\!/}
\newcommand{\aslash}{A\!\!\!/}

\def\ni{\noindent}

\begin{document}

\begin{center}

%------------------------------------------ Title --------------------
{\Large \bf Topics on the Quantum Dynamics of Chiral Bosons}
%---------------------------------------------------------------------
\vspace{6mm}

%-------------------------------------- Author(s) --------------------
{\Large E. M. C. Abreu$^{a}$ and C. Wotzasek$^{b}$}
%---------------------------------------------------------------------
\vspace{3mm}

%------------------------------------ Address(es) --------------------

\begin{large}

$^a$
{\it Departamento de F\'{\i}sica Te\'orica, Instituto de F\'{\i}sica, Universidade do Estado do Rio de Janeiro,\\
Rua S\~ao Francisco Xavier 524, Maracan\~a, 20550-013, Rio de Janeiro, RJ, Brazil}\\
{\sf E-mail: everton@if.uerj.br} \\[0pt]\bigskip
$^b$
{\it Departamento de F\'{\i}sica Te\'orica, Instituto de F\'{\i}sica, Universidade Federal do Rio de Janeiro,\\ 
Caixa Postal 68528, 21945-970, Rio de Janeiro, RJ, Brazil} \\
{\sf E-mail: clovis@if.ufrj.br}\\

\end{large}
%---------------------------------------------------------------------
%\vspace{4mm}
\bigskip

\today

%\vspace{14mm}

\end{center}

%\documentstyle[prd,tighten,aps]{revtex}

%\newcommand{\be}{\begin{equation}}
%\newcommand{\ee}{\end{equation}}
%\newcommand{\bn}{\begin{eqnarray}}
%\newcommand{\en}{\end{eqnarray}}
%\newcommand{\p}{\partial}
%\newcommand{\dslash}{\partial\!\!\!/}
%\newcommand{\Dslash}{D\!\!\!\!/}
%\newcommand{\aslash}{A\!\!\!/}

%\def\ni{\noindent}

%\documentstyle[aps,twocolumn]{revtex}
%\bibliographystyle{prsty} se for usar bibtex
%\begin{document}
% o comando \draft colocas os pacs numbers
%\draft
%\title{Chiral Bosons, Dual Projection and Interference Phenomena}
%\author{Everton M. C. Abreu$^a$ and Clovis Wotzasek$^b$}
%\address{$^a$Departamento de F\'{\i}sica Te\'orica, Instituto de F\'{\i}sica, 
%Universidade do %Estado do Rio de Janeiro,\\
%Rua S\~ao Francisco Xavier 524, Maracan\~a, 20550-013, Rio de Janeiro, RJ, Brazil\\
%{\sf E-mail: everton@if.uerj.br}\\
%$^b$Departamento de F\'{\i}sica Te\'orica, Instituto de F\'{\i}sica, 
%Universidade Federal do Rio %de Janeiro,\\ 
%Caixa Postal 68528, 21945-970, Rio de Janeiro, RJ, Brazil \\
%{\sf E-mail: clovis@if.ufrj.br}}
%\date{\today}
%\maketitle

\begin{abstract}
\noindent Chiral bosons are important building blocks in the study of supergravity, string theory and quantum Hall effect. Along the last two decades many different formulations have appeared trying to describe the dynamics and the quantization of these curious objects. However two of them have gain special attention among people working on this area: the gauge invariant formulation proposed by Siegel and the noninvariant one put forward by Floreanini and Jackiw. We call these distinct analysis as chiral bosonization schemes (CBS).
In this report we make a study of the relationships among many of these different chiral bosonization schemes. 
This is done in the context canonical framework with two different techniques known as soldering formalism and dual projection formalism. The first considers
the phenomenon of interference between chiral modes and the second is able to separate dynamics from the symmetry behavior in a quantum field theory.
While the soldering formalism discloses phenomena analogous to the double slit interference phenomena in classical-quantum physics with important consequences to the bosonization program, the dual projection, in particular, is able to disclose the presence of a noton, a nonmover field, in different formulations for chiral bosons.
The importance of this last result is that it proves the duality between Siegel and Floreanini-Jackiw model without invoking gauge-fixing: while the Floreanini-Jackiw component describes the dynamics, it is the noton that carries the symmetry contents, acquiring dynamics upon quantization and is fully responsible for the Siegel anomaly.
\end{abstract}
%\pacs{11.10.Ef,11.15.-q,11.30.Rd,11.40.Ex}
\bigskip


\newpage

{\small\tableofcontents}

\newpage

\section{Introduction}

The main motivation of this work is to report on the contributions made by the authors to the understanding of the bosonic chiral dynamics and discuss some new related aspects. Covariantization of the chiral constraints, separation of the bosonic chiral dynamics from the (chiral) diffeomorphism symmetry and mass generation resulting from chiral interactions are some of our original contributions \cite{CW1,aw,abw}.

To begin with, let us point, in a nutshell, some of the incitement involving this issue.
Chiral bosons appeared in the physical scenario in the 1980's in different contexts, both in the high energy string related research as well as in the research of the quantum Hall effect by the low energy condensed matter community.  Having such distinct origin, different formulations have been offered to describe the chiral bosonic dynamics. In supergravity models, the extension of the chiral boson to higher dimensions has naturally introduced the concept of the chiral $p$-forms. 
Chiral $p$-forms enter in the spectrum of type IIB superstring and it can be
shown that a string can couple directly to a chiral $p$-form \cite{schwarz}.
A chiral $p$-form has a number of properties which are very peculiar.
Although its excitations obey Bose statistics, it shares many of the
features of fermionic fields like a field equation which is linear in
derivatives; it leads to species doubling if one attempts to define it on a
lattice \cite{tang} and it gives rise to gravitational anomalies \cite{awebn}.

An analysis (classical and quantum) of chiral bosons coupled to Liouville gravity in a closed space was
performed by Banerjee, Chatterjee and Ghosh in \cite{ghosh1}.  In few words we can say that the authors show that
in the classic case, the chiral boson remains free nevertheless.  At the quantum level, the anomaly was introduced via
Wess-Zumino term and several versions of the chiral bosons were analyzed with interesting results.  The quantization
of the chiral bosons using the BFV method can be saw in \cite{ghosh2}.

A path integral point of view for the chiral bosons was accomplished in \cite{semenoff1}.  It provides the quantization of
chiral bosons which yields a bosonized theory of complex Weyl fermions.  Furthermore, the work deals with the conformal anomaly.
A geometrical interpretation of the chiral constraint with a manifestly local Lorentz invariant action as well as the 
demonstration of the equivalence of correlation functions of currents and energy-momentum
tensors can be found in \cite{semenoff2}.

We can describe chiral bosons by a $p$-form gauge potentials $B_p$. The $B_p$%
's curvatures $H_{p+1}=d B_p$ satisfy, as equations of motions, a Hodge
(anti) self-duality condition in a space-time with dimension $D=2(p+1)$. The, 
values of $p$ are restricted to even values by the self-consistency of such
equation in space-times with Minkowskian signature $\eta_{ab} =
(1,-1,-1,\ldots,-1)$. The restriction on $p$ makes $D=2,6,10,\ldots$, be the
relevant dimensions of the chiral boson theories.

Chiral bosons are important in superstring and supergravity theories, and
more recently $M$ theory. Two dimensional chiral bosons (scalar) are basic
ingredients in string theory. The six-dimensional ones belong to the
supergravity and tensor multiplets in $N=1, D=6$ supergravity theories. They
are necessary to complete the $N=2, D=6$ supermultiplet of the $M$-theory
five-brane. Finally a ten dimensional chiral bosons appears in $IIB, D=10$
supergravity.

In this work we depict our investigations over the years about chiral bosons.  
These studies are based on the following issues: constraint conversion that turns second-class constraints into first-class \cite{CW2,CW3}, the soldering formalism developed by M. Stone \cite{solda,clovis} and the dual projection concept, developed by us (based in the seminal idea of chiral decomposition by Mandelstam \cite{man}) in several works \cite{aw,baw2,everton,everton2,everton3}.  
The development of these formalisms has brought new insights to the investigation of the chiral boson dynamics and quantization.

Although they may, at the first sight, look quite disconnected, all the above topics are somehow correlated.  The use of the constraint conversion technique has solved the question of covariantization of the constraints put forward by the Floreanini-Jackiw formulation for the chiral boson \cite{fj}. This approach has revealed the idiosyncratic character of the chiral constraint that induces an infinite tower of new fields to produce the covariant form.  The analogous problem in the super-Yang-Mills action \cite{berkovits} has been treated by a similar technique. Later on the covariantization problem has been alternatively approached by Pasti, Sorokin and Tonin \cite{pst1} using a single field but nonlocal action.  We will not refer to this problem any further in this report and refer the interested reader to the original literature.  The dual projection technique was introduced in the chiral boson context in \cite{aw} to deal with the long standing problem of the equivalence between Siegel chiral boson formulation with the non-covariant second-class formulation proposed in \cite{fj}. This technique, strictly related to
canonical transformations, was first proposed in \cite{clovis3} to deal with the issue of classification of electromagnetic duality groups. The difference between both concepts being that 
the dual projection is performed at the level of the actions while the canonical transformation
is at the Hamiltonian level. However, in the following pages we will consider only
first-order actions so that the equivalence between the dual
projection and the canonical transformation becomes manifest \cite{bg}. The
most useful and interesting point in this dual projection procedure is that
it is not based on any specifically dimensional concept and may be extended
to  a higher dimensional situation \cite{baw2}. 
However, as long as this work is only about $D=2$ chiral bosons, we will not analyze this higher-dimensional feature here. In the above analysis, an important fact is that coupling chiral scalars to background
gravitational fields reveals the presence of notons \cite{hull,dgr}. Noton
is a nonmover field at classical level, carrying the representation of the
Siegel symmetry \cite{siegel}, that acquires dynamics upon quantization. At
the quantum level, it was shown \cite{aw} that its dynamics is fully
responsible for the Siegel anomaly. The proposal of \cite{aw} to the chiral equivalence was based on the ability of the dual projection to separate the proper variables carrying chiral dynamics from those carrying the representation of the Siegel algebra.

Concerning the non-Abelian chiral bosons theories, the Wess-Zumino-Witten-Novikov (WZWN) \cite{witten,novikov} model is a conformal field theory that has been used in the past to reproduce various two dimensional systems like Toda
field theories and low-dimensional black holes. Recently it was carried out the
connection of this model to a combination of three dimensional topological
BF and Chern-Simons gauge theories defined on a manifold with boundaries
using a direct application of non-Abelian T duality on the WZWN nonlinear
sigma model \cite{mohammed}. 

Finally, turning into the last issue mentioned above, we consider the problem of constructing a covariant multiplet out of independent self-dual modes, christened as soldering mechanism \cite{solda}. Based on this approach we proposed recently \cite{abw} a new interpretation 
for the dynamical mass generation known as Schwinger mechanism.  This study considered the interference of
right and left gauged FJ chiral bosons \cite{fj}.
The result of the chiral interference shows the presence of a massive
vectorial mode, for the special case
where the Jackiw-Rajaraman regularization parameter is $a=1$ \cite{jr}, which is the value where the chiral theories have only one massless excitation in the spectrum.
This clearly shows that the massive vector mode results from
the interference between two massless modes.

Incidentally, the FJ model is the chiral dynamical sector of the more general CBS proposed by Siegel \cite{siegel}.  
The Siegel modes (rightons and leftons) carry not only
chiral dynamics but also symmetry information. 
The symmetry content of the theory is well described by the Siegel algebra, a truncate
diffeomorphism, that disappears at the quantum level.
The chiral interference between gauged rightons and leftons should lead to
a massive vector mode, plus the symmetry of
the combined right-left Siegel algebras.
As we will show, the result of the chiral interference between gauged
Siegel modes is a Hull noton \cite{hull}. This represents only the symmetry part of the
expected result, disclosing the destructive interference of the massive mode \cite{clovis2}
resulting from the simultaneous soldering of dynamics and symmetry.

This work will be presented in the next two sections.  For pedagogic reasons the fusion of chiral modes will be considered first.  This will help to discuss the concept of duality projection in Section 3.  At the end of each section, we make the pertinent conclusions and considerations.

%%%%%%%%%%%%%%%%%%%%%%%%%%%%%%%%%%%%%%%%%%%%%%%%%%%%%%%%%%%%%%%%%%%%%%%%%%%%%%%%%%%%%%%%%%%%%%%%%%%%%%%%%%%%%%%%%%%%%%%%%%%%%%%%%%%%%%%%%%%%%%%%%%%%%%%%%%%

\section{The Soldering Formalism of Chiral Theories}

%\subsection{Introduction}

The process of fusing together different modes of chiral theories is very much related to the issue of duality, quite in vogue in string problem these days. The role of duality as a qualitative tool in the investigation of physical systems is very much appreciate at different
context and dimensions \cite{AG}.
Different aspects of duality symmetric actions have been examined.
A new technique, able to deal with distinct manifestations of the duality symmetry was proposed
many years ago by Stone \cite{solda}.
The Stone soldering mechanism for fusing together opposite aspects of duality symmetries, provides a new formalism that
includes the quantum interference effects between the independent components, and is dimension independent.
In fact the electromagnetic duality in 2D and 4D can be reexamined under the soldering point of view providing
interesting new results, both explictly dual and covariant as the result of the interference between
self and anti-self dual actions displaying opposite aspects of
the electromagnetic duality symmetry \cite{baw}.

Exciting new results in 2D and 3D bosonization were generated via soldering that could not be obtained by any of the
known existent bosonization techniques. These problems have been analyzed also for field theories defined on non-commutative base-space \cite{NC-ghosh}.
In 2D this technique allowed us to show how massless chiral fields combine to provide the gauge invariant massive mode
present in the $QED_2$, showing the Schwinger mechanism as the result of quantum interference between
right and left moving modes, very much like in the optical double slit Young's experiment.
In fact, by equipping the soldering technique with
gauge and Bose symmetry \cite{RB3}, it automatically selects, in a unique way, the massless sector of
the chiral models displaying the Jackiw-Rajaraman parameter that reflects the bosonization ambiguity by $a=1$.
In the 3D case, the soldering mechanism was used to show the result of fusing together two topologically massive modes
generated by the bosonization of two massive Thirring models with opposite mass signatures in the long
wave-length limit.  The bosonized modes, which are described by self and anti-self dual Chern-Simons models \cite{tpn,dj}, were
then soldered into the two massive modes of the 3D Proca model \cite{baw}.

\subsection{The Soldering Formalism and Chiral Bosons}

In this section we will follow basically the references \cite{clovis,adw} to make a short, but at the same time self-consistent review of the method of
soldering two opposite chiral versions of a theory. 

The soldering formalism gives an useful bosonization scheme for Weyl fermions, since a level one representation of LU(N) has an
interpretation as the Hilbert space for a free chiral fermion \cite{ps}.
However, only Weyl fermions can be analyzed in this way, since a $2D$ conformally invariant QFT has separated right and left current algebras. In
other words, it is trivial to make a (free) Dirac fermion from two (free)
Weyl fermions with opposite chiralities. The action is just the sum of two
Weyl fermion actions. It seems, however, non-trivial to get the action of
the WZW model from two chiral boson actions of opposite ``chiralities'',
because it is not the sum of the two.

To solve this problem, Stone \cite{solda} introduced the idea of soldering the
two chiral scalars by introducing a non-dynamical gauge field\footnote{In
\cite{harada2}, Harada proposed a physical interpretation for these
soldering fields.} to remove the
degree of freedom that obstructs the vector gauge invariance \cite{clovis}. This
is connected to the necessity that one must have more
than the direct sum of two fermions representations of the Kac-Moody algebra
to describe a Dirac fermion. In another way we can say that the equality for
the weights in the two representations is physically connected with the
necessity to abandon one of the two separate chiral symmetries, and accept
that the vector gauge symmetry should be maintained. This is the main
motivation for the introduction of the soldering field which makes possible
the fusion of dualities in all space-time dimensions. Besides, being just an
auxiliary field, it may posteriorly be eliminated in favor of the physically
relevant quantities. This restriction will force the two independent chiral
representations to belong to the same multiplet, effectively soldering them
together. We will see below, in a precise way, more details about the
physical significance of the soldering field.

In \cite{adw}, the authors promoted the soldering the two (Siegel)
invariant representations of opposite chiralities. The symmetry content of
each theory is well described by the Siegel algebra, a truncate
diffeomorphism, that disappear at the quantum level. The resulting action
is invariant under the full diffeomorphism group, which is not
a mere sum of two Siegel symmetries. As we will see later, the result can 
{\it also} be seen as a scalar field immersed in a gravitational background.

%Recently, there has been a great deal of interest in soldering distinct manifestations of duality. The procedure leads to new physical results including quantum contributions. For instance, these results provided the idea of an interference effect. However, this ``wave'' interpretation is not new. E. Witten, in \cite{witten}, associated the fields depending on only one chirality to left-moving or right-moving waves as being the $\gamma ^{5}$ eigenstates.

The present authors, with R. Banerjee \cite{abw}, have promoted the interference of
two chiral Schwinger models with opposite chiralities. As a result it was
obtained a new interpretation for the phenomenon of mass generation. The Bose symmetry fixed the
Jackiw-Rajaraman parameter ($a=1$) \cite{jr} so that in the spectrum only
massless harmonic excitations survived. The soldered action represents a vector 
Schwinger model which has a
massive particle in the spectrum. This behavior characterizes a constructive
interference with the arising of a mass term that is typical of the
right-left quantum interference \cite{goto}\footnote{%
The extension of this case to the four dimensional one was performed in \cite
{bw2}.}. 
In terms of degrees of freedom we can say that each (chiral) action
contributes with ``one half" degree of freedom of opposite signals. Hence, the
soldered action has one ``full" degree of freedom.  

It was shown later \cite{aw}, that in the soldering process of two Siegel's 
\cite{siegel} modes (lefton and righton) coupled to a gauge field \cite{gs},
this gauge field has decoupled from the physical field. The final action
describes a non-mover field (a noton) at the classical level. The noton
acquires dynamics upon quantization. This field was introduced by Hull \cite
{hull} to cancel out the Siegel anomaly. It carries a representation of the
full diffeomorphism group, while its chiral components carry the
representation of the chiral diffeomorphism. We will turn back to talk about the noton in the next section.

We can say in few words on the $3D$ and $4D$ cases. There the soldering mechanism was used to show the result of
fusing together two topologically massive modes generated by the
bosonization of two massive Thirring models with opposite mass signatures in
the long wavelength limit. The bosonized modes, which are described by self
and anti-self dual Chern-Simons models \cite{tpn,dj}, were then soldered into
the two massive modes of the $3D$ Proca model \cite{baw}. In the $4D$ case,
the soldering mechanism produced an explicitly dual and covariant action as
the result of the interference between two Schwarz-Sen \cite{ss} actions
displaying opposite aspects of the electromagnetic duality \cite{baw}.

Back to chiral bosons, one of us \cite{clovis3} has obtained the field theoretical analog of the
``quantum destructive interference'' phenomenon, by coupling the non-Abelian
chiral scalars to appropriately truncated metric fields known as chiral WZW
models, or non-Abelian Siegel models \cite{witten}. In fact, this effective
action does not contain either right or left movers, but can be identified
with the non-Abelian generalization of the bosonic non-mover action proposed
by Hull.

In a recent work \cite{ainw}, it was analyzed the restrictions posed by the
soldering formalism over a new regularization class that extends the
classification of the regularization ambiguity of $2D$ fermionic determinant
from four to a three-constraint class. This analysis results from the
interference effects between right and left movers, producing a massive
vectorial photon that constrains the regularization parameter to 
this three-constraints class. In other words, the new Faddeevian class of
chiral bosons proposed by Mitra \cite{mitra} has interfered constructively
to produce a massive vectorial mode.

\subsection{Description of the Formalism}

The basic idea of the soldering procedure is to raise a global Noether
symmetry of the self and anti-self dual constituents into a local one, but
for an effective composite system, consisting of the dual components and an
interference term. The objective in \cite{clovis} is to systemize the procedure
like an algorithm and, consequently, to define the soldered action.

An iterative Noether procedure was adopted to lift the global
symmetries. Therefore, assume that the symmetries in question are being
described by the local actions $S_{\pm}(\phi_{\pm}^\eta)$, invariant under a
global multi-parametric transformation

\begin{equation}  \label{ii10}
\delta \phi_{\pm}^\eta = \alpha^\eta\;\;,
\end{equation}
where $\eta$ represents the tensorial character of the basic fields in the
dual actions $S_{\pm}$ and, for notational simplicity, will be dropped from
now on. As it is well known, we can write,

\begin{equation}
\delta S_{\pm}\,=\,J^{\pm}\,\partial_{\pm}\,\alpha\;\;,
\end{equation}
where $J^{\pm}$ are the Noether currents.

Now, under local transformations these actions will not remain invariant,
and Noether counter-terms become necessary to reestablish the invariance,
along with appropriate auxiliary fields $B^{(N)}$, the so-called soldering
fields which have no dynamics.
%Nevertheless we can say that $B^{(N)}$ is an auxiliary field which
This makes a wider range of gauge-fixing conditions
available \cite{harada2}. In this way, the $N$-action can be written as,

\begin{equation}  \label{ii20}
S_{\pm}(\phi_{\pm})^{(0)}\rightarrow S_{\pm}(\phi_{\pm})^{(N)}=
S_{\pm}(\phi_{\pm})^{(N-1)}- B^{(N)} J_{\pm}^{(N)}\;\;.
\end{equation}
Here $J_{\pm}^{(N)}$ are the $N-$iteration Noether currents. For the self
and anti-self dual systems we have in mind this iterative gauging
procedure is (intentionally) constructed not to produce invariant actions
for any finite number of steps. However, if after N repetitions, the non
invariant piece end up being only dependent on the gauging parameters, but
not on the original fields, there will exist the possibility of mutual
cancelation if both self and anti-self gauged systems are put together.
Then, suppose that after N repetitions we arrive at the following
simultaneous conditions,

\begin{eqnarray}  \label{ii30}
\delta S_{\pm}(\phi_{\pm})^{(N)} \neq 0  \nonumber \\
\delta S_{B}(\phi_{\pm})=0\;\;,
\end{eqnarray}
with $S_B$ being the so-called soldered action 
\begin{equation}  \label{ii40}
S_{B}(\phi_{\pm})=S_{+}^{(N)}(\phi_{+}) + S_{-}^{(N)}(\phi_{-})+ %
\mbox{Contact Terms}\;\;,
\end{equation}
and the ``Contact Terms" being generally quadratic functions of the soldering
fields. 
Then we can immediately identify the (soldering) interference term as, 
\begin{equation}  \label{ii50}
S_{int}=\mbox{Contact Terms}-\sum_{N}B^{(N)} J_{\pm}^{(N)}\;\;.
\end{equation}
Incidentally, these auxiliary fields $B^{(N)}$ may be eliminated, for instance,
through its equations of motion, from the resulting effective action, in
favor of the physically relevant degrees of freedom. It is important to
notice that after the elimination of the soldering fields, the resulting
effective action will not depend on either self or anti-self dual fields $%
\phi_{\pm}$ but only in some collective field, say $\Phi$, defined in terms
of the original ones in a (Noether) invariant way

\begin{equation}
S_{B}(\phi _{\pm })\rightarrow S_{eff}(\Phi )\;\;.  \label{ii60}
\end{equation}
Analyzing in terms of the classical degrees of freedom, it is obvious that we have now a
bigger theory. Once such effective action has been established, the physical
consequences of the soldering are readily obtained by simple inspection.
This will progressively be clarified in the specific application to be given
next.

In order to present an example, we will analyze the Siegel chiral actions in the light
of the interference phenomenon \cite{adw}. First of all, we have to
describe the light-front variables used here as, 
\begin{eqnarray}  \label{lightfront}
x_{\pm}&=&{\frac{1 }{\sqrt{2}}} (x_0\pm x_1)\,=\,x^{\mp}\;\;,  \nonumber \\
\partial_{\pm}&=&{\frac{1 }{\sqrt{2}}}(\partial_0\,\pm\,\partial_1) \,=\, \partial^{\mp}\;\;, \\
A_{\pm}&=&{\frac{1 }{\sqrt{2}}}(A_0\,\pm\,A_1) \,=\, A^{\mp}\;\;,  \nonumber
\end{eqnarray}
and now we can work out our first example.

\subsubsection{An example: the Siegel action}

The original classical Lagrangian density for a chiral scalar field as introduced by Siegel \cite{siegel} for a left moving scalar (a lefton) is \cite{fs}
%The Siegel action for a left-moving chiral boson, a lefton (which will be
%explained below), is 
\begin{eqnarray}  \label{03}
{\cal L}_0^{(+)} &=& \partial_+\varphi\partial_-\varphi\,+\,\lambda_{++}
\partial_-\varphi\partial_-\varphi \nonumber \\
&=& {1 \over 2}\,\sqrt{-g}\,g^{\alpha\beta}\,\partial_{\alpha}\varphi\,\partial_{\beta}\varphi\;\;,
\end{eqnarray}
where the metric is given by
\bn
\sqrt{-g}g^{++}\,=\,0\:\:\:\: , \:\:\:\:\sqrt{-g}g^{+-}\,=\,1 \:\:\: , \:\:\: \sqrt{-g}g^{--}\,&=&\,2\,\lambda_{++}\;\;.
\en
The Lagrangian (\ref{03}) is invariant under Siegel gauge symmetry which is an invariance under the combined coordinate transformation and a Weyl rescaling of the form
\bn \label{11}
x_- \rightarrow \tilde{x}_-\,&=&\,x_-\,-\,\epsilon\,(x_+,x_-) \nonumber \\
\delta_w\,g_{\alpha\beta}\,&=&\,-\,g_{\alpha\beta}\,\partial_-\,\epsilon^-\;\;,
\en
where $\epsilon^{\pm}=\epsilon^{\pm}\,(x_{\pm})$.
The fields $\varphi$ and $\lambda_{++}$ transform under (\ref{11}) as follows:
\bn \label{12}
\delta\,\varphi\,&=&\,\epsilon^-\,\partial_-\varphi\;\;, \nonumber \\
\delta\,\lambda_{++}\,&=&\,-\,\partial_+\,\epsilon\,+\,\epsilon\,\partial_+\,\lambda_{++}\,-\,\lambda_{++}\,\partial_+\,\epsilon^-\;\;.
\en
In addition (\ref{03}) is invariant under the global axial transformation
\be
\varphi \rightarrow \tilde{\varphi}\,=\,\varphi\,+\,\bar{\varphi}\;\;,
\ee
along with their associated currents.  It is beyond our work to write explicitly these axial currents as well the conserved vector current.  These objects can be found in literature (see \cite{fs} for example).

The symmetry (\ref{12}) describes a lefton. This is the main difference between a
lefton (righton) and a left-moving (right-moving) FJ particle. The first is
provided with symmetry and dynamics, while the second is responsible only
for the dynamics of the theory. We will discuss in the next sections how the lefton (or righton)
carries the anomaly of the system \cite{aw}.
Similarly, one can gauge the semi-local affine symmetry 
\begin{eqnarray}  \label{06}
\delta\,\varphi\,&=&\,\epsilon^+\partial_+\varphi  \nonumber \\
\delta\,\lambda_{--}\,&=&\,-\partial_-\epsilon^+
+\epsilon^+\partial_+\,\lambda_{--}-\lambda_{--}\partial_+\epsilon^+\,.
\end{eqnarray}
to obtain the righton.  Next we will promote the fusion of the righton and the lefton obtaining the final soldered action.

%\subsubsection{The soldering procedure}

In fact, if we construct the righton and lefton chiral boson actions as 
\begin{equation}  \label{08}
{\cal L}_0^{(\pm)} = {\frac{1}{2}} J_{\pm}(\varphi)\partial_{\mp}\varphi
\end{equation}

\noindent with 
\begin{equation}  \label{09}
J_{\pm}(\varphi)=2\left(\partial_{\pm}\varphi
+\,\lambda_{\pm\pm}\partial_{\mp}\varphi\right)\;\;,
\end{equation}
\noindent it is easy to verify that these models are indeed invariant under
Siegel's transformations (\ref{12}) and (\ref{06}), using that 
\begin{equation}  \label{010}
\delta J_{\pm}= \epsilon_{\pm}\partial_{\mp}J_{\pm}\;\;.
\end{equation}

\noindent We can realize at this point that Siegel's actions for
leftons and rightons can be seen as the action for a scalar field immersed
in a gravitational background whose metric is appropriately truncated. In
this sense, Siegel symmetry for each chirality can be seen as a truncation
of the reparametrization symmetry existing for the scalar field action. We
should mention that the Noether current $J_+$ defined above is in fact the
non-vanishing component of the left chiral current $J_+ = J_{(L)}^-$, while $%
J_-$ is the non-vanishing component of the right chiral current $J_- =
J_{(R)}^+$, with the left and right currents being defined in terms of the
axial and vector currents as 
\begin{eqnarray}  \label{333}
J_\mu^{(L)}=J_\mu^{(A)}+J_\mu^{(V)}\,,  \nonumber \\
J_\mu^{(R)}=J_\mu^{(A)}-J_\mu^{(V)}\,.
\end{eqnarray}

\ni As our objective now is to depict the method, for the interested reader there exist specific conditions concerning the appropriate couplings of these currents \cite{remarks}.

Let us next consider the question of the vector gauge symmetry. We can use
the iterative Noether procedure described above to gauge the global U(1)
symmetry 
\begin{eqnarray}  \label{201}
\delta\varphi &=& \alpha\,,  \nonumber \\
\delta\lambda_{++}\, &=& 0
\end{eqnarray}

\noindent possessed by Siegel's model (\ref{03}). Under the action of the
group of transformations (\ref{201}), written now as a local parameter, the
action (\ref{03}) changes as 
\begin{equation}  \label{30}
\delta {\cal L}_0^{(+)} = \partial_-\alpha J_+
\end{equation}

\noindent with the Noether current $J_+=J_+(\varphi)$ being given as in (\ref
{09}). To cancel out this piece, we introduce the soldering field $B_-$
coupled to the Noether current, redefining the original Siegel's Lagrangian
density as 
\begin{equation}  \label{501}
{\cal L}_0^{(+)}\rightarrow {\cal L}_1^{(+)}={\cal L}_0^{(+)} +B_- J_+ \,,
\end{equation}

\noindent where the variation of the gauge field is defined conveniently as 
\begin{equation}  \label{601}
\delta B_-=-\partial_-\alpha\,.
\end{equation}

\noindent As the variation of ${\cal L}_1^{(+)}$ does not vanish modulo
total derivatives, we introduce a further modification as 
\begin{equation}  \label{701}
{\cal L}_1^{(+)}\rightarrow {\cal L}_2^{(+)}={\cal L}_1^{(+)}+
\lambda_{++}B_-^2
\end{equation}
\noindent whose variation gives 
\begin{equation}  \label{801}
\delta{\cal L}_2^{(+)} = 2 B_-\partial_+\alpha\,.
\end{equation}

\noindent This piece cannot be canceled by a Noether counter-term,
so that a gauge invariant action for $\varphi$ and $B_-$ does not exist.
%at least with the introduction of only one gauge field.
We observe, however,
that this action has the virtue of having a variation dependent only on $B_-$
and $\alpha$, and not on $\varphi$. Expression (\ref{801}) is a reflection of
the standard anomaly\footnote{%
The soldering analysis of the anomaly has been depicted in \cite{adas}.} that
is intimately connected with the chiral properties of $\varphi$.

Now, if the same gauging procedure is followed for a Siegel boson of
opposite chirality, say 
\begin{equation}  \label{90}
{\cal L}_0^{(-)} = \partial_+\rho\partial_-\rho + \lambda_{--}
\partial_+\rho\partial_+\rho
\end{equation}

\noindent subject to 
\begin{eqnarray}  \label{1001}
\delta\rho &=& \alpha\,,  \nonumber \\
\delta\lambda_{--} &=& 0\,,  \nonumber \\
\delta B_+&=& -\partial_+\alpha\,,
\end{eqnarray}

\noindent then one finds that the sum of the right and left gauged actions $%
{\cal L}_2^{(+)}+{\cal L}_2^{(-)}$ can be made gauge invariant if a contact
term of the form 
\begin{equation}  \label{110}
{\cal L}_C = 2 B_+ B_-
\end{equation}

\noindent is introduced. One can check that indeed the complete gauged
Lagrangian 
\begin{eqnarray}  \label{1201}
{\cal L}_{TOT} &=& \partial_+\varphi\partial_-\varphi + \lambda_{++}
\partial_-\varphi\partial_-\varphi + \partial_+\rho\partial_-\rho  
\,+\,\lambda_{--} \partial_+\rho\partial_+\rho + B_+J_-(\rho) + B_- J_+(\varphi) \nonumber \\
& &+\lambda_{--}B_+^2 +\,\lambda_{++}B_-^2 +2 B_- \, B_+
\end{eqnarray}

\noindent with $J_{\pm}$ defined in Eq. (\ref{09}) above, is invariant under
the set of transformations (\ref{201}), (\ref{601}) and (\ref{1001}). For
completeness, we note that Lagrangian (\ref{1201}) can also be written in the
form
\begin{eqnarray}  \label{121}
{\cal L}_{TOT} = D_+\varphi D_-\varphi + \lambda_{++} D_-\varphi D_-\varphi
\,+\,D_+\rho D_-\rho + \lambda_{--} D_+\rho D_+\rho
\,+\,\left(\varphi-\rho\right)E\,,
\end{eqnarray}

\noindent modulo total derivatives. In the above expression, we have
introduced the covariant derivatives $D_{\pm}\varphi=
\partial_\pm\varphi+B_\pm$, with a similar expression for $D_\pm\rho$, and $%
E\equiv\partial_+B_--\partial_-B_+$. In the form (\ref{121}), ${\cal L}_{TOT}$
is manifestly gauge invariant.

After solving the equations of motion for the soldering fields we can write, 
\begin{equation}  \label{133}
{\cal L}_g ={\frac{1}{2}} \sqrt{-g}g^{\alpha\beta}
\partial_\alpha\Phi\partial_\beta\Phi\;\;.
\end{equation}

\noindent where, in the above expression we have introduced the metric tensor
density 
\begin{eqnarray}  \label{134}
\sqrt{-g}g^{--}&=&-4{\frac{\lambda_{++}}{\Delta}}\,,  \nonumber \\
\sqrt{-g}g^{++}&=&-4{\frac{\lambda_{--}}{\Delta}}\,,  \nonumber \\
\sqrt{-g}g^{+-}&=&-{\frac{2}{\Delta}}(1+\lambda_{++}\lambda_{--}) \,,
\end{eqnarray}

\noindent where $\Delta=2(\lambda_{++}\lambda_{--}-1)$ and 
\begin{equation}  \label{135}
\Phi={\frac{1 }{\sqrt{2}}}(\rho-\varphi)\,.
\end{equation}

We observe that in two dimensions $\sqrt{-g}g^{\alpha\beta}$ needs only two
parameters to be defined in a proper way. As it should be, $det(\sqrt{-g}%
g^{\alpha\beta})=-1$. We also note that, because of conformal invariance, we
cannot determine $g_{\alpha\beta}$ itself. We could, therefore, think of $%
{\cal L}_{TOT}$ as an effective theory, which represents a scalar boson $\Phi
$ in a gravitational background. It can be shown \cite{adw} that the action (%
\ref{133}) can be made invariant under the full group of diffeomorphism.
Hence, we can easily see that, in terms of symmetry, the new theory is
bigger than the old one. This new theory can be interpreted as a
constructive interference of symmetries. However, solving the equations of
motion for the multipliers, we can see that, in fact, this field has no
dynamics. This characterizes a nonmover field, the noton, introduced by Hull 
\cite{hull} to cancel out the gravitational anomaly of the Siegel model.

In the next section we promote the soldering analysis of another gauged coupled chiral boson, the chiral Schwinger model using the very important concept of the Bose symmetry \cite{RB3}.

%%%%%%%%%%%%%%%%%%%%%%%%%%%%%%%%%%%%%%%%%%%%%%%%%%%%%%%%%%%%%%%%%%%%%%%%%%%%%%%%%%%%%%%%%%%%%%%%

\subsubsection{Chiral interference of the gauged Siegel modes}

In this subsection we study the interference of gauged rightons and leftons
in the framework of \cite{solda}\footnote{The results of 2D-soldering may, equivalently, be obtained
including a right-left fermionic interaction term, before bosonization, as shown first in \cite{dgr}.}.
Gates and Siegel \cite{gs} have examined the interactions of leftons and
rightons with external gauge fields including the supersymmetric and the
non-Abelian cases.  Let us write the Siegel action with an electromagnetic coupling \cite{gs} as,
\begin{equation}
\label{sie1}
S_{\mp}^{(0)} = <\partial_{\pm}\phi_{\pm}(\partial_{\mp}\phi_{\pm} + 2A_{\mp}) +
\lambda_{\pm\pm} 
(\partial_{\mp}\phi_{\pm}+A_{\mp})^{2}>.
\end{equation}
Here $\phi_{\pm}$ describes a lefton (righton) and $<...>$ means space-time integration. 
This action has been used in \cite{gs}
to bosonize a generalized Thirring model that was shown to be invariant
under the extended conformal transformation,
\begin{eqnarray}
\label{II1}
\delta \, \phi_{\pm} & = & \xi^\mp\,(\,\partial_\mp \phi_{\pm} \; + \; A_\mp) \nonumber \\
\delta \lambda_{\pm\pm} & = & -\,\partial_\pm\,\phi_{\pm} \;+\; 
\xi^\mp\,\stackrel{\leftrightarrow}{\partial}_\mp\,\lambda_{\pm\pm}\nonumber\\
\delta A_\mp &=& 0 \;\;.
\end{eqnarray}

\noindent In the soldering formalism, seen in the last section, one must examine the response of the system to the transformation \cite{solda,abw} 
\begin{equation}
\label{II2}
\phi_\pm \rightarrow \phi_\pm + \alpha
\end{equation}
and compute the corresponding Noether currents,
\begin{eqnarray}
\label{II3}
J_{\phi_\pm}^{\mp} = 2 \, [ \,\partial_{\pm}\phi_\pm \;+\; \lambda_{\pm\pm}\,
(\partial_{\mp}\phi_\pm \;+\; A_{\mp}) \,];\;\;\; J_{\phi_\pm}^{\pm} = 2 \, A_{\mp}.
\end{eqnarray}

\noindent Following the knowledge above, we construct an iterated action introducing the soldering field $B_\mu$,
\begin{equation}
\label{a1}
S_{\mp}^{(0)} \rightarrow  S_{\mp}^{(1)} = S_{\mp}^{(0)} -
<B_{\mu}\,J_{\phi_\pm}^{\mu}>+<\lambda_{\pm\pm} \, B^{2}_{\mp}>.
\end{equation}
This iterated action behave, under the axial gauge transformation (\ref{II2}) and, $\delta B_{\pm} = \partial_{\pm}\alpha$ as,

\begin{equation}
\label{var}
\delta S_{\mp}^{(1)} \;=\; - \, 2 \, <B_{\mp} \, \delta B_{\pm}>\;\;.
\end{equation}
We can see that $S^{(1)}_\mp$ are not invariant, but their variations
are independent of the chiral fields $\phi_\pm$ and the result
(\ref{var}) reflects the anomalous behavior of the chiral model
under gauge transformations.  However,  being dependent only on the fields
taking values on the gauge algebra, they might cancel out mutually.
Indeed, the sum of $S^{(1)}_+$ and $S^{(1)}_-$
together with a contact term of the form $2\,<B_{-}\,B_{+}>$
%\begin{equation}
%\label{II8}
%S_{cont}= 2\,<B_{-}\,B_{+}>
%\end{equation}
results in an invariant action, 
\begin{eqnarray*}
%\label{II9}
\lefteqn{S_{T}  =  <\partial_{+}\phi \, (\,\partial_{-}\phi +2A_{-}\,) 
+\lambda_{++} \, (\,\partial_{-}\phi+A_{-}\,)^{2}} \\ 
& & +  \partial_{-}\rho \, (\,\partial_{+}\rho +2A_{+}\,)+
 \lambda_{--} \, (\,\partial_{+}\rho+A_{+} \, )^{2}
- \; B_{\mu}\,J_{\phi}^{\mu}\\
& &   - B_{\mu}\,J_{\rho}^{\mu}+ \lambda_{--}\,B_{+}^{2} +  \lambda_{++}\,B_{-}^{2} + 
2\,B_{-}\,B_{+}>
\end{eqnarray*}
where we have used $\phi_+ = \phi$ and $\phi_- =\rho$ for clarity.  Eliminating $B_\mu$ from their field equations,
leads to an effective action that incorporates the effects of the
right-left interference
\begin{eqnarray}
\label{II11}
\lefteqn{S_{eff}  = <\Delta\left\{\left(1 + \lambda_{++}\lambda_{--}\right) 
\partial_{-}\Psi \partial_{+}\Psi\right.} \nonumber\\ 
&& \left. + \lambda_{++}\left(\partial_{-}\Psi\right)^2 + 
\lambda_{--}\left(\partial_{+}\Psi\right)^2\right\} > - 2<a_{-}A_{+}>
\end{eqnarray}
%eff} & = &<\Delta\left\{\left(1 + \lambda_{++}\lambda_{--}\right)
%\mbox{}^{}\partial_{-}\Psi \partial_{+}\Psi + \lambda_{++}\left(\partial_{-}\Psi\right)^{2} + \nonumber \\
%& + &\lambda_{--}\left( \partial_{+}\Psi \right)^2 \right\} > - 2<a_{-}A_{+}>\;\;, 
\noindent where $\Delta =\left(1- \lambda_{++}\lambda_{--}\right)^{-1}$ and
$\Psi = \rho - \phi$.  The soldering process is now completed.  We have succeeded in including the effects of
interference between rightons and leftons.  Consequently, these components have lost their
individuality in favor of a new, gauge invariant, collective field  that 
does not depend on $\phi$ or $\rho$ separately. 

The physical meaning of (\ref{II11}) can be appreciated by
solving for the multipliers and using the symmetry induced by the soldering \cite{clovis2}, showing that it represents the action for the
noton.  In fact (\ref{II11}) is basically the action
proposed by Hull \cite{hull} as a candidate for canceling the Siegel
anomaly. This field carries a representation of the full diffeomorphism group \cite{hull}
while its chiral (Siegel) components carry the representation of the chiral diffeomorphism.
Observe the complete disappearance of the dynamical sector due to the destructive interference between the leftons and the rightons.  This happens
because we have introduced only one soldering field to deal with both the dynamics and the symmetry.
To recover dynamics we need to separate these sectors and solder them
independently as we will do next.
%%%%%%%%%%%%%%%%%%%%%%%%%%%%%%%%%%%%%%%%%%%%%%%%%%%%%%%%

\subsubsection{Constructive interference of the Siegel modes}

We are now in position to apply the dual projection to
the actions (\ref{sie1}), and study the effect of their interference.
According to the discussion of the last section
these systems decompose in chiral, right and left,
scalars $\varphi$ and $\varrho$ coupled to gauge fields,
and two notons, $\sigma$ and $\omega$, carrying the
representation of the right and the left Siegel symmetry,
\begin{eqnarray}
S_{-}^{(0)} & = & <\left\{ {\varphi'}\,\dot{\varphi}-{\varphi'}^2+2\,\sqrt{2}\,A_{-}\,{\varphi'}-
A_{-}^2+ {a \over 2}\,A_{\mu}^2 \right\}>\,+\, <\left\{ -{\sigma'}\,\dot{\sigma}-\eta_{+}{\sigma'}^2\right\}>\nonumber \\
S_{+}^{(0)} & = & <\left\{-{\varrho'} \dot{\varrho}-{\varrho'}^2-2\,\sqrt{2}\,A_{+}\,{\varrho'}-A_{+}^2+ 
{b \over 2}\,A_{\mu}^2\right\}>\,+\,<\left\{\;{\omega'}\,\dot{\omega}-\eta_{-}{\omega'}^2\right\}>.  \nonumber \\
\end{eqnarray}

\noindent The coefficients $a$ and $b$ are the Jackiw-Rajaraman regularization
parameters.  Each sector, dynamics or symmetry, corresponds to a self-dual or antiself-dual aspect
of the chirality.  This sets the stage for the independent soldering.  Consider the behavior
of the above theories under the following axial gauge transformation,
\begin{eqnarray}
\left(\varphi, \varrho\right)\;\; & \longrightarrow &\;\;
\left(\varphi\;\;+\;\; \alpha, \varrho \;\;+\;\; \alpha\right)
\nonumber \\
\left(\sigma, \omega\right)\;\; & \longrightarrow & \;\;
\left(\sigma \;\;+\;\; \beta, \omega \;\;+\;\; \beta\right)
\end{eqnarray}
which introduces two independent soldering.  The corresponding Noether's currents,
${\cal J}_{\alpha}^{\mu(\pm)}$, with $\mu=0,1$ being the Lorentz index and $\pm$ indicating the chirality,  are,
\begin{eqnarray}
{\cal J}_{\alpha}^{0(+)} & = & {\cal J}_{\alpha}^{0(-)} = 0 \nonumber \\
{\cal J}_{\alpha}^{1(+)} & = & -\;\;2\,(\,\dot\varrho \;\;+\;\;\varrho' \;\;+\;\;\sqrt{2}\,A_+\,) \nonumber \\
{\cal J}_{\alpha}^{1(-)} & = & 2\,(\,\dot{\varphi} \;\;-\;\;{\varphi'} \;\;+\;\;\sqrt{2}\,A_-\,)
\end{eqnarray}
and ${\cal J}_{\beta}^{\mu(\pm)}$
\begin{eqnarray}
{\cal J}_{\beta}^{0(+)} & = & {\cal J}_{\beta}^{0(-)} = 0 \nonumber \\
{\cal J}_{\beta}^{1(+)} & = & 2\,(\,\dot\omega \;\;-\;\;\eta_-\,\omega'\,) \nonumber \\
{\cal J}_{\beta}^{1(-)} & = & -\;\;2\,(\,\dot\sigma \;\;+\;\;\eta_+\,\sigma'\,) \;\;.
\end{eqnarray}
Following Refs. \cite{solda,abw}, we get, after double iteration and elimination of the auxiliary vector soldering fields, the following effective action
\begin{equation}
S_{eff} = S_{+}^{(0)} \;\;+\;\;S_{-}^{(0)} \;\;+\;\; \frac{{\cal J}^2_{\alpha}}{8} \;\;+\;\; \frac{{\cal J}^2_{\beta}}{8\,\eta} 
\end{equation}
where
\begin{eqnarray}
\eta & = & \frac{\eta_- \;\;+\;\; \eta_+}{2} \nonumber \\
{\cal J}_{\alpha} & = &  {\cal J}_{\alpha}^{1(+)}+{\cal J}_{\alpha}^{1(-)} \nonumber\\
{\cal J}_{\beta} & = &  {\cal J}_{\beta}^{1(+)}+{\cal J}_{\beta}^{1(-)}\;\;.
\end{eqnarray}
After a tedious algebraic manipulation, we see that the final form of the action is independent of the individual fields, depending only on their gauge invariant combinations $\Phi$ and $\Psi$,
\begin{eqnarray}
S_{eff} & = &<\left\{ \frac{1}{2}\,\partial_{\mu}\,\Phi\,\partial^{\mu}\,\Phi+2\,\epsilon^{\mu \nu}\,A_{\mu}\,\partial_{\nu}\,\Phi 
+ \eta_{0}\,A_{\mu}^2\right\}> \nonumber \\
& + & <\left\{\frac{1}{2\,\eta}\,\dot{\Psi}^{2}-\frac{\eta_-\,\eta_+}{2\,\eta}\,{\Psi'}^2+\,\eta_{1}\dot{\Psi}\,\Psi'\right\}>,
\end{eqnarray}
where 
$$\eta_{0}= \frac{1}{2}\left(a+b-2\right),\:\: \eta_{1} = \frac{\eta_{+} -\eta_-}{2\,\eta}$$

\ni and the new collective fields are $\Phi = \varphi - \varrho$ and $\Psi =\sigma - \omega$.

In the case where $a=b=1$, the first part of this action leads to the Schwinger model.  The combination of the massless modes led to a massive vectorial mode as a consequence of the chiral interference. The noton action was shown to propagate neither to the left nor to the right \cite{hull}. This action has the
same form as the one defined by Hull to cancel the  Siegel anomaly.  Noticeably, this noton has not coupled to the gauge field of the theory.

The separation of the symmetry sectors (notons) followed by soldering has led to another new and interesting result.  The dynamical chiral sectors interfered to produce a vector massive mode \cite{abw}. The Siegel notons, on the other hand, led to a Hull noton.  However, while each component carry a representation of the Siegel algebra, the soldered (Hull) noton carries a representation of the full diffeomorphism group.  It is worth to stress that Siegel invariance is not a diffeomorphism sub-group so that the soldering gives more than the mere direct sum of the chiral algebras \cite{dgr}.

%\section{Conclusions}

The chiral bosonization process is not free of ambiguities.  This is different from the vectorial case, where the presence of the gauge symmetry leads to an exact and unambiguous result.  In particular, it is well known that the bosonization dictionary,
where  
$$\bar\psi i\partial\!\!\!/\psi\rightarrow\partial_+\phi\partial_-\phi$$
and  
$$\bar\psi\gamma_\mu \psi\rightarrow\frac{1}{\sqrt{\pi}}\epsilon_{\mu\nu}\partial^\nu \phi$$  

\ni cannot be applied in the chiral case.  This ambiguity has been characterized by Jackiw and Rajaraman \cite{jr} with an arbitrary mass parameter regulator.  To pass from the vectorial to the chiral case one has to destroy the right-left interference.  As seen in \cite{abw} the interference contributes to the mass term.  Therefore its absence
leaves the mass coefficient arbitrary.  This stands at the origin of the Jackiw-Rajaraman effect.

The arbitrariness in the process of chiral bosonization has led to different bosonization schemes \cite{siegel,fj,many}.  In this section we compared the two most successful proposals \cite{siegel,fj}, and showed that they are related by the presence of a noton.  We showed that the dual projection diagonalizes Siegel action into a dynamical and a symmetry carrying parts.  This gives a deeper insight in the composition of the Siegel mode.  We have also analyzed the quantum contents of the noton sector and showed that although it is a nonmover field classically, it acquires dynamics at the quantum level thanks to the gravitational
anomaly.

In the soldering formalism context the dual projection was important to clarify the reason why gauged rightons and leftons fail to produce a constructive interference pattern. Separating the symmetry carrying notons before soldering, a double constructive interference was made possible.  This process led to a massive vector mode in the dynamical sector.
The symmetry sector is described by a Hull noton, carrying the representation of the full diffeomorphism group.
It was the soldering that allowed for the construction of the full group in terms of its chiral parts by incorporating the interference term. The physical picture of the interference pattern disclosed here will be detailed discussed next, both in the Abelian as well as in the non-Abelian context.

%%%%%%%%%%%%%%%%%%%%%%%%%%%%%%%%%%%%%%%%%%%%%%%%%%%%%%%%%%%%%%%%%%%%%%%%%%%%%%%%%%%%%%%%%%%%%%%

\subsection{Bose Symmetry and Chiral Decomposition of 2D Fermionic Determinants}

That the decomposition of the two dimensional 
fermionic determinant poses an obstruction to gauge invariance is a recurrent point of analysis in the literature of the subject \cite{AAR}.  
In this section we discuss several aspects of this decomposition.  Moreover, contrary 
to the usual route, the inverse approach, whereby two chiral components 
are fused or soldered, is also examined in full details. A close correspondence 
between the splitting and the soldering processes is established.  
By following Bose symmetry it is possible to give explicit expressions for 
the chiral determinants which show, in both these procedures, that there is
no incompatibility with gauge invariance at the quantum level.
Two important consequences emerging from this analysis are the close
connection between Bose symmetry and gauge invariance, 
and a novel interpretation of the Polyakov-Wiegman identity \cite{pw}.  

The understanding the properties of
$2D$-fermionic determinants and the associated role of Bose symmetry is crucial 
because of several aspects. For instance, the precise form of the one cocycle 
necessary in the recent discussions on smooth functional bosonization
\cite{dns,dn} is only dictated by this symmetry \cite{RB}. 
Furthermore this cocycle, which is just the $2D$ anomaly, is known to be the origin of anomalies in higher dimensions by a set
of descent equations \cite{jackiw}.
%Incidentally, the anomaly phenomenon still defies a complete explanation.

To briefly recapitulate the problem of chiral decomposition,
 consider the vacuum functional,
\begin{eqnarray}
\label{10}
e^{iW[A]} &=& \int d\bar\psi \; d\psi \exp\{i \int d^2 x \:
\bar\psi(i \dslash + e \aslash )\psi\}\nonumber\\
&=&\det(i\dslash +e \aslash)
\end{eqnarray}

\noindent where the expression for the determinant follows immediately by
imposing gauge invariance,
\begin{equation}
\label{20}
W[A] = N\int d^2x\: A_\mu \Pi^{\mu\nu}A_\nu
\end{equation}

\noindent with 
$$\Pi^{\mu\nu}=g^{\mu\nu} - 
{\partial^\mu\partial^\nu\over \Box}\;\; ;\; \mu ,\nu=0,1\;$$

\ni being the transverse projector.  An explicit one loop calculation 
yields \cite{JS} 
$$N={e^2\over 2\pi}\;.$$

\ni Using the light-cone variables defined in (\ref{lightfront}), with the projector matrix given by,
\begin{equation}
\label{40}
\Pi^{\mu\nu}= {1\over 2}\left( 
\begin{array}{cc}
{-{\partial_-\over \partial_+}} & 1 \\
1 & {-{\partial_-\over \partial_+}} 
\end{array}
\right)
\end{equation}

\noindent it is simple to rewrite (\ref{20}) as,
\begin{equation}
\label{50}
W[A_+,A_-]=  -{ N\over 2} \int d^2x\: \{A_+ {\partial_-\over \partial_+}A_+ 
+ A_- {\partial_+\over \partial_-}A_- - 2 A_+ A_-\}
\end{equation} 

\noindent The factorization of (\ref{10}) into its chiral components yields,
\begin{equation}
\label{60}
\det(i\dslash +e \aslash)=\det(i\dslash +e \aslash_+)\det(i\dslash +e 
\aslash_-)
\end{equation}
where 
$${\aslash}_\pm ={\aslash} P_{\pm}$$ 
with the chiral projector defined as 
$$P_\pm ={1\pm\gamma_5\over 2}\;.$$

\ni The effective action for the vector theory in terms of the chiral
components is now obtained from (\ref{60}), leading to an effective action,
\begin{eqnarray}
\label{70}
W_{eff}&=&  -{ N\over 2} \int d^2x\: \{A_+ {\partial_-\over 
\partial_+}A_+ + A_- {\partial_+\over \partial_-}A_- \}\nonumber\\
& =& W[A_+,0]+W[0,A_-]
\end{eqnarray}

\noindent which does not reproduce the expected gauge invariant
result (\ref{50}).  The above factorization is therefore regarded
as an obstruction to gauge invariance.

It is important to notice that (\ref{70}) follows from (\ref{60}) only
if one naively computes the chiral determinants from the usual vector
case (\ref{50}) by substituting either $A_+=0$ or $A_-=0$.  This may
be expected naturally since $\det(i\dslash +e \aslash)=\det(i\dslash 
+e \aslash_+ + e \aslash_-)$.  But the point is that whereas the usual
Dirac operator has a well defined eigenvalue problem,
\begin{equation}
\label{80}
\Dslash \:\psi_n =(i\dslash + e\aslash)\psi_n=\lambda_n\psi_n
\end{equation}

\noindent with the determinant being defined 
by the product of its eigenvalues, this is not true for the chiral pieces 
in the RHS of (\ref{60}), which lacks a definite eigenvalue equation \cite
{GW, RB} because the kernels map from one chiral sector to the other,
\begin{equation}
\label{901}
\Dslash_\pm \psi_\pm = \lambda\psi_\mp
\end{equation}

\noindent with $\psi_\pm = P_\pm\psi $.  Consequently, it is not possible
to interpret, however loosely or naively, any expression obtainable from
$\det\Dslash$ by setting $A_\pm=0$, as characterizing $\det{\Dslash}_\mp$.

Since $\det{\Dslash}_\pm$ are not to be
regarded as $W[A_+,0]$ or $W[0,A_-]$ in (\ref{70}),
it is instructive to clarify the meaning of the latter expressions. 
Reconsidering $\det\Dslash$ as $\det(i\dslash +e \aslash_+ + e \aslash_-)$
it is easy to observe that the fundamental fermion loop decomposes into four pieces \cite{adw}.
%{\bf(see figure)}.

At the unregularized level there are different choices of 
interpreting these diagrams, depending on the location of the chiral 
projectors $P_\pm$.  In particular, by pushing one of these projectors 
through the loop and inserting it at the other vertex would yield vanishing 
contributions for the last two diagrams, since $P_+ P_- =0$.  It was shown 
earlier by one of us \cite{RB}, in a different context, that Bose symmetry
provided a 
definite guideline in manipulating such diagrams.  In other words, the 
position of the projectors is to be preserved exactly as appearing above, 
and the contributions explicitly computed from (\ref{20}) by appropriate 
replacements at the vertices.  This procedure implies a consistent way of
regularizing all four graphs.  Thus,
\begin{eqnarray}
\label{100}
W_{1(2)}&=& N\int d^2x\: A_\mu {\cal P}^{\mu\nu}_{+(-)}\Pi_{\nu\alpha}
{\cal P}^{\alpha\beta}_{-(+)}A_\beta\nonumber\\
W_{3(4)}&=& N\int d^2x\: A_\mu {\cal P}^{\mu\nu}_{+(-)}\Pi_{\nu\alpha}
{\cal P}^{\alpha\beta}_{+(-)}A_\beta
\end{eqnarray}

\noindent where
\begin{equation}
\label{1101}
{\cal P}^{\alpha\beta}_{+(-)}={1\over 2}
(g^{\alpha\beta}\pm\epsilon^{\alpha\beta}) \;\;\; ;\epsilon^{+-}=
\epsilon_{-+}=1
\end{equation}

\noindent Using (\ref{40}) it is easy to simplify (\ref{100}) as,
\begin{equation}
\label{120}
W_1=W[A_+,0]\;\; ,\;\; W_2=W[0,A_-]\;\; ,\;\; W_3=W_4={N\over 2}A_+A_-
\end{equation}

\noindent Adding all four terms exactly reproduces the
gauge invariant result (\ref{50}).  If, on the contrary,
Bose symmetry was spoilt in the last two graphs as indicated
earlier so that $W_3=W_4=0$, the gauge noninvariant structure
(\ref{70}) is obtained.  This shows the close connection between
Bose symmetry and gauge invariance.  Recall that the same is also
true in obtaining the ABJ anomaly from the triangle graph \cite{R,adler}. 
Furthermore $W[A_+,0]$ and $W[0,A_-]$ are now seen to correspond to
graphs $W_1$ and $W_2$, respectively, evaluated in a very specific
fashion.  It is also evident that the incorrect manner of abstracting
$\det(i\dslash +e \aslash_{\pm})$ from $\det(i\dslash +e \aslash_)$
violates Bose symmetry leading to an apparent contradiction between
chiral factorization and gauge invariance.  Consequently the possibility
of ironing out this contradiction exists by interpreting the chiral
determinants as,
\begin{eqnarray}
\label{130}
-i\ln \det(i\dslash +e \aslash_+)&=& W[A_+,0]+{N\over 2}\int d^2x\: A_+ A_-
\nonumber\\
-i\ln \det(i\dslash +e \aslash_-)&=& W[0,A_-] +{N\over 2}\int d^2x\: A_+ A_-
\end{eqnarray}

\noindent These expressions just reduce to the naive definitions if
the crossing graphs are ignored or, equivalently, Bose symmetry is violated.

\subsubsection{The soldering of the chiral Schwinger model}

To put (\ref{130}) on a solid basis it must be recalled that (\ref{60}), 
as it stands, is only a formal identity.  A definite meaning can be attached 
provided some regularization is invoked to explicitly define the determinants
 appearing on either side of the equation.  Using a regularization that 
preserves the vector gauge symmetry of the LHS of (\ref{60}) led to the 
expression (\ref{50}).  As is well known \cite{jackiw,jr} there is no 
regularization that retains the chiral symmetry of the pieces in the 
RHS of (\ref{60}).  An explicit one loop computation yields \cite{jackiw,RB1}, 
in a bosonized language,
\begin{eqnarray}
\label{140}
W_+[\varphi] &=& {1\over{4\pi}}\int d^2x\,\left(\partial_+
\varphi\partial_-\varphi +2 \, e\,A_+\partial_-\varphi + a\, 
e^2\, A_+ A_-\right)\nonumber\\
W_-[\rho]&=& {1\over{4\pi}}\int d^2x\,\left(\partial_+\rho\partial_-
\rho +2 \,e\, A_-\partial_+\rho
+ b\, e^2\, A_+ A_-\right)
\end{eqnarray}

\noindent where $a$ and $b$ are parameters manifesting regularization,
or equivalently, bosonization ambiguities.  It is simple to verify that
a straightforward application of the usual bosonization rules
%: $$\bar\psi i\dslash\psi \rightarrow \partial_+\varphi\partial_-\varphi$$ \ni and $$\bar\psi\gamma_\mu\psi\rightarrow{1\over\sqrt \pi}\epsilon_{\mu\nu} \partial^\nu\varphi\;,$$ 
%\ni which are valid {\it only} when the vector gauge symmetry is preserved,
would just reproduce (\ref{140}) with $a=b=0$.
Subsequently, by functionally integrating out the scalar fields $\varphi$
and $\rho$, exactly yields the two pieces $W[A_+,0]$ and $W[0,A_-]$ given
in (\ref{70}), which is what one obtains by simply putting $A_\pm=0$
directly into the expressions for the vector determinant.  This
reconfirms the invalidity of identifying the chiral determinants
by naively using rules valid for the vector case.

We now show precisely how two independent chiral components (\ref{140}) 
are soldered to yield the LHS of (\ref{60}).  Let us then
consider the gauging of the following global symmetry of (\ref{140})
\begin{eqnarray}
\label{150}
\delta \varphi &=& \delta\rho\;=\;\alpha\nonumber\\
\delta A_{\pm}&=& 0
\end{eqnarray}

\noindent Then it is found from (\ref{140}) that
\begin{eqnarray}
\label{160}
\delta W_+[\varphi] &=& \int d^2x\, \partial_-\alpha \;J_+
(\varphi)\nonumber\\
\delta W_-[\rho]&=& \int d^2x\, \partial_+\alpha \;J_-(\rho)
\end{eqnarray}

\noindent  where,
\begin{equation}
\label{170}
J_\pm(\eta)={1\over{2\pi}}(\partial_\pm\eta +\, e\,A_\pm)\;\;\; ; \;\;\eta=
\varphi , \rho
\end{equation}

\noindent  Next, introduce the soldering field $B_\pm$ so that,
\begin{equation}
\label{180}
W_\pm^{(1)}[\eta] = W_\pm[\eta] -\int d^2x\, B_\mp\, J_\pm(\eta)
\end{equation}

\noindent Then it is easy to verify that the modified action,
\begin{equation}
\label{190}
W[\varphi,\rho]= W_+^{(1)}[\varphi] + W_-^{(1)}[\rho]
 + {1\over{2\pi}} \int d^2x \, B_+ \,B_-
\end{equation}

\noindent is invariant under an extended set of transformations that 
includes (\ref{150}) together with,
\begin{equation}
\delta B_{\pm}= \partial_{\pm}\alpha
\label{191}
\end{equation}

\noindent  Using the equations of motion, the auxiliary soldering
field can be eliminated in favor of the other variables, 
\begin{equation}
\label{200}
B_\pm= 2\pi J_\pm
\end{equation}

\noindent so that the soldered effective action derived from (\ref{190}) reads,
\begin{equation}
\label{210}
W[\Phi]={1\over {4\pi}}\int d^2x\:\Big{\{}\Big{(}\partial_+
\Phi\partial_-\Phi + 2\,e\, A_+\partial_-\Phi - 2\,e\, A_-
\partial_+\Phi\Big{)} +(a+b-2)\,e^2\,A_+\,A_-\Big{\}}
\end{equation}

\noindent where,
\begin{equation}
\label{220}
\Phi=\varphi - \rho
\end{equation}

\noindent We may now examine the variation of (\ref{210}) under the lifted gauge transformations,
$$\delta\varphi=\delta\rho=\alpha$$ 

\ni and 
$$\delta A_\pm = \partial_\pm\alpha\;,$$

\ni induced by the soldering process. Note that this is just the usual
gauge transformation.  It is easy to see that the expression
in parenthesis (\ref{210}) in is gauge invariant, and by functionally
integrating out the $\Phi$ field one 
verifies that it reproduces (\ref{50}).  Thus,
the soldering process leads to a gauge invariant
structure for $W$ provided
\begin{equation}
\label{230}
a+b-2=0\,\,.
\end{equation}

\noindent It might appear that there is a whole one parameter class of
solutions.  However Bose symmetry imposes a crucial restriction.  Recall
that in the Feynman graph language this symmetry was an essential
ingredient in preserving compatibility between gauge invariance and
chiral decomposition.  In the soldering process, this symmetry, which
is just the left-right (or $+\, -$) symmetry in (\ref{140}), is preserved
with $a=b$.  Coupled with (\ref{230}) this fixes the parameters to unity
and proves our assertion announced in (\ref{130}).  It may be
observed that the soldering process can be carried through for the
non-Abelian theory as well, and a relation analogous to (\ref{210}) will be 
obtained later.

An alternative way of understanding the fixing of parameters is to recall 
that if a Maxwell term is included in (\ref{130}) to impart dynamics, then 
this corresponds to the chiral Schwinger model \cite{jr}.  It was 
shown that unitarity is violated unless $a$ (or $b$)$\geq 1$.  
Imposing (\ref{230}) immediately yields $a=b=1$ as the only valid answer,
showing that the bound 
gets saturated.  It is therefore interesting to note that
(\ref{230}) together with unitarity leads naturally to a Bose
symmetric parametrization. 
In other words, the chiral Schwinger model may have any 
$a\geq 1$, but if two such models with opposite chiralities are soldered to 
yield the vector Schwinger model, then the minimal bound is the unique
choice.  Interestingly, the case $a=1$ implies a massless mode in the
chiral Schwinger model.
The soldering mechanism therefore generates the massive mode of the 
Schwinger model from a fusion of the massless modes in the chiral Schwinger 
models.  

We have therefore explicitly derived expressions for the chiral determinants
(\ref{130}) 
which simultaneously preserve the factorization property (\ref{60}) and 
gauge invariance of the vector determinant.  It was also perceived that the 
naive way of interpreting the chiral determinants as $W[A_+,0]$ or 
$W[0,A_-]$ led to the supposed incompatibility of factorization with gauge 
invariance since it missed the crossing graphs.  Classically 
these graphs do vanish ($P_+ P_- =0$) so that it becomes evident that this 
incompatibility originates from a lack of properly accounting for the quantum 
effects.  It is possible to interpret this effect, as we will now show,
as a typical quantum mechanical 
interference phenomenon, closely paralleling the analysis in Young's double 
slit experiment.  As a bonus, we provide a new interpretation for 
the Polyakov-Wiegman \cite{pw}
identity. Rewriting (\ref{50}) in Fourier space as
\begin{eqnarray}
\label{240}
W[A_+,A_-] & = & -{N\over 2}\int d^2k\, \{A^*_+(k){k_-\over k_+}A_+(k)
+ A^*_-(k)
{k_+\over k_-}A_-(k) - 2A^*_+(k)A_-(k)\} \nonumber\\
&=& -{N\over 2}\int d^2k\, \mid\sqrt{k_-\over k_+}A_+(k) -
\sqrt{k_+\over k_-}A_-(k)
\mid^2
\end{eqnarray}

\noindent immediately displays the typical quantum mechanical interference 
phenomenon, in close analogy to the optical example,
\begin{equation}
\label{250}
W[A_+,A_-]= -{N\over 2}\int d^2k\,\left(\mid\psi_+(k)\mid^2
+\mid\psi_-(k)\mid^2 
+ 2\cos\theta \psi^*_+(k)\psi_-(k)\right)
\end{equation}

\noindent with $\psi_\pm (k)=\sqrt{k_\mp\over k_\pm}A_\pm(k)$ 
and $\theta=\pm\pi$, simulating the roles of the amplitude and the phase,
respectively.  Note that in one space dimension, these are 
the only possible values for the phase angle $\theta$ between the 
left and the right movers. The dynamically  
generated mass arises from the interference 
between these movers, thereby preserving gauge invariance. Setting either 
$A_+$ or $A_-$ to vanish, destroys 
the quantum effect, very much like closing one slit in the optical 
experiment destroys the interference pattern.   
Although this analysis was done for the Abelian theory, it is
straightforward to perceive that the effective action for a
non-Abelian theory can also be expressed in the form of an absolute
square (\ref{240}), except that there will be a repetition of
copies depending on the group index.
This happens because only the two-legs graph has
an ultraviolet divergence, leading to the interference (mass) term.
The higher legs graphs are all finite, and satisfy the naive
factorization property.

It is now simple to see that (\ref{240}) represents an abelianized
version of the Polyakov 
Wiegman identity by making a familiar change of variables, 
\begin{eqnarray}
\label{260}
A_+&=&{i\over e}U^{-1}\partial_+ U\nonumber\\
A_-&=&{i\over e}V\partial_- V^{-1}
\end{eqnarray}

\noindent where, in the Abelian case, the matrices $U$ and $V$ are given as,
\begin{equation}
\label{270}
U=\exp\{i\varphi\}\;\;\;\; ; \;\;\;\; V=\exp\{-i\rho\}\;\;\;\; ; \;\;\;\; 
UV =\exp\{i\Phi\}
\end{equation}

\noindent  with $\Phi$ being the gauge invariant soldered field
introduced in (\ref{220}).  It is possible to recast (\ref{240}),
in the coordinate space,  as
\begin{equation}
\label{280}
W[UV]=W[U] + W[V] + {1\over{2\pi}}\int d^2x\, \left(U^{-1}\partial_+ U\right)
\left(V\partial_- V^{-1}\right)
\end{equation}

\noindent which is the Polyakov-Wiegman identity, satisfying gauge invariance.
The result can be extended to the non-Abelian case since, as already mentioned,
the nontrivial interference term originates from the two-legs graph which
has been taken into account. It is now relevant to point out that the important crossing piece in
either (\ref{240}) or (\ref{280}) is conventionally \cite{pw,AAR}
interpreted as a contact (mass) term, or a counterterm, necessary to
restore gauge invariance. In our analysis, on the contrary, this term
was uniquely specified from the interference between the left and right
movers in one space dimension, automatically providing gauge invariance.
This is an important point of distinction.

To conclude, our analysis clearly revealed that no obstruction
to gauge invariance is posed by the chiral decomposition of the 2D
fermionic determinant. 
The claimed obstruction actually results from an incorrect interpretation 
of the chiral determinants. Bose symmetry gave a precise way of making sense 
of these determinants which were explicitly computed by considering the dual 
descriptions of decomposition
as well as soldering. The close interplay between Bose symmetry and
gauge invariance was illustrated in both these ways of looking at the
fermionic determinant. At the dynamical level it was also shown how this
symmetry is instrumental in fusing the massless modes of the left and right
chiral Schwinger models to yield
the single massive mode of the vector Schwinger model. Indeed it was
explicitly shown that this mass generation is the quantum interference effect
between the two chiralities, closely resembling the corresponding effect
in the double slit optical experiment. This led us to provide a novel
interpretation of the Polyakov-Wiegman identity.

Our analysis indicated that the $a=1$ regularization for
the determinant of the
Chiral Schwinger Model was important leading to interesting effects. 
This parametrization was also found to be useful in a different context
\cite{W}. On the other hand 
much of the usual analyzes is confined to the 
$a=2$ sector \cite{L}.

Finally, to put this study in a proper perspective it may be useful to once
again remind the importance of Bose symmetry. It is an essential
ingredient in getting the classic ABJ anomaly from the triangle
graph \cite{R,adler}. Just imposing gauge invariance on the
vector vertices does not yield the cherished result. Bose symmetry coupled
with gauge invariance does the job. This symmetry also played a crucial
role in providing a unique structure for the 1-cocycle that is mandatory
for smooth bosonization \cite{RB,dns}.  It is therefore not surprising that 
Bose symmetry provided the definite guideline in preserving the compatibility
between gauge invariance and chiral decomposition or soldering.

\subsubsection{Extension to the non-Abelian case}
 
Here we explicitly show the soldering mechanism in the non-Abelian context.
The expressions for the chiral determinants analogous to (\ref{140}) are
given by \cite{LR},
\begin{eqnarray}
W_+[g]&=& I_{wzw}^{(-)}[g] -\frac{ie}{2\pi}\int d^2x 
tr(A_+g^{-1}\partial_-g)-\frac{e^2a}{4\pi}\int d^2x tr(A_+A_-)\nonumber\\
W_-[h]&=& I_{wzw}^{(+)}[h] -\frac{ie}{2\pi}\int d^2x 
tr(A_-h^{-1}\partial_+h)-\frac{e^2b}{4\pi}\int d^2x tr(A_+A_-)
\label{a10}
\end{eqnarray}
where the Wess-Zumino-Witten functional at the critical point $(n=\pm 1)$
is given by (for details and the original papers, see
\cite{AAR}),
\begin{equation}
I_{wzw}^\pm[k]=\frac{1}{4\pi}\int d^2x tr(\partial_+k\partial_-k^{-1})
\mp\frac{1}{12\pi}\Gamma_{wz}[k]\,\,\,\,\, ;k=g, h
\label{a11}
\end{equation}
with the familiar Wess-Zumino term defined over a 3D manifold with the 
two-dimensional Minkowski space-time as its boundary,
\begin{equation}
\Gamma_{wz}[k]=\int d^3x \epsilon^{lmn}\,tr(k^{-1}\partial_l k\, 
k^{-1}\partial_m k\, k^{-1}\partial_n k)
\label{a12}
\end{equation}
In the above equations $g$ and $h$ are the elements of some compact Lie group
and the parameters $a$ and $b$ manifest the regularization or bosonization
ambiguities. Let us next consider the gauging of the global right and left
chiral symmetries analogous
to (\ref{150}),
\begin{eqnarray}
\delta g&=&\omega g\nonumber\\
\delta h&=&h\omega\nonumber\\
\delta A_{\pm}&=&0
\label{a13}
\end{eqnarray}
where $\omega$ is an infinitesimal element of the algebra of the corresponding
group. Note the order of $\omega$ which occurs once from the left and once
from the right to properly account for the two chiralities. In the Abelian
example, this just commutes and the ordering is unimportant leading to a
unique transformation in (\ref{150}). Under (\ref{a13}), the relevant
variations are found to be,
\begin{eqnarray}
\delta W_+[g]&=&\int d^2x tr(\partial_-\omega J_+(g))\nonumber\\
\delta W_-[h]&=&\int d^2x tr(\partial_+\omega J_-(h))
\label{a14}
\end{eqnarray}
where,
\begin{equation}
J_\pm=\frac{-1}{2\pi}\Big(\partial_\pm kk^{-1}+iekA_\pm k^{-1}\Big)
\label{a15}
\end{equation}
Now introduce the soldering field $B_\pm$ which transforms as,
\begin{equation}
\delta B_\pm=\partial_\pm\omega -[B_\pm, \omega]
\label{a16}
\end{equation}
whose Abelian version just corresponds to (\ref{191}). Then it may be
checked that the following effective action,
\begin{equation}
W[g, h]=W_+[g]+W_-[h] -\int d^2x tr\Big(B_-J_+(g)+B_+J_-(h)
+\frac{1}{2\pi}B_+B_-\Big)
\label{a17}
\end{equation}
is invariant under the complete set of transformations. The auxiliary 
soldering field is eliminated, as usual, in favor of the other variables,
by using the equations of motion,
\begin{equation}
B_\pm=-2\pi J_\pm(k)
\label{a18}
\end{equation}
The soldered effective action directly follows from (\ref{a17}) on substituting
this solution,
\begin{eqnarray}
W[G]&=&I_{wzw}^+[G]+\frac{ie}{2\pi}\int d^2xtr\Big(A_-G^{-1}\partial_+G
-A_+\partial_-GG^{-1}\Big)\nonumber\\
&-&\frac{e^2}{2\pi}\int d^2x tr\Big(A_+GA_-G^{-1}-\frac{a+b}{2}A_+A_-\Big)
\label{a19}
\end{eqnarray}
where $G=g^{-1}h$. Once again gauge invariance under the conventional set
of transformations in which $A_\pm$ changes as a potential, is recovered
only if $a+b=2$, exactly as happened in the Abelian case. Imposing Bose
symmetry leads to the unique choice $a=b=1$, completely determining the
structure for the separate chiral components. 
Incidentally, by including the Yang-Mills term to impart dynamics so that these
models become chiral $QCD_2$, it was found that unitarity could be preserved
only for $a, b\geq 1$ \cite{LR}. Coupled 
with the above noted restriction, this leads
to the Bose symmetric parametrization. 
It is easy to see that
(\ref{a19}) reduces to the Abelian result (\ref{210}) by
setting $G=\exp(i\Phi)$. Observe that the soldering
was done among the chiral components having opposite critical points. Any
other combination would fail to reproduce the gauge invariant result. 
Indeed the gauge invariant effective action, being a functional of 
$G=g^{-1}h$, can be obtained
by soldering effective actions (which are functionals of $g$ and $h$)
with opposite criticalities since changing $g\rightarrow g^{-1}$ converts
the Wess-Zumino-Witten functional from one criticality to the other.
The relevance of opposite criticality was also noted in another context
involving smooth non-Abelian bosonization \cite{dn}.

This non-Abelian exercise, however, clearly reveals 
that the physics of the problem
of chiral soldering (or decomposition) and the role of Bose symmetry
is contained in the Abelian sector. The rest is a matter of 
technical detail. Indeed, following similar steps, it is also
possible to discuss chiral decomposition for the non-Abelian case and obtain
identical conclusions.

Now we will use the soldering formalism in many other chiral boson's formulations.

%%%%%%%%%%%%%%%%%%%%%%%%%%%%%%%%%%%%%%%%%%%%%%%%%%%%%%%%%%%%%%%%%%%%%%%%%%%%%%%%%%%%%%%%%%%%%%%%%%%%%%%%%%%%%%%%%%%%%%%%%%%%%%%%%%%%%%%%%%%%%%%%%%%%%%%%%%%

\subsection{The Master Action}

In this section we will discuss a master action which represents, as a function of arbitrary parameters, several theories for the Siegel gauged model. In the sequel we go on to discuss the soldering of opposite chiral versions of this
master action and applied the final result, i.e., the 
soldered action, on several models for the self-dual theory to make an interference analysis
of the covariance of the new theories. This will reveal the full power of the formalism in dealing with the fusion of chiral modes in different contexts.

\subsubsection{The Generalized Gauged Siegel Model}

Let us now construct a class of generalized actions for Abelian chiral bosons coupled to a gauge field for each chirality, i.e., for the coupled leftons (${\cal L}_L$) and rightons (${\cal L}_R$). We will call it the
Generalized Gauged Siegel Model (GGSM) \cite{ad}, 
%\begin{mathletters}
\begin{eqnarray}
{\cal L}^{(0)}_{L}\,&=&\,(\,\partial_+\,\phi\,+\,a_1\,A_+\,)\,(\,\partial_-\,%
\phi\,+\,a_2\,A_-\,)\,+\, \lambda_{++}\,(\,\partial_-\,\phi\,+\,a_3\,A_-\,)^2
\label{eqa}
\end{eqnarray}
\begin{eqnarray}
{\cal L}^{(0)}_{R}\,&=&\,(\,\partial_+\,\rho\,+\,b_1\,A_+\,)\,(\,\partial_-\,%
\rho\,+\,b_2\,A_-\,)\,+\,\lambda_{--}\,(\,\partial_+\,\rho\,+\,b_3\,A_+\,)^2 \;\;,  
\label{eqb}
\end{eqnarray}
%\end{mathletters}
where $a_i,\,b_i\,(i=1,2,3)$ are parameters that define the theory studied
and $A_\pm$ are the vector field components. We will see below that making simple
substitutions of these parameters we can obtain several gauged forms of the
Siegel theory that appear in the literature.  It is important to observe the 
difference between the vector fields $A_\pm$ above and the soldering fields $B_\pm$ of 
equations (\ref{121}).  The $A$-fields are external (or background) fields and hence one does not consider the variation (and extrema) of the actions under the variations of these fields.
The last are the auxiliary fields, as mentioned above, which helps 
in the soldering process and will be naturally eliminated by solving its equations of motion.

Following the steps of the soldering formalism studied in the last section,
we can start considering the variation of the Lagrangians under the usual
transformations, 
\begin{equation}  \label{31}
\delta\,\phi\,=\,\delta\,\rho\,=\,\alpha \;\; \mbox{and} \;\;
\delta\,A_{\mu}\,=\,\partial_{\mu}\,\alpha
\end{equation}
where $\mu=+,-$. Now, consider that this symmetry is a global one with, obviously, a global parameter $\alpha$ so that the above transformations take the form
\begin{equation}  \label{32}
\delta\,\phi\,=\,\delta\,\rho\,=\,\alpha \;\; \mbox{and} \;\;
\delta\,A_{\mu}\,=\,0\;\;.
\end{equation}
Remember that the soldering process consists in lifting the gauging of a global symmetry to its local version.  Hence we will consider from now on the transformations (\ref{32}) as local.
Let us continue with the procedure writing only the main steps of the procedure.

In terms of the Noether currents we can construct 
\begin{equation}
\delta {\cal L}^{(0)}_{L,R}\,=\,J^{\mu}_{\phi,\rho}\,\partial_{\mu}\,\alpha%
\;\;,
\end{equation}
where 
\begin{eqnarray}
J^{+}_{\phi}&=&a_2\,A_-  \nonumber \\
J^{-}_{\phi}&=&2\,\partial_+\,\phi\,+\,a_1\,A_+\,+\,2\,\lambda_{++}\,(\,%
\partial_-\,\phi\,+\,a_3\,A_-\,)  \nonumber \\
J^{+}_{\rho}&=&b_2\,A_-  \nonumber \\
J^{-}_{\rho}&=&2\,\partial_-\,\rho\,+\,b_2\,A_-\,+\,2\,\lambda_{--}\,(\,%
\partial_+\,\rho\,+\,b_3\,A_+\,) \;\;.
\end{eqnarray}

\noindent The next iteration, as seen above, can be performed introducing
auxiliary fields, the so-called soldering fields 
\begin{equation}
{\cal L}^{(1)}_{L,R}\,=\,{\cal L}^{(0)}_{L,R}\,-\,B_{\mu}\,J^{\mu}_{\phi,%
\rho}\;\;,
\end{equation}

\noindent and one can easily see that the gauge variation of the GGSM is
\begin{equation}  \label{34}
\delta {\cal L}^{(1)}_{L,R}\,=\,-\,B_{\mp}\,\delta\,B_{\pm}\,-\,\lambda_{\pm%
\pm}\,\delta\, B^2_{\mp}\;\;.
\end{equation}
Let us define the variation of $B_{\pm}$ as 
\begin{equation} \label{341}
\delta B_{\pm}=\partial_{\pm}\alpha\;\;,
\end{equation}
and we see that the variation of ${\cal L}^{(1)}_{L,R}$ does not depend
neither on $\phi$ nor $\rho$. Hence, as explained in the last section, we
can construct the final (soldered) Lagrangian as 
\begin{eqnarray}  \label{351}
{\cal L}&=&\,{\cal L}_{L}\,\oplus\,{\cal L}_{R}  \nonumber \\
&=&{\cal L}^{(1)}_{L}\,+\,{\cal L}^{(1)}_{R}\,+\,2\,B_+\,B_-\,+
\,\lambda_{++}\,B^2_-\, +\,\lambda_{--}\,B^2_+  \nonumber \\
&=&\,(\,\partial_+\,\phi\,+\,a_1\,A_+\,)\,(\,\partial_-\,\phi\,+\,a_2\,A_-%
\,)\,+\, \lambda_{++}\,(\,\partial_-\,\phi\,+\,a_3\,A_-\,)^2  \nonumber  \\
&+&(\,\partial_+\,\rho\,+\,b_1\,A_+\,)\,(\,\partial_-\,\rho\,+\,b_2\,A_-\,)\,
+\, \lambda_{--}\,(\,\partial_+\,\rho\,+\,b_3\,A_+\,)^2  \\
&-&B_{\mu}\,J^{\mu}_{\phi}\,-\,
B_{\mu}\,J^{\mu}_{\rho}\,+\,2\,B_+\,B_-\,+\,\lambda_{++}\,B^2_-\,+\,\lambda_{--}\,B^2_+ \;\;, \nonumber                  
\end{eqnarray}
which remains invariant under the combined transformations (\ref{32}) and (%
\ref{341}). Following the steps of the algorithm depicted in the last
section, we have to eliminate the soldering fields solving their equations
of motion which results in 
\begin{equation}
B_{\pm}\,=\,\frac{J^{\mp}\,-\,\lambda_{\pm\pm}\,J^{\pm}}{2\,(1\,-\,\lambda)}%
\;\;,
\end{equation}
where $\lambda=\lambda_{++}\,\lambda_{--}$ and $J^{\pm}=J^{\pm}_{\phi}+J^{%
\pm}_{\rho}$.

Substituting it back in (\ref{351}) we have the final soldered action
\begin{eqnarray}  \label{371}
{\cal L}\,&=&\,{\frac{1 }{2}}\,\sqrt{-g}\,g^{\mu\nu}\,\partial_{\mu}\,\Phi\,%
\partial_{\nu}\,\Phi     \nonumber \\
&+&\,{\frac{1}{{1-\lambda}}}\left\{\,
(a_1\,+\,b_1\,\lambda\,-\,2\,\lambda\,b_3)\,\partial_-\,\Phi\,A_+\, \right. \nonumber \\
&+& \left. (\,2\,\lambda\,a_3\,-\,a_2\,\lambda\,-\,b_2)\,\partial_+\,\Phi\,A_-\, \right.
\nonumber \\
&+& \left.
\,\lambda_{++}\,(2\,a_3\,-\,a_2\,-\,b_2)\,\partial_-\,\Phi\,A_-\, \right. \nonumber \\
&+& \left. \lambda_{--}(a_1\,+\,b_1\,-\,2\,b_3)\,\partial_+\,\Phi\,A_+ \right. 
\nonumber \\
&+& \left.
\,C_1\,\lambda_{++}\,A^2_-\,+\,C_2\,\lambda_{--}\,A^2_+\,+\,C_{\lambda}%
\,A_+\,A_- \,\right\}
\end{eqnarray}
where the new compound field are defined as $\Phi=\phi\,-\,\rho$. The new
parameters are 
\begin{eqnarray}
C_1&=&a_3^2\,-\,b_2\,a_3\,+\,{\frac{1 }{4}}(a_2\,+\,b_2)^2  \nonumber \\
C_2&=&b_3^2\,-\,b_3\,a_1\,+\,{\frac{1 }{4}}(a_1\,+\,b_1)^2  \nonumber \\
C_{\lambda}&=&({\frac{1 }{2}}\,-\,\lambda)\,a_1\,a_2\,+\,({\frac{1 }{2}}%
\,-\,\lambda)\,b_1\,b_2\,-\, {\frac{1 }{2}}\,(a_1\,b_2\,+\,b_1\,a_2) 
\nonumber \\
&+&\,[(a_2\,+\,b_2)\,b_3\,+\,(a_1\,+\,b_1)\,a_3\,-\,
2\,a_3\,b_3]\,\lambda\, \nonumber \\
&-&\,a_2\,a_3\,\lambda_{++}\,-\,b_1\,b_3\,\lambda_{--}%
\;\;.
\end{eqnarray}
and the metric is 
\begin{equation}  \label{metrica}
{\frac{1 }{2}}\,\sqrt{-g}\,g^{\mu\nu} = {\frac{1 }{{2\,(1-\lambda)}}}%
\,\left( 
\begin{array}{cc}
2\lambda_{--} & 1+\lambda \\ 
1+\lambda & 2\lambda_{++}
\end{array}
\right)
\end{equation}
which reminds the gravitational feature of the soldered action of the two
Siegel modes. We can note that the action (\ref{371}) is covariant. Hence, in
this case, we have that the covariance of the generalized gauged Siegel
action is maintained.  This general action form will allow us to apply it to the
various gauged theories for the chiral boson with second order constraint.
This will be accomplished next.

%\subsubsection{The self-dual models}

Next we will analyze four kinds of theories in the light of the
soldering formalism. The first of them is the Siegel action 
\cite{siegel}, studied above. But now it will be used to demonstrate the
validity of the general soldered action (\ref{371}). The second example, 
which is not a new result also, will be a coupling of the chiral boson with a gauge
field. We are talking, in this case, about the Gates and Siegel gauged
action \cite{gs}. The new results will appear with the next three models. 
We will
use two models well known in the literature: the massless Bellucci, Golterman and Petcher model and the Frishman
and Sonnenschein model.

%\subsubsection{Siegel's model}

{\bf Siegel's model}: It is easy to see that to obtain the expression (\ref{03}) we have to fix
the parameters with the following values: 
\begin{equation}
a_{i}\,=\,b_{i}\,=\,0
\end{equation}
where $i=1,2,3$. Hence, substituting these values in the expression (\ref{371}%
) it follows that 
\begin{eqnarray} \label{lalala}
{\cal L}_{TOT}\, &=&\,{\frac{1}{{1-\lambda }}}\left\{ \left( 1+\lambda
_{++}\lambda _{--}\right) \partial _{-}\Phi \partial _{+}\Phi \,+\,\lambda
_{++}\left( \partial _{-}\Phi \right) ^{2}\,+\,\lambda _{--}\left( \partial
_{+}\Phi \right) ^{2}\right\}   \nonumber \\
&=&{\frac{1}{2}}\,\sqrt{-g}\,g^{\mu \nu }\,\partial _{\mu }\,\Phi \,\partial
_{\nu }\,\Phi 
\end{eqnarray}
where 
$${\frac{1}{2}}\,\sqrt{-g}\,g^{\mu \nu }\,\,,$$ 

\ni from now on, is written like in (\ref{metrica}). As we have stressed above, the expression (\ref{lalala}) represents the noton action.

%\subsubsection{Gates and Siegel's model}

{\bf Gates and Siegel's model}: In \cite{gs} the authors have studied the interactions of leftons and
rightons with external vector fields including the supersymmetric and the
non-Abelian cases. The soldering of this model has been obtained already
in \cite{aw}, but as a further test for our GGSM, let us write 
\begin{eqnarray}
{\cal L}_{GS}^{\phi}\,&=&\,(\,\partial_-\,\phi\,+\,2\,A_-\,)\,(\,\partial_+\,%
\phi\,)\,+\, \lambda_{++}\,(\,\partial_-\,\phi\,+\,A_-\,)^2  \nonumber \\
{\cal L}_{GS}^{\rho}\,&=&\,(\,\partial_+\,\rho\,+\,2\,A_+\,)\,(\,\partial_-\,%
\rho\,)\,+\, \lambda_{--}\,(\,\partial_+\,\rho\,+\,A_+\,)^2 \nonumber \\
\end{eqnarray}
and the correspondence with (\ref{eqa}) and  (\ref{eqb}) is direct 
\begin{equation}
a_2\,=\,b_1\,=\,2 \;\; ; \;\; a_1\,=\,b_2\,=\,0 \;\;;\;\;
a_3\,=\,b_3\,=\,1\;.
\end{equation}

The soldered action is, using (\ref{371}), 
\begin{equation}  \label{II111}
{\cal L}_{TOT}\,=\,{\frac{1 }{2}}\,\sqrt{-g}\,g^{\mu\nu}\,\partial_{\mu}\,%
\Phi\,\partial_{\nu}\,\Phi \,-\,2\,A_{-}A_{+}\;\;,
\end{equation}
confirming the result in \cite{aw}. We can note that the covariance has not
been broken.

The physical meaning of (\ref{II111}) can be appreciated by eliminating the
multipliers and using the symmetry induced by the soldering formalism \cite{clovis3},
showing that it represents the action for the noton. In fact (\ref{II111}) is
basically the action proposed by Hull \cite{hull} as a candidate for
canceling the Siegel anomaly. This field carries a representation of the
full diffeomorphism group \cite{hull} while its chiral (Siegel) component
carry the representation of the chiral diffeomorphism. Observe the complete
disappearance of the dynamical sector due to the destructive interference
between the leftons and the rightons. This happens because we have
introduced only one soldering field to deal with both the dynamics and the
symmetry. To recover dynamics we need to separate these sectors and solder
them independently, as stressed in \cite{aw}.

%\subsubsection{The gauged massless Bellucci, Golterman and Petcher model}

{\bf The gauged massless Bellucci, Golterman and Petcher model}: The form of this gauged chiral boson action is 
\begin{eqnarray}
{\cal L}_{BGP}^{\phi}\,=\,(\,\partial_+\,\phi\,)\,(\,\partial_-\,\phi\,+\,e\,A_-\,)\,+\, \lambda_{++}\,(\,\partial_-\,\phi\,+\,e\,A_-\,)^2 
\nonumber \\
{\cal L}_{BGP}^{\rho}\,=\,(\,\partial_-\,\rho\,)\,(\,\partial_+\,\rho\,+\,e\,A_+\,)\,+\, \lambda_{--}\,(\,\partial_+\,\rho\,+\,e\,A_+\,)^2                            
\end{eqnarray}
hence 
\begin{equation}
a_1\,=\,b_2\,=\,0 \;\; ; \;\; a_2\,=\,a_3\,=\,b_1\,=\,b_3\,=\,e\;\;.
\end{equation}
and the final action reads, 
\begin{eqnarray}
& &{\cal L}_{TOT}\,=\,{\frac{1 }{2}}\,\sqrt{-g}\,g^{\mu\nu}\,\partial_{\mu}\,%
\Phi\,\partial_{\nu}\,\Phi\, \nonumber \\
&+&\, {\frac{1 }{{1-\lambda}}}\,\left[\,e\,\lambda%
\,(\partial_+\,\Phi\,A_-\,-\,\partial_-\,\Phi\,A_+\,) \right.  \nonumber \\
&+&\left.
e\,\lambda_{++}\,\partial_-\,\Phi\,A_-\,-\,e\,\lambda_{--}\,(1\,+\,\lambda)\,%
\partial_+\,\Phi\,A_+\, \right. \nonumber \\
&+& \left. \, {\frac{5}{4}}\,e^2\,(\,\lambda_{++}\,A_-^2\,+\,%
\lambda_{--}\,A_+^2\,) \right.  \nonumber \\
&-& \left. e^2\, ({\frac{1}{2}}\,-\,\lambda_{++}\,-\,\lambda_{--}\,)\,A_+%
\,A_-\,\right]\;\;,
\end{eqnarray}
it is easy to see that the last two terms break the covariance.  Hence, in this case
we have clearly a destructive interference of covariances.

%\subsubsection{The Frishman and Sonnenschein model}

{\bf The Frishman and Sonnenschein model}: The chiral actions developed in \cite{fs} are 
\begin{eqnarray}
{\cal L}_{FS}^{\phi}\,&=&\,(\,\partial_+\,\phi\,)\,(\,\partial_-\,\phi\,)\,+\,
\lambda_{++}\,(\,\partial_-\,\phi\,)^2\,+\,\partial_+\,\phi\,A_-\,-\,\partial_-\,\phi\,A_+,  \nonumber \\
{\cal L}_{FS}^{\rho}\,&=&\,(\,\partial_-\,\rho\,)\,(\,\partial_+\,\rho\,)\,+\,
\lambda_{--}\,(\,\partial_+\,\rho\,)^2\,+\,\partial_-\,\rho\,A_+\,-\,\partial_+\,\rho\,A_-\;\;,
\end{eqnarray}
and identifying the parameters, 
\begin{equation}
a_1\,=\,b_2\,=\,-\,1 \;\; ; \;\; a_2\,=\,b_1\,=\,1 \;\; ;\;\;
a_3\,=\,b_3\,=\,0\;\;,
\end{equation}
we can construct the soldered action as 
\begin{eqnarray}
{\cal L}_{TOT}\,&=&\,{\frac{1 }{2}}\,\sqrt{-g}\,g^{\mu\nu}\,\partial_{\mu}\,%
\Phi\,\partial_{\nu}\,\Phi\,+\,
\epsilon^{\mu\nu}\,\partial_{\mu}\Phi\,A_{\nu}\,+\,{\frac{{1-2\lambda} }{{1-\lambda}}}\,A_+\,A_-\;\;,
\end{eqnarray}
where $\epsilon^{+-}=1$.  

Now we have a constructive interference of covariance, since, the soldered action
is explicitly covariant.

%%%%%%%%%%%%%%%%%%%%%%%%%%%%%%%%%%%%%%%%%%%%%%%%%%%%%%%%%%%%%%%%%%%%%%%%%%%%%%%%%%%%%%%%%%%%%%%%%%%%%%%%%%%%%%%%%%%%%%%%%%%%%%%%%%%%%%%%%%%%%%%%

\subsection{Destructive Interference of Dualities}

%The role of duality as a qualitative tool in the investigation of physical systems is being gradually disclosed at different context and dimensions \cite{AG}. In two space-time dimensions in particular, we face the intriguing situation where chirality also plays the role of duality. This enables the investigation of the former to be performed using the techniques developed for the latter. The techniques of fusion or soldering, introduced by Stone \cite{solda}, as viewed above, helps to investigate some new aspects of dualities at different space-time dimensions, and to study the physical consequences of their combination by the soldering process. As we have mentioned before, in the 3D case, the soldering mechanism was used to show the result of fusing together two topologically massive modes generated by the bosonization of two massive Thirring models with opposite mass signatures in the long wave-length limit.  The bosonized modes, which are described by self and anti-self dual Chern-Simons models \cite{tpn,dj}, were then soldered into the two massive modes of the 3D Proca model \cite{baw2}. In the 4D case, the soldering mechanism produced an explicitly dual and covariant action as the result of the interference between two Schwarz-Sen \cite{ss} actions displaying opposite aspects of the electromagnetic duality \cite{baw2}.  It is our intention in this section to show the physical consequences of combining actions possessing truncate diffeomorphism invariance and opposite chiralities using the fusion of dualities technique.

In this section we show that it is also possible to obtain the field
theoretical analog of the
``quantum destructive interference" phenomenon, by coupling the chiral
scalars to appropriately truncated metric fields, known as
chiral WZW models, or non-Abelian Siegel models.  By soldering the 
two (Siegel) invariant representations
of the chiral WZW model \cite{siegel} of opposite chiralities, the effective
action that results from this process is
shown to be invariant under the full diffeomorphism group, which is
not a mere sum of two Siegel symmetries. In fact, this effective action
does not contain either right or left movers, but can be
identified with the non-Abelian generalization of the
bosonic non-mover action proposed by Hull \cite{hull},
thanks to the richer symmetry structure
induced over it by the soldering mechanism.

To begin with, let us review some facts about the
non-Abelian Siegel model \cite{fs}.  The action for
a left mover chiral scalar (using again the light-cone variables and $\tilde g = g^{-1}$ denotes the inverse
matrix.), is given as
\begin{equation}
\label{leftzero1}
S_0^{(+)}(g)=\int\;d^2x\; tr\left(\partial_+g\:
\partial_-\tilde g+\lambda_{++}\partial_-g\:\partial_-\tilde g\right)
+\Gamma_{WZ}(g)\, ,
\end{equation}

\noindent where $g\in G$ is a matrix-valued field taking values on some compact
semi-simple Lie group $G$, with an algebra ${\hat G}$ and $\tilde{g}=g^{-1}$. 
The term $\Gamma_{WZ}(g)$ is the topological Wess-Zumino functional,
as defined in Ref. \cite{pw}.  It is invariant under a chiral
diffeomorphism known as Siegel transformation where,
\begin{equation}
\label{siegel}
\delta\lambda_{++} = -\partial_+\epsilon^- + \lambda_{++}
\partial_-\epsilon^- +\epsilon_-\partial_-\lambda_{++}\, ,
\end{equation}

\noindent and $g$ transforming as a scalar.  This action can be
seen as the
WZW action, immersed in a gravitational background, with an appropriately
truncated metric tensor,

\begin{equation}
\label{leftzerograv1}
S_0^{(+)}(g)={1\over 2}\int d^2 x \sqrt{-\eta_+}\; 
\eta_+^{\mu\nu}\:tr\left(\partial_\mu g\:
\partial_\nu \tilde g\right) +\Gamma_{WZ}(g)\, ,
\end{equation}

\noindent with $\eta^+ = det(\eta^+_{\mu\nu})$ and
\begin{equation}
\label{metric+}
{1\over 2} \sqrt{-\eta_+}\: \eta_+^{\mu\nu}=\left(
\begin{array}{cc}
0 & {{1\over 2}}\\
{{1\over 2}} & {\lambda_{++}}
\end{array}
\right)\, .
\end{equation}

Next, let us compute the Noether currents for the axial, vectorial and the
right and left chiral transformations.  The variation of the Siegel-WZW
action (\ref{leftzero1}) or (\ref{leftzerograv1}) gives,
\begin{equation}
\label{variation}
\delta S_0^{(+)}(g)=\left\{
\begin{array}{lll}
\int d^2x\:tr\Big\{\delta g \tilde g\: 
2\Big[\partial_+\left(\partial_- g\tilde g\right)
\:+\:\partial_-\left(\lambda_{++}\partial_- g\tilde g\right)\Big]\Big\}\\
\mbox{}\\
\int d^2x\:tr\Big\{\tilde g\delta g \: 
2\Big[\partial_-\left(\tilde g\partial_+ g\right)
\:+\:\partial_-\left(\lambda_{++}\tilde g\partial_- g\right)\Big]\Big\}\, .
\end{array}\right.
\end{equation}

\noindent  From (\ref{variation}) and the axial transformation
($g\rightarrow kgk $) we obtain,
\begin{eqnarray}
\label{axialcurr}
J_A^+&=& 2g\partial_-\tilde g\nonumber\\
J_A^-&=& -2\Big[\tilde g\partial_+ g +\lambda_{++}
(\tilde g\partial_- g +\partial_-g \;\tilde g)\Big]\, ,
\end{eqnarray}

\noindent where $k \in K $ take their values in some subgroup $K\subset G$.  From the transformation
($ g \rightarrow \tilde k g k$) we obtain the vector current,
\begin{eqnarray}
\label{vectcurr}
J_V^+&=& 2g\partial_-\tilde g\nonumber\\
J_V^-&=& 2\Big[\tilde g\partial_+ g +\lambda_{++}(\tilde g\partial_- g -
\partial_-g \;\tilde g)\Big]\, .
\end{eqnarray}

\noindent Incidentally, it should be observed that the axial and the vectorial
currents (\ref{axialcurr}) and (\ref{vectcurr}) are dual to each other
only if the following extended definition is adopted,
\begin{equation}
\mbox{}^*T^\mu=\sqrt{-\eta_+} \; \eta_+^{\mu\nu}
\epsilon_{\mu\lambda}T^\lambda\, ,
\end{equation}

\noindent and use of the following relations is made,
\begin{eqnarray}
J_+ &=& J^- -2 \lambda_{++}J^+\nonumber\\
J_- &=& J^+\, ,
\end{eqnarray}

\noindent which is valid for all currents.
Similarly, the chiral currents can be obtained  from the left
($g\rightarrow gk$) and
right ($g\rightarrow \tilde k g$) transformation. The result is,
\begin{eqnarray}
J_L^{(+)} &=& 0\nonumber\\
J_L^{(-)} &=& 2\left(\tilde g \partial g +\lambda_{++} 
\tilde g\partial_- g\right)\, ,
\end{eqnarray}

\noindent and
\begin{eqnarray}
\label{direita}
J_R^{(+)} &=& -2 g\partial_-\tilde g\nonumber\\
J_R^{(-)} &=& -2\lambda_{++} g \partial_-\tilde g\, .
\end{eqnarray}

\noindent  It is crucial to notice that out of the two affine invariances
of the original WZW model, only one is left over due to the chiral constraint
$\partial_- g\approx 0$.  Indeed, the affine invariance is only present
in the left sector since $J_L^{(+)}=0$ and $\partial_-J_L^{(-)}=0$,
which implies $J_L^{(-)}=J_L^{(-)}(x^+)$,
while $J_R^{(-)}\neq 0$ and $J_R^{(+)}\neq J_R^{(+)}(x^-)$.

Next we work out the details for the right chirality action,
\begin{eqnarray}
\label{rightzero}
S_0^{(-)}(h)&=&\int \;d^2x\; tr\left(\partial_+ h
\partial_-\tilde h +\lambda_{--}\partial_+ h
\partial_+\tilde h\right) - \Gamma_{WZ}(h)\nonumber\\
&=& {1\over 2} \int d^2x\; \sqrt{-\eta_-} \;\eta_-^{\mu\nu}
\:tr\left(\partial_\mu h \partial_\nu 
\tilde h\right)- \Gamma_{WZ}(h)\, ,
\end{eqnarray}

\noindent  where
\begin{equation}
{1\over 2}\sqrt{-\eta_-}\:\eta_-^{\mu\nu}=\left(
\begin{array}{cc}
{\lambda_{--}} & {{1\over 2}}\\
{{1\over 2}} & 0
\end{array}
\right).
\end{equation}

\noindent  Notice that $S_0^{(+)}(g)$ and $S_0^{(-)}(h)$
are chosen at opposite critical points, otherwise they will
not carry different chiralities, a crucial condition
for the soldering to be performed.
The set of axial, vector and chiral Noether currents is similarly
obtained,
\begin{eqnarray}
J_A^{(+)}(h)&=& 2\Big[\tilde h\partial_- h +\lambda_{--}
\left(\tilde h\partial_+ h +\partial_+ h \tilde h\right)\Big]\nonumber\\
J_A^{(-)}(h)&=&2\:\partial_+ g\tilde g\\
J_V^{(+)}(h)&=&2\Big[\tilde h\partial_- h +\lambda_{--}
\left(\tilde h\partial_+ h -\partial_+ h\:\tilde h\right)\Big]\nonumber\\
J_V^{(-)}(h)&=&-2\:\partial_+ h\:\tilde h\\
J_L^{(+)}(h)&=& 2\:\partial_+ h\:\tilde h\nonumber\\
J_L^{(-)}(h)&=&2\:\lambda_{--}\:\partial_+ h\:\tilde h\\
J_R^{(+)}(h)&=&2\left(\tilde h\partial_- h +\lambda_{--}
\:\tilde h\partial_+ h\right)\nonumber\\
J_R^{(-)}(h)&=&0\, ,
\end{eqnarray}

\noindent with the corresponding interpretation analogous to
that following Eq.(\ref{direita}).  Since  the actions (\ref{leftzero1})
and (\ref{rightzero}) do correspond to opposite aspects of a symmetry (chirality), the stage is set
for the soldering.

Next, let us discuss the gauging procedure to be adopted in the soldering
of the right and the left chiral WZW actions just reviewed.  
As well known at this point, the basic idea of the soldering procedure is to lift a global Noether symmetry
present at each individual chiral component into a local symmetry for the
composite system that, consequently, defines the soldered action.
It is of vital importance to notice that
the coupling with the (auxiliary) soldering gauge field is only consistent
if use is made of the correspondent chiral current.
Otherwise the equations of motion, after the gauging, will result being incompatible with the covariant chiral constraint, by the presence of an anomaly.  Anomalies can certainly, be accommodate into the theory, but not at the expense of violating
the consistence between equations of motion and gauge constraints.  
We will use the iterative Noether procedure to lift the global
(left) chiral symmetry of (\ref{leftzero1}),
\begin{eqnarray}
\label{transf.1}
g&\rightarrow & gk\nonumber\\
\lambda_{++}&\rightarrow & \lambda_{++}\nonumber\\
A_-&\rightarrow & \tilde k A_-k + \tilde k\partial_- k
\end{eqnarray}

\noindent  into a local one.  To compensate for the non-invariance
of $S_0^{(+)}$, we introduce the coupling term,
\begin{equation}
S_0^{(+)}\rightarrow S_1^{(+)} =S_0^{(+)} +A_- J_L^-(g)\, ,
\end{equation}

\noindent  along with the soldering gauge field $A_- $, taking values
in the subalgebra ${\hat K}$ of $K$, whose transformation properties
are being defined in (\ref{transf.1}).  Using such transformations,
it is a simple algebra to find that,
\begin{equation}
\delta\left(S_1^{(+)}-\lambda_{++} A_-^2\right)=2\;\partial_+\omega\; A_-\, ,
\end{equation}

\noindent with $\omega \in {\hat K}$ being an infinitesimal element of
the algebra.   One can see that,
\begin{equation}
S_2^{(+)}=S_1^{(+)} -\lambda_{++}\;A_-^2
\end{equation}
cannot be made gauge invariant by additional Noether counter-terms, but
it has the virtue of being independent of the transformation
properties of $g$ while depending
only on the elements of the gauge algebra $\hat K$.  
Similarly, for the right chirality we find,
\begin{equation}
\delta S_2^{(-)}=-\;2\;A_+\;\partial_-\omega
\end{equation}

\noindent for
\begin{equation}
S_2^{(-)}(h)=S_0^{(-)}(h)-A_+J_R^{+}(h)-\;\lambda_{--}\;A_+^2
\end{equation}

\noindent  when the basic fields transform as,
\begin{eqnarray}
\label{transf.2}
h&\rightarrow &hk\nonumber\\
A_+&\rightarrow & kA_+\tilde k + k\partial_+\;\tilde k\nonumber\\
\lambda_{--} &\rightarrow & \lambda_{--}\, .
\end{eqnarray}

\noindent  It is important to observe that the action for the right
sector depends functionally on a different field, namely $h \in H$. 
Although the gauged actions for each chirality could not
be made gauge invariant separately,  with the inclusion of a contact
term, the combined action,
\begin{equation}
\label{eff}
S_{eff}=S_2^{(+)} + S_2^{(-)} + 2 A_+ \; A_-\, ,
\end{equation}

\noindent is invariant under the set of transformations
(\ref{transf.1}) and (\ref{transf.2}) simultaneously.

As explained in the section above, we eliminate the (non dynamical) gauge
field $ A_\mu $.  From the equations of motion one gets,
\begin{equation}
{\cal J} = 2 {\bf M} {\cal A}\, ,
\end{equation} 

\noindent where we have introduced the following matricial notation,
\begin{equation}
{\cal J} =\left(
\begin{array}{c}
J_L^-(g)\\
J_R^+(h)
\end{array}
\right)
\end{equation}

\begin{equation}
{\cal A} =\left(
\begin{array}{c}
{A_+}\\
{A_-}
\end{array}
\right)
\end{equation}

\noindent and

\begin{equation}
{\bf M} =\left(
\begin{array}{cc}
1 & {\lambda_{++}}\\
{\lambda_{--}} & 1
\end{array}
\right)\, .
\end{equation}

\noindent  Bringing these results into the effective action (\ref{eff}) gives,
\begin{eqnarray}
S_{eff} &=& S_0^{(+)}(g) + S_0^{(-)}(h) +\nonumber\\
&+& \int d^2x {1\over{1-\lambda^2}} 
tr\left\{ 2\left[\tilde g\partial_+ g \;\tilde h\partial_- h + 
\lambda^2 \;\tilde g\partial_- g\;\tilde h\partial_+ h 
+\right.\right.\nonumber\\
&+& \left. \lambda_{++}\;\tilde g\partial_- g \;\tilde h\partial_- h 
+\lambda_{--}\;\tilde g\partial_+ g\;\tilde h\partial_- h \right] +\nonumber\\
&+& \lambda_{--}\;\left(\partial_+ g\;\partial_+\tilde g + 2\lambda_{++}
\;\partial_+ g\;\partial_-\tilde g +\lambda_{++}^2\;\partial_- g\;
\partial_-\tilde g\right) +\nonumber\\
&+& \left. \lambda_{++}\;\left(\partial_- h\;\partial_-\tilde h + 
2 \lambda_{--}\;\partial_+ h\;\partial_+\tilde h 
+\lambda_{--}^2\;\partial_+ h\;\partial_+\tilde h\right)\right\}\, .
\end{eqnarray}

\noindent where $\lambda^2=\lambda_{++}\lambda_{--}$. 
It is now a simple algebra to show that this effective action
does not depend on the fields $g$ and $h$ individually, but
only on a gauge invariant combination of them, defined below
(\ref{sss}). 
This effective action corresponds to that of a (non-chiral) WZW model
coupled minimally to an effective metric built out of the Lagrange multipliers
of the original Siegel actions,
\begin{equation}
\label{ss}
{1\over 2}\sqrt{-\eta}\; \eta^{\mu\nu}= {1\over{1-\lambda^2}}\;\left(
\begin{array}{cc}
{\lambda_{--}} & {{{1+\lambda^2}\over 2}}\\
{{{1+\lambda^2}\over 2}} & {\lambda_{++}}
\end{array}
\right)
\end{equation}

\noindent and a new (effective) field,
\begin{equation}
\label{sss}
{\cal G}= g\tilde h
\end{equation}

\noindent and reads,
\begin{equation}
\label{s}
S={1\over 2}\int \:d^2x\:\sqrt{-\eta}\:\eta^{\mu\nu}\:tr\left(\partial_\mu 
{\cal G}\;\partial_\nu\tilde{\cal G}\right) +\Gamma_{WZ}({\cal G})\, .
\end{equation}

\noindent Here we have used the well known property of the Wess-Zumino
functional $\Gamma_{WZ}(h)\:=\:-\;\Gamma_{WZ}(\tilde h)$, and the
Polyakov-Weigman identity \cite{pw}.

It is interesting to notice that the original chiral transformations
(\ref{transf.1}) and (\ref{transf.2}) are now hidden, since the effective action is
composed of only the gauge invariant objects (\ref{ss}) and (\ref{sss}).  To unravel the physical
contents of the effective soldered action (\ref{s}), it is important
to study the new set of symmetries
of the composite theory.  We first observe that under 
diffeomorphism the metric transform as a symmetric tensorial density,
\begin{eqnarray}
\label{138}
\delta\lambda_{++}&=&-\partial_+\epsilon^-+\lambda^2_{++}\partial_-\epsilon^+
+(\partial_+\epsilon^+-\partial_-\epsilon^-+\epsilon^+\partial_+
+\epsilon^-\partial_-)
\lambda_{++}\nonumber\\
\delta\lambda_{--}&=&-\partial_-\epsilon^++\lambda^2_{--}\partial_+\epsilon^-
+(\partial_-\epsilon^--\partial_+\epsilon^++\epsilon^+\partial_+
+\epsilon^-\partial_-)
\lambda_{--}
\end{eqnarray}
while the ${\cal G}$ transforms as a scalar.
It is important to observe that if we
restrict the diffeomorphism to just one sector,
say by requiring $\epsilon^+=0$, 
we reproduce the original Siegel
symmetry for the sector described by the pair ${\cal G}\,,\lambda_{++}$
in the same way as it appears in the original chiral theory (\ref{leftzero1}).
However,
under this restriction,
$\lambda_{--}$ transforms in a non-trivial
way as,
\begin{equation}
\label{c1}
\delta\lambda_{--}=\lambda^2_{--}\partial_+\epsilon^-+
\left(\partial_-\epsilon^-+\epsilon^-\partial_-\right)\lambda_{--}\,.
\end{equation}
The original Siegel symmetry, therefore, is 
not a subgroup of the diffeomorphism group but 
it is only recovered if we also make a 
further truncation, by imposing  that $\lambda_{--}=0$. 
The existence of the residual symmetry (\ref {c1}) seems to be related
to a duality symmetry satisfied by the effective action (\ref{s}) when the
metric is parametrized as in (\ref{ss}).
Under the discrete transformation,
\begin{equation}
\label{c2}
\lambda_{\pm\pm}\rightarrow{1\over\lambda_{\mp\mp}}  \,,
\end{equation}

\noindent the residual transformation (\ref{c1}) swaps to
(\ref{siegel}) while that becomes the residual symmetry for
the opposite chiral sector.  Indeed we see that the
classical equations of motion remain invariant
under (\ref{c2}) while the effective action changes its signature,
very much like in the original electromagnetic duality transformation.
This is obviously
related to the interchange symmetry between the right and the
left moving sectors
of the theory, and seems to be of general validity \cite{baw}. 
Also notice that the gauged Lagrangian in one sector, either $S_2^+$ or
$S_2^-$, cannot be written in a diffeomorphism invariant manner. Therefore,
gauging in one of the sectors breaks Siegel invariance. However,
let us note that if we integrate out either the $A_-$ or the $A_+$
field, the Siegel theory changes
chirality with the identification provided by (\ref{c2}), that is again
related to the discrete duality symmetry.

By solving the equations of motion
and setting the $\lambda_\pm$ to zero by invoking the diffeomorphism invariance
discussed above, it is simple to see that the composite field (\ref{sss})
of the effective
action (\ref{s}) describes a non-mover field, as first proposed by Hull \cite{hull}. 
The right and the left moving modes have therefore
disappeared from the spectrum.
The soldering procedure has clearly produced a destructive interference between
left and right movers of the original chiral components.  Moreover,
it can be easily seen that the coupling of chiral scalars to a dynamical gauge field
before soldering, as done above, will decouple the gauge sector
from the effective soldered action.  This seems to be a
natural result since a non-mover field cannot couple to either right or left components of
the vector gauge field.  This is a
distinctive result produced by the presence of the full group of diffeomorphism
resulting from the soldering process, that constrains the matter
scalar field into the non-moving sector, quite in opposition to the 
constructive interference result that comes from soldering the noninvariant
models.

%%%%%%%%%%%%%%%%%%%%%%%%%%%%%%%%%%%%%%%%%%%%%%%%%%%%%%%%%%%%%%%%%%%%%%%%%%%%%%%%%%%%%%%%%%%%%%%%%%%%%%%%%%%%%%%%%%%%%%%%%%%%%%%%%%%%%%%%%%%%%%%%%%%%%%%%%%%

\section{The Dual Projection}

In the preceeding Section we described a self-contained algorithm whereby distinct and independent chiral massless modes of propagation are fused together into a new multiplet that carries both representation of the chiral diffeomorphism being, therefore, vectorial.  In this Section we discuss a related subject which is in a sense ``to take the opposite route" and reorganize the field-variables in a given model in order to better display both the dynamical and symmetry contents of it.  However, differently form the analysis presented above, the models under scrutiny here need not be symmetric regarding its basic constituents. This feature will provide us with new and surprising results.

The dual projection technique is an approach closely related to a canonical transformation. Its basic goal is to transform a given set of field-variables into a new set such that, in terms of them, the action acquires a diagonal form. Different parts of the new action may display distinct and specialized properties of the original model. For instance, sometimes it is possible to diagonalize the action such that a subset of the field-variables is pure gauge while the other set possess no symmetry whatsoever.  These ideas will be illustrated below with a series of interesting and physically motivated examples.

As we know, the FJ model displays the chiral dynamics the more general chiral bosonization schemes (CBS) proposed by Siegel \cite{siegel}.  
The Siegel modes (rightons and leftons) carry not only
chiral dynamics but also symmetry information. 
The symmetry content of the theory is well
described by the Siegel algebra, a truncate
diffeomorphism, that disappears at the quantum level.

%The chiral interference between gauged rightons and leftons should lead to a massive vector mode, plus the symmetry of the combined right-left Siegel algebras. As we will see in this section, the result of the chiral interference between gauged Siegel modes is a Hull noton \cite{hull}. This represents only the symmetry part of the expected result, disclosing the destructive interference of the massive mode \cite{clovis2} resulting from the simultaneous soldering of dynamics and symmetry.

The main purpose of this investigation is to recover the dynamics
of the chiral interference.  To this end
we introduce the new idea of dual projection that separates
the chiral sector from the symmetry sector in the Siegel CBS. 
Under this field redefinition, the Siegel action can
be reexpressed as a FJ action carrying the chirality, plus a noton,
carrying the representation of the Siegel symmetry.
This proposal shows that the difference between these
CBS's is given by the presence of a noton.
We stressed that the noton is a nonmover field at classical level that
acquires dynamics upon quantization.  To find out that the noton is
already contained in the Siegel's theory was a new and interesting
result. In this section we will examine the contribution of the noton to the coefficient of the Schwinger term
in the energy-momentum tensor current algebra, where it is shown that
its quantum dynamics is fully responsible for the Siegel anomaly.
With the dynamical and the symmetry sectors isolated,
we might introduce independent soldering to avoid the destructive interference of the chiral modes.

\subsection{Dual Projection, Noton and Anomaly}

We will demonstrate in this section that different CBS's are related by the presence of a noton. 
We show that a Siegel mode may be decomposed in a FJ mode, responsible for the dynamics, and a 
nonmover, carrying the representation of the symmetry group, as mentioned above.  
This is done introducing a {\it dynamical redefinition} (a canonical transformation \cite{bg})
in the phase space of the model.
We stress that these fields are independent as they originate from completely different actions, and the presence of a Siegel noton was not
pointed in the literature.
This new result is complementary to the established knowledge, where the FJ action is interpreted as a gauge fixed Siegel action \cite{siegel}. 
Under this new point of view we look at the gauge fixing process as the condition that sets the noton field to vanish.
These points will now be clarified.
   
Let us begin with the Siegel action for a left-mover scalar,
\begin{equation}
\label{1order}
S = <\pi\,\dot{\phi}- \frac{{\phi'}^2}{2} -\frac{1}{2} \frac {\left(\pi-\lambda_{++}{\phi'}\right)^2}{1-\lambda_{++}}
-\frac{\lambda_{++}}{2}{\phi'}^2>.
\end{equation}
One can fix the value of the multiplier as 
$\lambda_{++} \rightarrow 1$ to reduce
it to its FJ form. The phase space of the model is correspondingly reduced to 
$$\pi \rightarrow \phi'\:.$$

\ni The third term in (\ref{1order}) reduces to 
$$(1-\lambda_{++}){\phi'}^2\rightarrow 0$$

\ni as $\lambda_{++}$ approaches its unit value.
This reduces the symmetry of the model, leaving behind its dynamics described by a FJ action.
The above behavior suggests the following canonical transformation \cite{bg},
\begin{eqnarray}
\label{theresult}
\phi = \varphi + \sigma\;\;\;\mbox{and}\;\;\;\pi = \varphi' - \sigma',
\end{eqnarray}

\noindent which is our cherished result.  The lefton $\phi$ is
related to the FJ chiral mode $\varphi$ by the presence of a noton
$\sigma$. Such a decomposition immediately diagonalizes (\ref{1order}) as,
\begin{equation}
\label{III2}
S  =  <{\varphi'}\,\dot{\varphi}\;-\;{\varphi'}^2>\;+\;
 <-\;\sigma'\dot{\sigma}\;-\;\eta_+\,\sigma'^2>\;\;,
\end{equation}
and 
$$\eta_{\pm} = \frac{1\;\;+\;\;\lambda_{\pm\pm}}{1\;\;-\;\;\lambda_{\pm\pm}}\:.$$

\ni In this form, the chiral information is displayed by the FJ field ${\varphi}$ while the noton $\sigma$ carries the symmetry of the original model.
The reduction of the phase space is attained by letting the noton
$\sigma$ approach zero as the multiplier $\eta_+$ diverges.  This eliminates the symmetry carrying sector leaving behind only the FJ mode. 

%\subsection{The Noton and the Anomaly} 

To disclose the meaning of the symmetry, we need to study the noton invariances, imposed by the constraint $\sigma^2\approx 0$.  Following symplectic formalism \cite{annals}, we obtain  from (\ref{III2}), the following symplectic matrix   
\begin{equation}
\label{III4}
f = 2 \left( \begin{array}{cc}
               1 & {\sigma_y}' \\
              {\sigma_x}' & 0
             \end{array} \right)\;\delta'(\,x\,-\,y\,)
\end{equation}
whose single zero-mode displays the searched symmetry. 
\begin{eqnarray}
\label{III5}
\delta\,\sigma  &=&  \epsilon\,\sigma';\;\;\; \nonumber \\
\delta\,\eta  &=&  \dot{\epsilon} + \eta\,\epsilon' - \epsilon\,\eta'.
\end{eqnarray}
As claimed, $\sigma$ carries a representation of the Siegel algebra.  Solving the equations of motion
and making use of (\ref{III5}), we find that $\sigma$ is indeed a nonmover.  
In this way we have realized, in a deeper level, the decomposition of the Siegel's chiral boson in terms of dynamics and symmetry
and identified the associated fields.

When the quantization process is accomplished the noton acquires dynamics through the gravitational anomaly \cite{Quant}.  To see this we examine its quantum contents and show that it contributes fully to the Siegel anomaly.
This is done by computing the Schwinger terms of the energy-momentum tensor current algebra in the noton action.
We begin with a separation of the field operator in creation and annihilation parts \cite{Fuji}, $A = A_+\, + \,A_-$
%\begin{equation}
%\label{III6}
%A = A_+\, + \,A_-
%\end{equation}
and define the vacuum state by $A_-\,|0> = 0$.
Define the positive and negative frequency projector as,
\begin{equation}
\label{III8}
A_{\pm}\,(x) = \int\,dz\,\delta_{\pm}\,(x-z)\,A(z)
\end{equation}

\noindent where the chiral delta functions are defined as
\begin{equation}
\label{III10}
\delta_{\pm} \,(x)= \mp\, {i\over{2\,\pi}}{1\over{x\,\mp\,i\,\epsilon}}
\end{equation}

\noindent and satisfy the following property,
\begin{equation}
\label{III11}
\left(\,\delta'_{+}\,(x)\,\right)^2 - \left(\, \delta'_{-} \,(x)\right)^{2} = {i\over{12\pi}} \delta'''\,(x) .
\end{equation}

\noindent To compute the Schwinger term, we examine the energy-momentum tensor $T\,(x) = [\,\sigma'(x)\,]^2$,
whose classical algebra is,
\begin{equation}
\label{III15}
\{\,T(x),T(y)\,\} = ( \,T(x)\,+\,T(y)\,)\,\delta'(x\,-\,y)
\end{equation}

%\begin{equation}
%\label{III14}
%T\,(x) = [\,\sigma'(x)\,]^2
%\end{equation}

\noindent Then, upon quantization, the presence of a Schwinger term \cite{JS} is completely disclosed by normal ordering
the energy-momentum operator. Call $\sigma'(x)\,=\,\xi(x)$,
with $\xi_+$ and $\xi_-$
being the creation and annihilation operators respectively.
Using that,
\begin{equation}
\left[\,\xi(x)\,,\,\xi(y)\,\right] \;=\; \frac{i\,\hbar}{2}\,\delta'(\,x\,-\,y\,),
\end{equation}

\noindent it is easy to check that the following results are obeyed
\begin{eqnarray}
\left[\,\xi_{\pm}(x)\,,\,\xi(y)\,\right] \;&=&\; \frac{i\,\hbar}{2}\,\delta'_{\pm}(\,x\,-\,y\,),\nonumber\\
\left[\,T(x)\,,\,\xi_{\pm}(y)\,\right] \;&=&\; i\,\hbar\,\xi(x)\,\delta'_{\mp}(\,x\,-\,y\,)\;\;.
\end{eqnarray}

\noindent The current-current commutator,
\begin{eqnarray}
{\left[\,T(x)\,,T(y)\right] = i\,\hbar\,\left(T(x)+T(y)\right)\,\delta'\,(\,x\,-\,y\,)}\,+\, \frac{i\,\hbar^2}{24\,\pi}\,\delta'''\,(\,x\,-\,y\,)
\end{eqnarray}
is identical to the well known current algebra for the energy-momentum tensor operator for the Siegel model \cite{Son} with the correct value
for the central charge.  This shows that the noton 
present in the Siegel formulation completely takes care of the symmetries,
both classically and quantically.  This is a new and outstanding result.  It explains why the Hull's mechanism for canceling
the Siegel anomaly works by including a properly normalized external noton.

%%%%%%%%%%%%%%%%%%%%%%%%%%%%%%%%%%%%%%%%%%%%%%%%%%%%%

\subsection{Coupling Chiral Bosons to Gravity}

Few years back there was a great deal of attention devoted to
the quantization of chiral scalar fields.  The main motivation
is that chiral bosons are the basic objects of  two of the
most interesting opened problems of present days theoretical physics.
%To remember the motivation,
These object appear in the construction of many string models \cite{string}
where some symmetries are
manifest before chiral boson fermionization, but not after.
Technically, it is advantageous to keep chiral bosons, instead of their
fermionic counterparts, because it suffices to compute lower loop graphs on
the world-sheet.  Furthermore, in the description of quantum Hall effect
\cite{QHE},
chiral bosons play an important role since there they appear as the
edge-states of the Hall fluid, which are believed to be the only gapless
excitation of the sample.

Using a constraint context to obtain a chiral boson, one usually eliminates one half of the degrees of
freedom
from the scalar field by means of a chiral constraint $\partial_{\pm}
\phi \approx 0$.  The problem in following this route is that the chiral
constraint is second-class by Dirac's classification scheme
\cite{Dirac}.  Therefore one
is not allowed to gauge away
its associated Lagrange multiplier field that therefore acquires a
dynamical character.
Siegel \cite{siegel}  proposed to covariantize the chiral constraint, i.e.,
to transform
it from second to first-class, by squaring it, or what is equivalent,
setting one chiral component of the energy-momentum tensor to zero
as a constraint.
Siegel's action has a reparametrization symmetry, at tree level, that
becomes anomalous at one-loop level (i.e., becomes second-class again after
quantization) \cite{anomaly1,anomaly2} due
to the existence of a central extension in the conformal algebra of the
energy-momentum tensor.  Later on, Hull \cite{hull} has shown how to cancel the
conformal anomaly of Siegel's model introducing auxiliary fields on the
zero-mode sector, the so called no-movers fields.
Nevertheless, to square a second-class constraint to
make it first-class results in a theory presenting (infinitely) reducible
constraints \cite{restucia}.  Independently, Floreanini and Jackiw
\cite{fj}  proposed an action
where the chiral constraint appears from the equations of
motion and therefore does not involve any Lagrange multiplier field.  This
(first-order) chiral boson formulation introduces however some spurious
solutions to (second-order) field equations that need careful boundary
conditions adjustments to be eliminated.
Another drawback in FJ proposal is the lack of manifest Lorentz covariance,
which makes the coupling to gauge and gravitational fields difficult
\cite{Son}.

After that, it was shown that one may have an action
containing infinite scalar fields, coupled by combinations of right and left chiral
constraints, carefully adjusted to be first-class, possess the spectrum of
a single chiral boson \cite{MWY} or by setting the second-class system of constraints to first-class \cite{CW1} by iterative conversion\footnote{The idea of using infinite auxiliary scalar
fields to covariantize second-class constraints has been introduced earlier
in the literature by Mikovic {\it et al} \cite{mik} in the context of the
relativistic super-particle.}.
In fact, MWY-W and FJ are just two different
representations of the same chiral boson theory \cite{clovis4}, each of which displaying
a different feature of the very same problem: in the MWY-W side not only
the Lorentz covariance is
manifest, but it also displays a symmetry that is hidden in the other
representation, while the FJ side, on the other hand, presents the
spectrum in a simpler manner\footnote{The equivalence between \cite{MWY} and \cite{CW1} was provided in \cite{CW-RC}.}.  Depending on one's interests, one
can pass from one representation to the other either transforming,
iteratively, the second-class constraint of the FJ model into
first-class, \`a la Faddeev-Shatashvili, with the
introduction of infinite Wess-Zumino fields, or
by resolving iteratively the MWY-W constraints by means of the
Faddeev-Jackiw technique for first-order constrained systems.

In this section \cite{bw1} we use the manifest covariance of the MWY-W
representation to couple it to background gravity.  The results 
obtained are shown to be consistent with the ones previously obtained
for the FJ representation by Sonnenschein \cite{Son} and by
Bastianelli and van Nieuwenhuizen \cite{BN}, which will be quickly
derived later, in a form slightly different then their
original formulation.

The MWY-W action is a representation of a  chiral boson in terms of
a sum over infinite scalar fields
\begin{equation}
\label{mwy}
S^{MWY-W}=\int d^2 x \sum_{k=0}^{\infty}(-)^k {1 \over 2} \partial_{\mu} \phi_k
\partial^{\mu}\phi_k
\end{equation}

\noindent chirally coupled to each other by a set of (infinite)
irreducible constraints $T_k^{(+)}$ or $T_k^{(-)}$ each one corresponding
to one of the chiralities
\begin{equation}
\label{mwy-constraint}
T_k^{(\pm)} = \Omega_k^{(\pm)}- \Omega_{k+1}^{(\mp)}
\end{equation}

\noindent with $\Omega^{(\pm)}_k$ being  right and left chiral constraints
$$\Omega_k^{(\pm)}= \pi_k \pm \phi'$$ 
that satisfy two uncoupled Kac-Moody algebra:
\begin{eqnarray}
\label{kac-moody}
\left\{\Omega_k^{(\pm)}(x),\Omega_m^{(\pm)}(y)\right\} & = & (\pm)2 \delta_{km}
\delta'(x-y)\nonumber\\
\left\{\Omega_k^{(+)}(x),\Omega_m^{(-)}(y)\right\} & = & 0
\end{eqnarray}

\noindent Here the curly brackets $\left\{A(x),B(y)\right\}$ represents the
Poisson bracket
of the fields $A(x)$ and $B(x)$.  One can verify that MWY-W constraints
$T_k^{(\pm)}(x)$ closes an (infinite) first-class Abelian algebra
under the Poisson bracket operation
\begin{equation}
\label{firstclass}
\left\{T_k^{(\pm)}(x),T_m^{(\pm)}(y)\right\}=0
\end{equation}

In view of the manifest covariance presented by the model, coupling to
gravity is straightforward, and reads
\begin{eqnarray}
\label{mwy-gravity}
S^{MWY-W}& = &{1\over 2}\int d^2 x \sqrt{-g}\sum_{k=0}^{\infty}(-)^k g^{\mu
\nu} \partial_{\mu}\phi_k\partial_{\nu}\phi_k\nonumber\\
& = & {1\over 2}\int d^2 x \sum_{k=0}^{\infty}
(-)^k\sqrt{-g} g^{00} \left\{\dot\phi^2_k +
2{g^{01} \over g^{00}}\dot\phi_k\phi'_k +{g^{11}\over g^{00}} \phi'^2_k
\right\}
\end{eqnarray}

\noindent where $g_{\mu \nu}$ is the background metric tensor and $g=\det
g_{\mu \nu}=g_{00}g_{11}-g_{01}^2$.  During this section we adopt the usual notation where
$\dot\phi=\partial_0 \phi$ and $\phi' =\partial_1\phi$, and $x^0=\tau$ and
$x^1=\sigma$ are the two-dimensional world-sheet variables.
To obtain the gauged
Floreanini-Jackiw counterpart we have to reduce the MWY-W-constraints
(\ref{mwy-constraint}) as explained above.  In order to effect such a
reduction we make use of the Faddeev-Jackiw symplectic technique.  To this
end we rewrite the MWY-W action in its first-order form, by introducing
the momentum $\pi_k $ conjugate to $\phi_k$,
\begin{equation}
\label{1st-mwy-gravity}
S_{\pm}^{MWY-W} = \int d^2 x \sum_{k=0}^{\infty}\left\{\pi_k \dot\phi_k -
{\cal H}^{MWY-W}+ \lambda_k T_k^{(\pm)}\right\}
\end{equation}

\noindent where ${\cal H}^{MWY-W}$ is the canonical Hamiltonian density
\begin{equation}
\label{hamiltonian}
{\cal H}^{MWY-W}={1\over
2}\sum_{k=0}^{\infty} \left[{(-)^k \over{g^{00}\sqrt{-g}}}(\pi_k^2 +\phi_k'^2)
-{g^{01}\over g^{00}}\pi_k\phi_k'\right]
\end{equation}

\noindent Now we eliminate the momentum $\pi_k$, in an iterative fashion,
 making use of
the MWY-W-constraints.  Implementing the first constraint,
$\pi_0=\pi_1\mp\phi'_0\mp\phi'_1$ results, after its substitution, in the
following Lagrangian density
\begin{eqnarray}
\label{1st-iteration}
{\cal L}_{\pm}^{MWY-W}  &=& \mp\dot\phi_0\phi'_0 - {\cal G_{\pm}} \phi_0'^2 -
\phi'_1\left(\dot\phi_0 +{\cal G_{\pm}}\phi'_0\right)\,+\,  \pi_1\left[\dot\phi_0+\dot\phi_1+{\cal {G_{\pm}}}
(\phi'_0+\phi'_1) \right]    \nonumber\\
& &\mbox{} + \sum_{k=2}^{\infty}\left[\pi_k \dot\phi_k -{1\over
2}{(-)^k \over{g^{00}\sqrt{-g}}}(\pi_k^2 +\phi_k'^2) +{g^{01}\over
g^{00}}\pi_k\phi_k' \right]
\end{eqnarray}

\noindent where
\begin{equation}
\label{couplings}
{\cal G_{\pm}}={1 \over g^{00}}\left({1 \over
\sqrt{-g}}\pm g^{01}\right)
\end{equation}

\noindent We repeat this procedure for the constraint 
$$\pi_1=\pi_2\mp\phi'_1\mp\phi'_2\:,$$ 

\ni in this way eliminating the momentum $\pi_1$, and
so on.  After all the remaining constraints have been implemented we find the
following effective action
\begin{equation}
\label{iterated1}
{\cal L}_{\pm}^{MWY-W}=\sum_{k=0}^{\infty}\left[\mp\dot\phi_k\phi'_k-
{\cal G_{\pm}}
\phi_k'^2 -2\left(\dot\phi_k +{\cal G_{\pm}}\phi'_k\right)
\sum_{m=k+1}^{\infty}\phi'_m\right]
\end{equation}

\noindent It is a simple algebraic manipulation to rewrite this action as
\begin{equation}
\label{iterated2}
{\cal L}_{\pm}^{MWY-W}=\sum_{k=0}^{\infty}\left[\mp\dot\phi_k\phi'_k-
{\cal G_{\pm}}
\phi_k'^2 -2\phi'_k\sum_{m=1}^{k-1} \left(\dot\phi_m +
{\cal G_{\pm}}\phi'_m\right)\right]
\end{equation}

\noindent which shows that all the MWY-W scalar fields have decoupled from
each other.  To make this point clearer, we rewrite the action
(\ref{iterated2}) as a double series function
\begin{equation}
\label{doubleseries}
{\cal S}_{\pm}^{MWY-W}= \sum_{k=0}^{\infty}\sum_{m=0}^{\infty}\int d^2x
\left(\mp\dot\phi_k\phi'_m - {\cal
G_{\pm}}\phi'_k\phi'_m \right)
\end{equation}

\noindent and introduce a new (collective) variable as
\begin{equation}
\label{collective}
\Phi = \sum_{k=0}^{\infty} \phi_k
\end{equation}

\noindent The MWY-W action for the collective field $\Phi$ assumes the form
of a Floreanini-Jackiw action coupled to the gravitational field
by the factor  ${\cal G_{\pm}}$
\begin{equation}
\label{florjack}
{\cal S}_{\pm}^{MWY-W} = \int d^2x\left(\mp\dot\Phi\Phi' -
{\cal G_{\pm}}\Phi'^2\right)
\end{equation}
and we know that it is a noton.

An interesting feature of the interacting action
(\ref{florjack}) is the reparametrization symmetry
that the coupling  ${\cal G_{\pm}}$ induces on the system, that reads
\begin{eqnarray}
\label{symmetry}
\delta_{\epsilon} \Phi &=& \epsilon \Phi'\nonumber\\
\delta_{\epsilon} {\cal G_{\pm}} &=& (\dot\epsilon + \epsilon {\cal G_{\pm}}' -
\epsilon' {\cal G_{\pm}})
\end{eqnarray}

It is also interesting to note that for each of the chiral fields,
$\phi_+$ or $\phi_-$, there will exist a class of metrics
for which they propagate as in flat space.
This will be the case whenever ${\cal G}_+ = 1$ or ${\cal G}_- = 1 $,
respectively.
These two conditions corresponds, as can be seen from (\ref{couplings}),
to respectively, 
$$g^{00} + g^{11} \pm 2 g^{01}=0\:,$$ 
or in terms of light-cone coordinates (\ref{lightfront}), to:
\begin{eqnarray}
\label{conditions}
g^{--} &=& 0\nonumber\\
g^{++} &=& 0.
\end{eqnarray}

\noindent A symmetric two dimensional metric can be written in terms of
the light cone components in the general form
\begin{equation}
\label{metric}
g^{\mu\nu}={1\over 2} \left(
\begin{array}{cc}
g^{++} + g^{--}+2 g^{+-} & g^{++} - g^{--} \\
g^{++} - g^{--} & g^{++} + g^{--} -2 g^{+-}
\end{array}
\right)
\end{equation}

\noindent Therefore, in metrics of the form
\begin{equation}
\label{metric+1}
g^{\mu\nu} ={1 \over 2} \left(
\begin{array}{cc}
g^{++} + g^{+-} & g^{++} \\
g^{++} & g^{++} - g^{+-}
\end{array}
\right)
\end{equation}

\noindent which gives ${\cal G_+}=1$, the chiral field $\phi_+$ propagates
as in flat space.  Similarly, for metrics whose general form reads
\begin{equation}
\label{metric-}
g^{\mu\nu} = \left(
\begin{array}{cc}
g^{--} + g^{+-} & -g^{--} \\
-g^{--} & g^{--} - g^{+-}
\end{array}
\right)
\end{equation}

\noindent implying ${\cal G_-}=1$, the chiral field $\phi_-$ remain
uncoupled.   Under conditions (\ref{conditions}) the
metrics (\ref{metric+1}) and (\ref{metric-}) become, in a sense,
``chiral metrics''.  Consequently when immersed in a chiral
background with chirality opposite to its own, chiral boson just do
not experiment the curvature. It should be noted that, contrarily to
the case of gauge fields, where the coupling is additive, when
conditions (\ref{conditions})are not satisfied each of the chiralities
of the field $\phi$ couples to the corresponding ${\cal G_{\pm}}$ that
depends on the whole metric and not on (\ref{metric+1}) or (\ref{metric-})
and now it is a noton.

Conditions (\ref{conditions}) correspond to a sort of selfduality
and anti-selfduality over the metric tensor, justifying us to call
them as chiral metrics.  Indeed, if we define the ``dual metric''
with respect to one index, as
\begin{equation}
\label{dual}
\mbox{}^{*}g^{\mu\nu}=\epsilon^{\mu\lambda}g_{\lambda}^{\nu}
\end{equation}

\noindent with $\epsilon^{10}=-\epsilon^{01}=1$, and take the trace, then
conditions (\ref{conditions}) will read
\begin{equation}
\label{trace}
Tr\,\left(g^{\mu\nu}\right)=\pm Tr\left(\mbox{}^{\ast} g^{\mu\nu}\right)
\end{equation}

\noindent as claimed

Concluding, we have shown explicitly in this section that the MWY-W representation for
the chiral boson can be coupled to an external background metric
in a standard way, making use of the manifest covariance.
The incorporation of the series of infinite chiral constraints in the model
was shown to lead to the same result as the one previously obtained for
the FJ model.
We have also pointed out the classes of metrics for which one of the
chiralities propagate as in free space.

\subsubsection{The dual projection}

In reference \cite{BN} it was shown that a version of the Floreanini Jackiw
action, coupled to gravity can be obtained beginning with a scalar field
coupled covariantly to gravity then writing it as a first order action
and imposing a (non covariant) constraint that selects one of the
(chiral) solutions of the classical equation of motion. 
Now we will trace the inverse way established before using the dual projection 
for a scalar field to decompose into notons.  Let us begin, as \cite{BN}, with a scalar
field coupled covariantly to gravity:
\begin{equation}
\label{scalar}
{\cal L} = {1 \over 2} \sqrt{-g} g^{\mu\nu} \partial_{\mu} \phi
\partial_{\nu}\phi
\end{equation}

\noindent The introduction of the auxiliary variable $p$ makes it possible to
write this Lagrangian density in a first order form
\begin{equation}
\label{scalarfirst}
{\cal L} = {\sqrt{-g}\over 2} \left[ -g^{00} p^2
+ 2 g^{00} p \dot\phi
+ 2g^{01} \dot\phi \phi^{\prime} +  g^{11} \phi^{\prime}\phi^{\prime}
\right]
\end{equation}

\noindent Performing the canonical transformation,
\begin{equation}
\label{mandelstan}
\phi = \phi_+ + \phi_-
\end{equation}

\noindent one is able to associate each of this fields with one
of the (chiral) solutions of the classical equation of motion by imposing \cite{clovis5}:
\begin{equation}
\label{mandelstan2}
p ={g^{01}\over g^{00}} ( \phi_+^{\prime} + \phi_-^{\prime} )
+ {1\over \sqrt{-g} g^{00}} ( \phi_-^{\prime} - \phi_+^{\prime})
\end{equation}

\noindent Inserting (\ref{mandelstan}) and (\ref{mandelstan2}) in (\ref{scalarfirst}) we get
\begin{eqnarray}
\label{scalarmandelstan}
{\cal L}  \,=\, -\dot\phi_+ \phi_+^{\prime} -
{\cal G_+} \phi_+^{\prime} \phi_+^{\prime}\,+\, \dot\phi_- \phi_-^{\prime} -
{\cal G_-} \phi_-^{\prime} \phi_-^{\prime}
\end{eqnarray}

\noindent showing explicitly the decomposition of the scalar field in
notons, each of them coupling with the metric exactly as in \cite{BN}.

%%%%%%%%%%%%%%%%%%%%%%%%%%%%%%%%%%%%%%%%%%%%%%%%%%%%%%%%%%%%%%%%%%%%%%%%%%%%%%%%%%%%%%%%%%%%%%%%%%%%%%%%%%%%%%%%%%%%%%%%%%%%%%%%%%%%%%%%%%%%%%%%%%%%

\subsection{The Linear Constraint Self-Dual Boson}

Let us stress again that a self-dual field in two dimensions is a scalar field which satisfies the
self-dual constraint (self-dual condition) $(\eta^{\mu\nu}+\epsilon^{\mu%
\nu})\partial_{\nu}\phi=0$ or $\dot{\phi}=\phi^{\prime}$.   In the formulation of
Floreanini and Jackiw \cite{fj}, the space derivative of the field instead
of the field itself satisfies the self-dual condition, i.e., $(\partial_0 -
\partial_1)\partial_1 \phi=0$, and the field violates the microcausality
postulate \cite{ggrs2}.

Trying to overcome these difficulties, Srivastava \cite{srivastava}
introduced an auxiliary vector field $\lambda_{\mu}$ coupled with a linear
constraint and constructed a Lorentz-invariant Lagrangian for a scalar
self-dual field. Although Harada \cite{harada} and Girotti {\it et al} \cite
{ggr} have pointed out consistency problems with the Srivastava model at the
quantum level, the linear formulation strictly describes a chiral boson from
the point of view of equations of motion at the classical level. In the view of that, some
methods were used to quantize the theory \cite{quantization}. The extension
to $D=6$ was accomplished in \cite{mm} as well as its supersymmetric case 
\cite{deri}.

Furthermore, an important ingredient in the study of such kind of systems
are the so called Wess-Zumino (WZ) terms \cite{wz}, which are introduced in
the theory in order to recover the gauge invariance \cite{fs2}. In \cite
{miao2}, it was proposed a new way of the derivation of the WZ counterterm.
It was based on the generalized Hamiltonian formalism of Batalin and Fradkin 
\cite{bf} which have suggested a kind of quantization procedure for
second-class constraint systems to which anomalous gauge theory belong \cite
{fs,faddeevkss}. The final action obtained, dependent on an arbitrary
parameter, has been constructed in order to become the Srivastava model
gauge invariant. The Lorentz invariance requirement has fixed the parameter
in two possible values which generates two possible WZ terms. The result,
with one of the WZ terms, after a kind of chiral decomposition, was that the
SCB spectrum is composed of two opposite FJ's chiral bosons. The
conclusion, however, was that the WZ term so obtained have {\it added} a new
physical degree of freedom, an antichiral boson, to the spectrum and
therefore changes the self-dual field into a massless scalar. Besides, in
another similar paper, Miao and Chen \cite{miao} have asserted that it is
impossible to apply the soldering formalism \cite{solda,harada2} to solder two
opposite chiral aspects of the model proposed by Srivastava, as was
successfully accomplished in the Siegel and Floreanini-Jackiw theories \cite
{adw}. It was pointed out that the method was invalid in the linear
formulation because of the inequivalence of Srivastava's and Siegel and
FJ's. Hence, to promote the fusion, it was constructed a chiral counterterm 
\cite{miao2} for the linear formulation of the chiral bosons. This
counterterm was the same Wess-Zumino term mentioned above.

In this section \cite{ad2} we have demonstrated that both conclusions are not really true.
We have applied successfully the soldering formalism and showed that the
interference on-shell of two SCB results in a massless scalar field. As
another result, we performed essentially a canonical transformation
(CT) \cite{djt,bg} and the outcome showed, in an exact way,
that the spectrum is already composed of two FJ's chiral bosons with the
same chirality confirming the well known result that the SCB has two degrees
of freedom thanks to the linear constraint structure \cite{bazeia}. Besides,
we have showed that the WZ term introduced in \cite{miao2} is in fact a
scalar field, i.e., it is composed of two FJ with opposite chiralities. So,
it is obvious that the WZ terms introduced naturally these particles since
the spectrum of the SCB is a vacuum-like one \cite{clo}.

\subsubsection{The soldering of two Srivastava's self-dual bosons}

The Srivastava action for a left-moving chiral boson, is 
\begin{equation}  \label{01}
{\cal L}^{(0)}_{\phi} = \partial_+\phi\partial_-\phi\,+\,\lambda_{+}
\partial_-\phi\;\;,
\end{equation}
where we have used the light-front variables (\ref{lightfront}) and 
$\lambda_{\pm}=\lambda_0 \pm\lambda_1$.

Following the steps of the soldering formalism studied in the last sections,
we can start considering the variation of the Lagrangians under the
transformations, %\begin{eqnarray}  \label{31}
$\delta\,\phi\,=\,\alpha$ and $\delta\,\lambda_{+}\,=\,0$. %\end{eqnarray}
We will write only the main steps of the procedure.

In terms of the Noether currents we can construct 
\begin{equation}
\delta {\cal L}^{(0)}_{\phi}\,=\,J^{\mu}_{\phi}\,\partial_{\mu}\,\alpha\;\;,
\end{equation}
where $\mu=+,-$, $J^{+}_{\phi}\,=\,0$ and $J^{-}_{\phi}\,=\,2\,\partial_+\,%
\phi\,+\,\lambda_{+}$. %\end{equation}

The next iteration, as seen in the last section, can be performed
introducing auxiliary fields, the so-called soldering fields 
\begin{equation}
{\cal L}^{(1)}_{\phi}\,=\,{\cal L}^{(0)}_{\phi}\,-\,B_{\mu}\,J^{\mu}_{\phi}%
\;\;,
\end{equation}

\noindent and one can easily see that the gauge variation of ${\cal L}%
^{(1)}_{\phi}$ is
\begin{equation}  \label{34111}
\delta {\cal L}^{(1)}_{\phi}\,=\,-\,2\,B_{-}\,\delta\,B_{+}\;\;,
\end{equation}
where we have defined the variation of $B_{\pm}$ as $\delta
B_{\pm}=\partial_{\pm}\alpha$, and we see that the variation of ${\cal L}%
^{(1)}_{\phi}$ does not depend on $\phi$. It is the signal to begin the
process with the other chirality, which is given by 
\begin{equation}
{\cal L}^{(0)}_{\rho} = \partial_+\rho\partial_-\rho\,+\,\lambda_{-}
\partial_+\rho\;\;,
\end{equation}
and again, let us construct the basic transformations $\delta\,\rho\,=\,%
\alpha$ and $\delta\,\lambda_{-}\,=\,0$.

The Noether\'{}s currents are $J^{+}_{\rho}\,=\,2\,\partial_+\,\rho\,+\,%
\lambda_{-}$ and $J^{-}_{\rho}\,=\,0$ %\end{equation}
and the variation of the final iteration is 
$$\delta {\cal L}^{(1)}_{\rho}\,=\,-\,2\,B_{-}\,\delta\,B_{+}$$.

Now we can see that the variation of ${\cal L}^{(1)}_{\phi,\rho}$ does not
depend neither on $\phi$ nor $\rho$. Hence, as explained before, we can
construct the final (soldered) Lagrangian as 
\begin{eqnarray}  \label{35}
{\cal L}_{TOT}&=&\,{\cal L}_{L}\,\oplus\,{\cal L}_{R}  \nonumber \\
&=&{\cal L}^{(1)}_{\phi}\,+\,{\cal L}^{(1)}_{\rho}\,+\,B_+\,B_-  \nonumber \\
&=&{\cal L}^{(0)}_{\phi}\,+\,{\cal L}^{(0)}_{\rho}\,-\,B_+\,J^+\,-\,B_-\,J^-
\,+\,B_+\,B_-
\end{eqnarray}
which remains invariant under the combined symmetry transformations for $%
(\phi,\rho)$ and $(\lambda_{+},\lambda_{-})$, i.e., $\delta\,{\cal L}%
_{TOT}\,=\,0$.

Following the steps of the algorithm depicted in section $2$, we have
to eliminate the soldering fields solving their equations of motion which
result in $B_{\pm}\,=\,J^{\mp}$ where $J^{\pm}=J^{\phi,\rho}$.

Substituting it back in (\ref{35}) we have the final effective Lagrangian
density 
\begin{eqnarray}  \label{37}
{\cal L}_{TOT}&=&\,(\,\partial_-\,\phi\,-\,\partial_-\,\rho\,)\,
(\,\partial_+\,\phi\,-\,\partial_+\,\rho\,)\,+\,\lambda_{+}\,(\,\partial_-\,\phi-\partial_-\,\rho\,) \nonumber \\
&-&\lambda_{-}\,(\,\partial_+\,\phi-\partial_+\,\rho\,) -{\frac{1}{2}}%
\,\lambda_{+}\,\lambda_{-}  \nonumber \\
&=&\,\partial_-\,\Phi\,\partial_+\,\Phi+\lambda_{+}\,\partial_-\,\Phi -
\lambda_{-}\,\partial_+\,\Phi\, -{\frac{1}{2}}\,\lambda_{+}\,\lambda_{-}\;\;.
\end{eqnarray}
where the new compound field are defined as $\Phi=\phi\,-\,\rho$.

As we can see we have a second order term in the Lagrange multipliers.
Solving the equations of motion for the multipliers, we obtain that, 
\begin{eqnarray}  \label{24}
\lambda_{-}\,=\,2\,\partial_-\,\Phi \qquad \mbox{and} \qquad
\lambda_{+}\,=\,-\,2\,\partial_+\,\Phi\;\;.
\end{eqnarray}

Substituting the equations (\ref{24}) in (\ref{37}) we have 
\begin{eqnarray} \label{ab0}
{\cal L}_{TOT}\,=\,-\,{\frac{1}{2}}\,\partial_{\mu}\,\Phi\,\partial^{\mu}\,%
\Phi,
\end{eqnarray}
which represents the massless scalar field action.

Hence, we have demonstrated in a precise way that it is possible to use the
soldering formalism to promote the fusion of two opposite SCB, in
contradiction with the assertion done in \cite{miao}. Finally, one can
conclude that, starting from these inconsistent Lagrangian densities, it is
recovered, in the soldering procedure, a consistent model which is, in fact,
the free scalar field. However, this result was not the expected one, but we
will come back to this issue later.

In the next section we will investigate the spectrum of the Srivastava model
constructing a canonical transformation \cite{bg}, i.e., using the dual projection 
\cite{aw}. The objective is to analyze
the result obtained by Miao {\it et al} \cite{miao2} previously with the
alternative construction of the Wess-Zumino term of the Srivastava theory.

\subsubsection{The dual projection of the Srivastava model}

In the Hamiltonian formulation, canonical transformations can be sometimes
used to decompose a composite Hamiltonian into two distinct pieces. A
familiar example \cite{djt}, is the decomposition of the Hamiltonian of a
particle in two dimensions moving in a constant magnetic field and quadratic
potential. It can be shown that this Hamiltonian can be separated into two
pieces corresponding to the Hamiltonians of two one dimensional oscillators
rotating in a clockwise and a anti-clockwise directions, respectively. Let
us now make a canonical transformation analysis of the SCB. In this case,
that the theory is already a chiral one, we will promote the dual projection, 
i.e., the theory will be decomposed in its dynamical
and symmetry parts. If the theory is not invariant, the result will show
only the dynamics of the system. To perform this we have to make a canonical
transformation \cite{bg} in (\ref{01}) using the Faddeev-Jackiw first-order
procedure.

At this point, some interesting comments are in order. The inconsistencies
of the SCB model at the quantum level, discussed in some works \cite
{harada,ggr}, can be verified from another point of view. This is done by
comparing the Lagrangian density of the SCB in Minkowski space, i.e., 
\begin{eqnarray}  \label{llag}
{\cal L} &=&{\frac{1}{2}}\,\partial _{\mu }\phi \,\partial ^{\mu }\phi
\,+\,\lambda _{\mu }\,(g^{\mu \nu }\,-\,\epsilon ^{\mu \nu }\,)\,\partial
_{\nu }\,\phi  \nonumber \\
&=&{\frac{1}{2}}\,(\,\dot{\phi}^{2}\,-\,{\phi ^{\prime }}^{2}\,)\,+\,\lambda
\,(\,\dot{\phi}\,-\,{\phi ^{\prime }}\,)
\end{eqnarray}
where $\lambda = \lambda _{+}$, with that of the bosonized version of the
CSM 
\begin{eqnarray}
{\cal L}\,&=&\,{\frac{1}{2}}\,(\partial _{\mu })^{2}\,+\,e\,(\,g^{\mu \nu
}\,-\,\epsilon ^{\mu \nu }\,)\,\partial _{\mu }\,\phi \,A_{\nu }\,+\,
\frac{a\,e^{2}}{2}\,A_{\mu }^{2}\,-\,{\frac{1}{4}}\,F_{\mu \nu }^{2}
\end{eqnarray}
and to note that the former is in fact a particular case of the latter,
where one should take care of the identifications: $a=0$ and $A_{\mu }\to
\lambda _{\mu }$, an external field with vanishing field strength. Now, one
can relate the inconsistency of the SCB with that of the CSM with the
regularization ambiguity parameter $a=0$, as shown by Girotti {\it et al}  
\cite{giro}. Now let us recover the discussion on the SCB, by doing the
dual projection and then discussing how and why the WZ terms
introduced in \cite{miao2} recover its quantum consistency.

The canonical momentum is defined by $\pi \,=\,\dot{\phi}\,+\,\lambda ,$ and
substituting it back in (\ref{llag}) to obtain the first-order form we have 
\begin{equation}  \label{02}
{\cal L}=\pi \,\dot{\phi}-{\frac{1}{2}}\,\pi ^{2}+\pi \,\lambda \,-\,{\frac{1%
}{2}}\,\lambda ^{2}-{\frac{1}{2}}\,{\phi ^{\prime }}^{2}-\lambda \,{\phi }%
^{\prime }\;\;,
\end{equation}

Now, as we have mentioned before, we have to do the following canonical
transformation 
\begin{equation}  \label{05}
\phi=\eta\,+\,\sigma \qquad \mbox{and} \qquad
\pi=\eta^{\prime}\,-\,\sigma^{\prime}\;\;.
\end{equation}
Notice that $\phi$ is a chiral field already. So, in this way, this canonical transformation will
allow us to know exactly what is the Srivastava chiral boson. Hence,
substituting (\ref{05}) in (\ref{02}) we have as a result 
\begin{eqnarray}
{\cal L}_{DP}={\eta^{\prime}}\,\dot{\eta} - {\eta^{\prime}}^2 -{%
\sigma^{\prime}} \,\dot{\sigma} - {\sigma^{\prime}}^2\, - 2\,\lambda\,{%
\sigma^{\prime}} - {\frac{1 }{2}}\,\lambda^2\;\;.  \nonumber
\end{eqnarray}
Again, solving the equations of motion for the $\lambda$-field we have $%
\lambda\,=\,-\,2\,{\sigma^{\prime}}$, and, substituting back, 
\begin{eqnarray}  \label{final}
{\cal L}_{DP}\,=\,{\eta^{\prime}}\,\dot{\eta}\,-\,{\eta^{\prime}}^2\,-\,({%
\sigma^{\prime}}\,\dot{\sigma}\,-\,{\sigma^{\prime}}^2)\;\;.
\end{eqnarray}

We can see clearly that this action represents two Floreanini-Jackiw's
chiral bosons. Each one with the same chirality. This is caused by the fact
that the Lagrange multiplier has acquired dynamics because of the linear
constraint form. In fact, we are demonstrating that (\ref{llag}) has two
degrees of freedom, represented in (\ref{final}) by $\eta$ and $\sigma$,
differently from Siegel's approach, where $\lambda$ is a pure gauge degree
of freedom. This result corroborates the one found by Bazeia in \cite{bazeia}
analyzing the linear constraint chiral boson quantum mechanics. We can say
that both particles in (\ref{final}) act like a Gupta-Bleuler's pair so that
each chiral excitation destroys the other and the Hilbert space is composed
of vacuum. This result confirms the one found in \cite{clo}.

Hence, in the soldering process of the SCB, each FJ's chiral boson interact
with its opposite chiral partner, so that the final result represents a
scalar field. We can observe also that the linear constraint formulation of
the chiral boson does not contain the Hull noton \cite{hull}, which is expected since the
SCB is not gauge invariant.

The result (\ref{final}) contradicts the result obtained in \cite{miao2} in
the following way. There, firstly it was built a final action composed of
the Srivastava action plus a WZ term with an arbitrary parameter. The
Lorentz invariance fixed the parameter in two possible values which
originated two different WZ terms. Hence, one of the actions obtained, after
a kind of chiral decomposition, is shown to have two FJ's particles of
opposite chiralities. Besides, the final Lagrangian obtained contain the BF
fields \cite{bf} used to construct the WZ term \cite{fik}. The conclusion
was that the WZ term constructed have added a new degree of freedom to the
theory in the form of an antichiral boson. %The action (\ref
%{final}), which does not have any auxiliary field, contradicts directly this
%conclusion. We have demonstrated that the SCB spectrum is already composed
%of two FJ's particles with opposite chiralities. 
%The WZ term obtained by
%Miao {\it et al} only have disclosed the other degree of freedom and not
%added a new one, as asserted in \cite{miao2}.
On the other hand, we can see what is really happening through a careful
analysis of the two WZ terms introduced in \cite{miao2}. It is not difficult
to see that the first WZ term defined in \cite{miao2}, i.e., 
\begin{equation}
{\cal L}_{WZ}^{(1)}\,=\,-\,{\frac{1}{2}}\,(\,\dot{\theta}^{2}\,+\,3\,{\theta
^{\prime }}^{2}\,)\,-\,\lambda \,(\,\dot{\theta}\,+\,\theta ^{\prime
}\,)\,-\,{\frac{1}{2}}\,\lambda ^{2}\;\;,
\end{equation}
where $\theta$ is the BF field, once integrated in the $\lambda $ field, a
chiral boson is recovered. Besides, if one takes the second WZ term
introduced in \cite{miao2}, 
\begin{equation}
{\cal L}_{WZ}^{(2)}\,=\,-\,\dot{\theta}\,\theta ^{\prime }\,-\,{\theta
^{\prime }}^{2}\,-\,\lambda \,(\,\dot{\theta}\,+\,\theta ^{\prime }\,)\,-\,{%
\frac{1}{2}}\,\lambda ^{2},
\end{equation}
and perform again the integration $\lambda $, one gets nothing but the
Lagrangian density of the free scalar boson. This result signalizes that the
WZ term obtained by Miao {\it et al} \cite{miao2} is already composed of two
opposite FJ's particles. Obviously it really introduces the degree of
freedom, because it is already there, in the WZ term, but it does not change
the physics of the SCB model, since, as we saw above, this last is composed
of vacuum.

Analyzing the interference aspects, we can apply the soldering formalism
again, but now, let us do it using two actions of the type of (\ref{final}),
i.e., 
\begin{eqnarray} \label{ab1}
{\cal L}_1 &=& \dot{\eta}\,\eta^{\prime}\,-\,{\eta^{\prime}}^2\,-\,(\dot{%
\sigma}\,\sigma^{\prime}\,-\,{\sigma^{\prime}}^2) 
\en
\bn \label{ab2}
{\cal L}_2 &=& -\,\dot{\xi}\,\xi^{\prime}\,-\,{\xi^{\prime}}^2\,-\,(-\,\dot{%
\omega}\,\omega^{\prime}\,-\,{\omega^{\prime}}^2)
\end{eqnarray}
where $\eta,\; \sigma,\; \xi$ and $\omega$ are all FJ's particles. We can see in
Eqs. (\ref{ab1}) and (\ref{ab2}) that the fields $(\eta,\xi)$ and $(\sigma,\omega)$ form
opposite chiralities particles pairs .

Performing the soldering procedure, one can easily see that the result is 
\begin{equation} \label{ab3}
{\cal L}\,=\,{\frac{1}{2}}\,(\partial_{\mu}\,\Psi)\,-\,{\frac{1}{2}}%
\,(\partial_{\mu}\,\Lambda)
\end{equation}
where $\Psi=\eta-\xi$ and $\Lambda=\sigma-\omega$. This is the expected
result, and not (\ref{ab0}), since we know that the SCB have a vacuum-like
spectrum. The soldering procedure in (\ref{ab3}) discloses the same behavior as
shown in (\ref{final}).

The result (\ref{ab3}) is quite different as the one shown in (\ref{ab0}). Since it is
well known that the soldering of two opposite FJ chiral bosons is a massless
scalar field, we should expect that the fusion of two SCB would be two
opposite scalar fields with the final vacuum-like spectrum. This difference
was explained before as we note that now, in each action of the Eqs.
(\ref{ab1}) and (\ref{ab2}), we have two fields, i.e., the action can be separated in two
different sectors, representing the FJ particles with the same chirality.
So, in the interference process (soldering) each sector of each action
interfere with its opposite partner. To obtain (\ref{ab0}), note that we have only
one sector in each action. In the interference process we have lost the
information about the other sector, like a destructive interference. This
does not occur in (\ref{ab3}).

To conclude this section, we can say that it is well known that the SCB has consistency problems. 
We have used the soldering formalism to show that the interference on-shell of two
Srivastava's chiral bosons resulted in a scalar field. The other aspect of
this result is that the soldering method recover the consistency of the SCB
model, i.e., the fusion of opposites chiralities of the model results in a
consistent theory. Again, we stress that this contradicts the conclusion published in the
literature, which asserts that it is impossible to apply the soldering
procedure to the SCB due to the inequivalence of this model with relation to
Siegel and Floreanini-Jackiw's models.

%%%%%%%%%%%%%%%%%%%%%%%%%%%%%%%%%%%%%%%%%%%%%%%%%%%%%%%%%%%%%%%%%%%%%%%%%%%%%%%%%%%%%%%%%%%%%%%%%%%%%%%%%%%%%%%%%%%%%%%%%%%%%%%%%%%%%%%%%%%%%%%%%%

\subsection{The Non-Abelian, PST and Supersymmetric Formulations of Hull's Notons}

The Wess-Zumino-Witten-Novikov (WZWN) \cite{witten,novikov} model is a conformal field theory that 
has been used in the past to reproduce various two dimensional systems like Toda field theories, 
black holes and others. Recently it was carried out the
connection of this model to a combination of three dimensional topological
BF and Chern-Simons gauge theories defined on a manifold with boundaries
using a direct application of non-Abelian T duality on the WZWN nonlinear
sigma model \cite{mohammed}. 

Now we will discuss the dual projection of
non-Abelian chiral fields coupled to external gravitational backgrounds
which is a result from the diagonalization of the first-order form of an action that
describes the chiral WZWN model.

The Lorentz covariant approach to describe chiral bosons, proposed by Pasti,
Sorokin and Tonin (PST) \cite{pst1}, has been studied in \cite
{lechner,blnpst,lechner2}. The quantum behavior of the PST models in the
framework of the algebraic renormalization \cite{psorella} was analyzed in \cite
{dps}. Differently of \cite{varios} it introduces only one scalar auxiliary
field, but in a non-polynomial way.

The PST procedure has important roles on the construction of a covariant
effective action for the $M$-theory five-brane \cite{blnpst}, covariant
actions for supersymmetric chiral bosons \cite{lechner2} and to rederive the
gravitational anomaly for the chiral bosons \cite{lechner}.

In this section we disclose the presence of the noton in four models using the dual
projection \cite{everton2}. In one we corroborate, in another way, the result found in 
\cite{clovis2} and depicted in the section $2.6$. Hence, we proposed three new formulations for the Hull
noton: two different Lorentz invariant versions of PST Siegel's chiral boson 
and a supersymmetric formulation.

\subsubsection{The Principal Chiral Model}

As a non-Abelian example, let us next extend the separability condition, due
to the dual projection, discussed above to the non-Abelian bosons following
only the main steps given in \cite{bw}. 

The most obvious choice would be to
consider an action given by a bilinear gradient of a matrix-valued field $g$
taking values on some compact Lie group $G$, which would be the natural
extension of the free scalar Abelian field. This is the action for the
Principal Chiral Model (PCM) that reads 
\begin{equation}  \label{pcm}
{\cal S}_{PCM}(g)= {\frac{1}{2}} \int d^2 x \,tr \left( \partial_\mu g
\;\partial^\mu \tilde{g} \right)\;\;.
\end{equation}

\noindent Here $g:R^{1,1}\rightarrow G$ is a map from the 2 dimensional
Minkowski space-time to $G$ and let us write for simplicity $\tilde{g}=g^{-1}$, as before. 
This action, however,
puts some difficulties. First of all, by examining its field equation we
learn that, it does not represent a free field. The Jacobian of the field
redefinition does not involve a time-derivative and can be reabsorbed in the
normalization of the partition function. Let us write the PCM in its
first-order form as 
\begin{equation}  \label{WZW1}
{\cal S}_{PCM}(g,P)\,=\, \int d^2 x\, tr \left( \partial_\tau g\, P \right) \,+\, {\frac{ 1}{2 }} \int d^2 x \,tr \left( PgPg \right) \,+\, {\frac{ 1}{2}} \int
d^2 x\, tr \left( \tilde{g}\partial_\sigma g\, \tilde{g}\partial_\sigma g
\right)
\end{equation}

\noindent and redefine the fields $g$ and $P$, which is an auxiliary field, 
through non-Abelian canonical transformations 
\begin{equation}  \label{nonabemand}
g \,=\, A \, B \qquad \mbox{and} \qquad P \,=\, \varepsilon\left(\tilde{B}%
\partial_\sigma \tilde{A} -\partial_\sigma \tilde{B} \tilde{A}\right)
\end{equation}
where, remembering our notation, we have that: $\tilde{A},\tilde{B}%
=A^{-1},B^{-1}$ respectively. The action for the PCM now reads 
\begin{equation}  \label{pcm2}
{\cal S}_{PCM}\left(A,B\right)\,=\, {\cal S}_{\varepsilon}\left(A\right)+ 
{\cal S}_{-\varepsilon}\left(B\right) \,+\, \varepsilon \int d^2x tr\left[%
\tilde{A} \left(\partial_\tau A \partial_\sigma B - \partial_\sigma A
\partial_\tau B\right)\tilde{B}\right]
\end{equation}

\noindent where 
\begin{equation}
{\cal S}_{\varepsilon}\left(A\right)=\int d^2x tr\left(\varepsilon
\partial_\sigma \tilde{A}\partial_\tau A -\partial_\sigma \tilde{A}%
\partial_\sigma A\right)\;\;,
\end{equation}
and ${\cal S}_{-\varepsilon}\,(B)$ is analogous.  We see that due to the non-Abelian nature of the fields, the cross-term cannot be eliminated, so that a complete separation cannot be
achieved. This fact should be expected. In the canonical approach, we have
for their field equations the pair 
\begin{equation}  \label{hje}
\partial_\tau I \,=\, \partial_\sigma J \qquad \mbox{and} \qquad
\partial_\tau J \,=\, \partial _\sigma I - \left[I,J\right]\;\;,
\end{equation}

\noindent where $I= \partial_\tau g\;\tilde{g}$ and $J=\partial_\sigma g\;%
\tilde{g} $. The second equation is a sort of Bianchi identity, which is the
integrability condition for the existence of $g$. Looking at this pair of
equations one can appreciate that chirality is not well defined in this
model. However, this picture changes drastically with the inclusion of the
Wess-Zumino topological term, i.e., when we consider the WZWN model \cite
{witten,novikov}. The first equation in (\ref{hje}) changes to $%
\partial_\tau I = \partial_\sigma J + \rho \left[I,J\right]$, with $\rho \in
Z$, producing a more symmetric set of equations. In particular for $%
\rho=\pm 1$ it is known that the currents above describe two independent
affine Lie algebras. One expects then that with the introduction of the
topological term, one would be able to obtain an identical mixing term, such
that in the total action they could cancel each other. With the topological term 
\begin{equation}
\Gamma_{WZ}(g)={\frac{1}{3}}\int d^3 x\,\epsilon^{ijk}\,tr\left[\,\tilde{g}%
\partial_i g \;\tilde{g}\partial_j g \;\tilde{g}\partial_k g \,\right]\;\;,
\end{equation}
we have now the complete WZWN model defined on the group manifold ${\cal M}%
_G$, based on the Lie algebra $G$, where $g \in {\cal M}_G$ and ${\cal M}$
is a three dimensional ball whose the boundary is the two dimensional
surface $\partial {\cal M}$ \cite{mohammed}.

Using the first of the field redefinitions (\ref{nonabemand}), it is a
lengthy but otherwise straightforward algebra to show that 
\begin{equation}  \label{topol}
\Gamma_{WZ}(A,B) \,=\,\Gamma_{WZ}(A)+\Gamma_{WZ}(B) \,+\, \int d^2x \,tr\left[%
\tilde{A}\left(\partial_\tau A \partial_\sigma B - \partial_\sigma A
\partial_\tau B\right)\tilde{B}\right]\;\;.
\end{equation}

Next, we can bring results (\ref{pcm2}) and (\ref{topol}) into the
WZWN action \cite{witten,novikov}, which is described by 
\begin{equation}  \label{WZW}
{\cal S}_{WZWN}(g) = {\frac{1}{{\lambda^2}}}{\cal S}_{PCM}(g)+{\frac{n}{{4\pi%
}}} \Gamma_{WZ}(g)\;\;.
\end{equation}

\noindent We mention the appearance of an extra parameter, both in the
action and in the canonical formalism, playing the role of coupling
constant. In terms of the chiral variables, the WZWN model reads 
\begin{eqnarray}  \label{wzw2}
{\cal S}_{WZWN}\left(A,B\right)&=&\left[{\frac{1}{\lambda^2}} {\cal S}%
_{\varepsilon}(A)+{\frac{n}{4\pi}}\Gamma_{WZ}(A)\right] +\left[{\frac{1}{%
\lambda^2}}{\cal S}_{-\varepsilon}(B)+{\frac{n}{4\pi}} \Gamma_{WZ}(B)\right]
\nonumber \\
&+& \left({\frac{\varepsilon}{\lambda^2}}+{\frac{n}{4\pi}}\right) \int d^2x
tr\left[\tilde{A}\left(\partial_\tau A\partial_\sigma B -\partial_\sigma
A\partial_\tau B\right)\tilde{B}\right]\;\;.
\end{eqnarray}

\noindent We can appreciate that the separability condition is only achieved
at the critical points, as expected. But also that our choice of $%
\varepsilon $ is now dependent on which of the critical points we choose: $%
4\pi\varepsilon=- \lambda^2 n$. The result is the non-Abelian version of the
dual projection, and corresponds to the sum of two Lagrangians describing
non-Abelian chiral bosons of opposite chiralities, each one having the form
proposed by Sonnenschein \cite{Son}. We will see that a change of
the critical point automatically switches the chirality of $A$ and $B$ by
changing the sign of $\varepsilon$. Indeed, in order to obtain separability,
we must have either 
\begin{equation}  \label{sepcon1}
(i)\:\:\:\:\:\:\:{\frac{{\lambda^2n}}{{4\pi}}}=-\varepsilon = 1
\end{equation}

\noindent or 
\begin{equation}  \label{sepcon2}
(ii)\:\:\:\:\:\:\:{\frac{{\lambda^2n}}{{4\pi}}}=-\varepsilon = -1\;\;.
\end{equation}

\noindent In the first case we find the set of chiral equations as 
$$\partial_x\left(\tilde{A}\partial_+A\right)=0 \qquad \mbox{and} \qquad \partial_-\left(\tilde{B}%
\partial_xB\right)=0$$ 

\ni whose solution reads respectively 
$$A\,=\,A_-(x^-)h_A(t) \qquad \mbox{and} \qquad B\,=\,h_B(t)B_+(x^+)\;\;.$$

\ni In the second case, the chiral equations are 
$$
\partial_x\left(\tilde{A}\partial_-A\right)=0 \qquad \mbox{and} \qquad \partial_+\left(\tilde{B}
\partial_xB\right)=0
$$

\ni and the solutions read 
$$
A\,=\,A_+(x^+)h_A(t) \qquad \mbox{and} \qquad B\,=\,h_B(t)B_-(x^-)\;\;.
$$ 

\ni The arbitrary functions of time $h_A(t)$ and $h_B(t)$ 
represent, in fact, the zero modes of the solutions of the chiral equations.  
Note that in both cases the general solution for $%
g(x^+,x^-) $ is given as 
$$
g=A_\varepsilon(x^\varepsilon)h_A(t)h_B(t)B_{-%
\varepsilon}(x^{-\varepsilon})\;\;.
$$ 

\ni As in the Abelian case, the constraint $%
h_A(t)=h_B^{-1}(t)$ becomes necessary in order that $g=AB$ satisfies the
equation of motion for the WZWN model. This constraint is the only memory
left for the chiral bosons stating that they belong to the same non-chiral
field. For the interested reader, more details about the dual projection of PCM and the coupling with
gravity can be seen in \cite{bw}.

\subsubsection{The new formulations of Hull's noton}

As has been said so far, there are indications that a deeper understanding of such issues as string
dynamics and fractional quantum Hall effect phenomenology can be achieved by
treating the chiral sectors in a more independent way. However, coupling
chiral fields to external gauge and gravitational fields is problematic. As
we said above (it was discussed in \cite{bw}) the coupling of chiral
(Abelian) fields to external gravitational backgrounds can be achieved by
diagonalization (dual projection) of the first-order form of a covariant
scalar action. The theory reduces then to a sum of a left and a right FJ's
actions \cite{fj}, circumventing the problems caused by the lack of manifest
Lorentz invariance. 

In this section we intend to supply the literature with new formulations of
Hull's noton: the chiral WZWN model (using the dual projection), with two different 
versions of PST formulations of the chiral bosons and with the supersymmetric case \cite{everton2}. 
We hope that the use of the dual projection formalism and consequently the disclosure 
of a Hull's noton mode inside these models may
help to gain a new insight into the structures of these theories. The
redefinition of the fields in the first-order form of the action naturally
reveals the two-dimensional internal structure hidden into the theory.

Note that in the last subsection, what we showed was the dual projection
applied in a nonchiral model. Now the interesting touch is to see what will
be disclosed in a well defined chiral model.

\subsubsection{The chiral WZWN model}

We know that the action for the chiral WZWN model \cite{witten,novikov} for
a particle which moves to the left, the so-called left mover is given by 
\begin{equation}  \label{leftzero}
S\,(g)=\int\;d^2x\; tr\left(\partial_+g\:\partial_-\tilde
g\,+\,\lambda\partial_-g\:\partial_-\tilde g\right) +\Gamma_{WZ}(g)
\end{equation}
where, for convenience we write $\lambda=\lambda_{++}$ and $g \in G$ was
defined in the last section. This action can be seen as the WZWN action
immersed in a gravitational background, with a characteristic truncated
metric tensor: 
\begin{equation}  \label{leftzerograv}
S\,(g)={\frac{1}{2}}\int d^2 x \sqrt{-\eta_+}\;
\eta_+^{\mu\nu}\:tr\left(\partial_\mu g\:\partial_\nu \tilde g\right)
+\Gamma_{WZ}(g)
\end{equation}

\noindent with $\eta^+ = det(\eta^+_{\mu\nu})$ and

\begin{equation}  \label{metric++}
{\frac{1}{2}} \sqrt{-\eta_+}\: \eta_+^{\mu\nu}=\left( 
\begin{array}{cc}
0 & {{\frac{1}{2}}} \\ 
{{\frac{1}{2}}} & {\lambda}
\end{array}
\right)
\end{equation}
where $\lambda$ transforms like in the Abelian case and $g$ as a scalar.

In Lorentz coordinates we can write (\ref{leftzero}) as 
\begin{eqnarray}
S\,(g)\,=\,\int\,d^2\,x\,tr \left[ {\frac{1 }{2}}\,(1+\lambda)\dot{g}\dot{%
\tilde{g}}\, \,-\,{\frac{1 }{2}}\,(1-\lambda)\,g^{\prime}\tilde{g}%
^{\prime}\,-\,{\frac{\lambda }{2}}(\,\dot{g}\,\tilde{g}^{\prime}\,+\,g^{%
\prime}\,\dot{\tilde{g}}\,) \right] \,+\,\Gamma_{WZ}(g)
\end{eqnarray}

This action can be written in its Faddeev-Jackiw's first-order form making
the following transformation which introduces an auxiliary field ${\cal P}$, 
\[
{\frac{1}{2}}\,\dot{g}\,\dot{\tilde{g}} \rightarrow {\cal P}\,\dot{g}\,+\,{%
\frac{1 }{2}}\,{\cal P}\,g\,{\cal P}\,g 
\]
where ${\cal P}=\dot{\tilde{g}}$.

Now we can write that 
\begin{eqnarray}  \label{firstorder}
S\,(g)&=&\int\,d^2\,x\,tr \left[ (1\,+\,\lambda){\cal P}\,\dot{g}\,+\,{\frac{%
1 }{2}}\,(1+\lambda)\,{\cal P}\,g\,{\cal P}\,g \,-\,{\frac{1}{2}}%
(1-\lambda)\,g^{\prime}\,\tilde{g}^{\prime} \,-\,{\frac{\lambda }{2}}\,(\,%
\dot{g}\,\tilde{g}^{\prime}\,+\,g^{\prime}\,\dot{\tilde{g}}\,) \right] 
\nonumber \\
&+&\,\Gamma_{WZ}(g)\;\;.
\end{eqnarray}
For convenience, let us promote the following transformation of variables,

\[
{\cal P}=\frac{1}{1+\lambda}\,{\cal P}\;\;, 
\]

\noindent and substituting in (\ref{firstorder}) we have that, 
\begin{eqnarray}  \label{firstorder2}
S\,(g)\,&=&\,\int\,d^2\,x\,tr \left[ {\cal P}\,\dot{g}\,+\,{\frac{1 }{2}}\,{%
\frac{1 }{(1+\lambda)}}\,{\cal P}\,g\,{\cal P}\,g \,-\,{\frac{1}{2}}%
(1-\lambda)\,g^{\prime}\,\tilde{g}^{\prime} \,-\,{\frac{\lambda }{2}}\,(\,%
\dot{g}\,\tilde{g}^{\prime}\,+\,g^{\prime}\,\dot{\tilde{g}}\,) \right] 
\nonumber \\
&+&\,\Gamma_{WZ}(g)\;\;.
\end{eqnarray}

\noindent Making again another transformation, i.e., 
\[
{\cal P} \rightarrow {\cal P}\,+\,\lambda\,\dot{\tilde{g}} 
\]
hence, 
\begin{equation}  \label{firstorder3}
S\,(g)\,=\,\int\,d^2\,x\,tr \left[ {\cal P}\,\dot{g}\,+\,{\frac{1 }{2}}\,%
\frac{1}{1+\lambda}\,({\cal P}\,g\,{\cal P}\,g\,-\,2 \lambda {\cal P}
g^{\prime}) \,-\,{\frac{1 }{2}}{\frac{1 }{(1+\lambda)}}\,g^{\prime}\,\tilde{g%
}^{\prime}\right]  \nonumber \\
\,+\,\Gamma_{WZ}(g)\;\;.
\end{equation}

Let us redefine the fields $g$ and ${\cal P}$ through the following canonical
transformation, 
\begin{eqnarray}  \label{nove}
g\,=\,N\,h \qquad \Rightarrow \qquad
g^{\prime}\,=\,N^{\prime}\,h\,+\,N\,h^{\prime}\;\;,
\end{eqnarray}
where $N$, as before, will play the role of a Hull's nonmover field, the
noton. 

Using (\ref{nove}) to justify a first choice for ${\cal P}$, 
let us construct it with two numerical (or not) coefficients which will be
determined in a further analysis, 
\begin{equation}
{\cal P}\,=\,a\,\tilde{h}\,\tilde{N}^{\prime }\,+\,b\,{\tilde{h}}^{\prime }\,%
\tilde{N}\;\;,  \label{noveb}
\end{equation}
but note the non-Abelian feature of this different construction. An
important observation is that the coefficients $a$ and $b$ incorporate the
roles of the separability points described in the last section. We will show
below that this form of the field ${\cal P}$ is the only possible general
form in order to promote the dual projection.

As we mentioned above, we showed that the dual projection of the
Wess-Zumino term, using the redefinition (\ref{nove}) has the form 
\begin{eqnarray}  \label{dez}
\Gamma_{WZ}(N,h) = \Gamma_{WZ}(N)\,+\,\Gamma_{WZ}(h)\,+\,\int\,d^2\,x\,tr
\left( \tilde{N}\,N^{\prime}\dot{h}\,\tilde{h}\, -\,\tilde{N}\,\dot{N}%
\,h^{\prime}\,\tilde{h}\right)\;\;.
\end{eqnarray}
We can see that the dual projection of (\ref{dez}) brings an extra term that
can not be split.  So, this term has to be eliminated. To perform this
elimination it is easy to see that the cast of the field ${\cal P}$ has to
have some term proportional to the extra term in (\ref{dez}). Other forms of such field
do not afford the elimination of this term, as we said before.

The canonical transformations (\ref{nove}) and (\ref{noveb}) lead us to an
action with fields taking values in the internal space. Substituting the
redefinitions (\ref{nove}), (\ref{noveb}) and (\ref{dez}) into (\ref
{firstorder3}) we can show that, 
\begin{eqnarray}  \label{onze}
& &S(N,h)\,=\, \Gamma_{WZ}\,(N)\,+\,\Gamma_{WZ}(h) \,+\, \int\,d^2\,x\,tr
\left\{ a\,\tilde{N}^\prime\,\dot{N}\,+\,b\,\tilde{h}^\prime\,\dot{h}
\,-\,(a-1)\,\tilde{N}\,\dot{N}\,\dot{h} \tilde{h} \right.  \nonumber \\
&-&\left. (b+1)\,\tilde{N}\,\dot{N}\,h^\prime \tilde{h}\, \,-\,{\frac{1 }{{%
1+\lambda}}}\left({\frac{a^2 }{2}}\,+\,\lambda a\,+\,{\frac{1 }{2}}\right)\,%
\tilde{N}^\prime\,{N}^\prime \,-\,{\frac{1 }{{1+\lambda}}}\left({\frac{b^2 }{%
2}}\,+\,\lambda b\,+\,{\frac{1 }{2}}\right)\,\tilde{h}^\prime\,{h}^\prime
\right.  \nonumber \\
&+&\left. 2\,(ab+1)\,\tilde{N}^\prime\,{N}\,h\,\tilde{h}^\prime \right\}\;\;.
\end{eqnarray}

\noindent To eliminate the extra terms we have to find $a$ and $b$ solving
the following very simple system 
\begin{equation}
a\,-\,1 \,=\, 0 \qquad \mbox{and} \qquad b\,+\,1 \,=\, 0\;\;,
\end{equation}
which solution is 
\begin{equation}
a\,=\,1 \qquad \mbox{and} \qquad b\,=\,-\,1 \qquad \Longrightarrow \qquad
a\,b\,=\,-\,1\;\;.
\end{equation}

Instead of applying directly these solutions, let us promote a short analysis making 
$a=-b=\epsilon$. With this new form for the solution, the canonical transformations are 
\begin{equation}
g\,=\,N\,h \qquad \mbox{and} \qquad p\,=\,\epsilon\,\tilde{h}\tilde{N}%
^\prime\,-\,\epsilon\,\tilde{h}^\prime\,\tilde{N}\;\;,
\end{equation}
and substituting it in (\ref{onze}) we have the action given by 
\begin{eqnarray}
S(N,h)&=&\int\,d^2\,x\,tr \left[ \epsilon\,\dot{N}\,\tilde{N}^{\prime}\,-\, 
\frac{\epsilon^2\,+\,2 \lambda \epsilon\,+\,1}{2(1+\lambda)}\,N^{\prime}\,%
\tilde{N}^{\prime} \right]\,+\,\Gamma_{WZ}\,(N)  \nonumber \\
&-&\int\,d^2\,x\,tr \left[ \epsilon\,\dot{h}\,\tilde{h}^{\prime}\,-\, \frac{%
\epsilon^2\,-\,2 \lambda \epsilon\,+\,1}{2(1+\lambda)}\,h^{\prime}\,\tilde{h}%
^{\prime} \right]\,+\,\Gamma_{WZ}\,(h) \;\;,
\end{eqnarray}
and finally we can say that 
\begin{eqnarray}  \label{notons}
S(N,h)_{\epsilon=1}&=&\int\,d^2\,x\,tr \left(\dot{N}\,\tilde{N}%
^{\prime}\,+\,N^{\prime}\,\tilde{N}^{\prime} \right)\,+\,\Gamma_{WZ}\,(N) 
\nonumber \\
&-&\int\,d^2\,x\,tr \left( \dot{h}\,\tilde{h}^{\prime}\,+\,\eta\,h^{\prime}\,%
\tilde{h}^{\prime} \right)\,+ \,\Gamma_{WZ}\,(h) \;\;,  \nonumber \\
S(N,h)_{\epsilon=-1}&=&-\int\,d^2\,x\,tr \left(\dot{N}\,\tilde{N}%
^{\prime}\,+\,\eta\,N^{\prime}\,\tilde{N}^{\prime} \right)
\,+\,\Gamma_{WZ}\,(N)  \nonumber \\
&-&\int\,d^2\,x\,tr \left(\dot{h}\,\tilde{h}^{\prime}\,+\,h^{\prime}\,\tilde{%
h}^{\prime} \right)\,+ \,\Gamma_{WZ}\,(h) \;\;,
\end{eqnarray}
where 
\[
\eta\,=\,\frac{1-\lambda}{1+\lambda}\;\;. 
\]
Since we are working in a $D=2$ dimensions, i.e., $0$-form $(D=2(2p+1))$,
each reduced phase space carries the representation for half the number of
degrees of freedom of the original system. This is a feature different from $%
D=4p$, where the duality symmetric systems maintains the phase space
structure intact \cite{baw2}. For the interested read, an analysis of the
WZWN duality groups is depicted in \cite{aal}.

We see in (\ref{notons}) that in accordance with the value of $\epsilon$, $N$ and $%
h$ change roles. For $\epsilon=1$, $N$ is the chiral field and $h$ is the
non-Abelian Hull's noton.   For $\epsilon=-1$, vice versa. This shows a
certain dependence of the dual projection on a certain parameter, but it is
immaterial since what we want to show is the presence, in the chiral WZWN,
of a chiral mode and mainly, the presence of a non-Abelian Hull's noton.

This result confirms the one found in section $2.6$. However, there,
the noton was found using the soldering technique of two
non-Abelian chiral bosons, showing a destructive interference.  This section 
describes a different way to obtain the same result.  As the WZWN model has a chiral mode
and an algebraic mode, in the fusion (soldering) of opposite non-Abelian
chiral models, the opposite chiral particles interfere destructively
disappearing from the spectrum. Only the Hull noton survives, which as
having no dynamics, does not interact destructively.  
In other words, we show that the Hull noton was already there, hidden 
on the chiral WZWN model.

\subsubsection{The PST self-dual formulation}

It can be shown \cite{pst2} that one of the possibilities to recover the
manifest Lorentz invariance of the FJ model, as in the case of Maxwell theory, is introducing an
unit-norm time-like auxiliary vector field $u_{m}\,(x)$ in the
FJ action, and to write the action in the form 
\begin{equation}
S\,=\,\int d^{2}x\,\left( \,\partial _{+}\phi \,\partial _{-}\phi
\,-\,u^{m}{\cal F}_{m}u^{n}{\cal F}_{n}\,\right)  \label{pst}
\end{equation}
where ${\cal F}_{m}=\partial _{m}\phi \,-\,\epsilon _{mn}\partial ^{n}\phi $%
. Because of its properties, $u_{m}$ contains only one independent
component. It was proved in \cite{pst2} that (\ref{pst}) reduces to the Siegel action \cite
{siegel} and also that $u_m$ is the gradient of a scalar.  It will be used below.

In Minkowski space we can write the Lagrangian density of (\ref{pst}) as 
\begin{equation}
{\cal L}\,=\,\dot{\phi}^{2}\,-\,{\phi ^{\prime }}^{2}\,-\,(u^{0}\,+%
\,u^{1})^{2}\,(\dot{\phi}\,+\,\phi ^{\prime })^{2}
\end{equation}
and following the procedure to promote the dual projection in first-order actions 
we have, after a little algebra, that 
\begin{equation}
{\cal L}\,=\,\pi \,\dot{\phi}\,-\,\frac{\left[ \pi
\,+\,2(u^{0}+u^{1})^{2}\phi ^{\prime }\right] ^{2}}{4\left[
1-(u^{0}+u^{1})^{2}\right] }\,-\,[1\,+\,(u^{0}+u^{1})^{2}]\,{\phi ^{\prime }}%
^{2}  \label{pst2}
\end{equation}
where $\pi$ is the canonical momentum.

Now, as before, we can use the convenient canonical transformations,  
\begin{equation}
\phi \,=\,\pm {\frac{1}{\sqrt{2}}}\,(\rho \,+\,\sigma )\qquad \mbox{and}%
\qquad \pi \,=\,\pm \sqrt{2}\,(\rho ^{\prime }\,-\,\sigma ^{\prime })\;\;.
\end{equation}
where one of these fields will represent the FJ particle and the other, of course, 
the noton, as we will see.

Substituting these transformations in (\ref{pst2}) and after an algebraic
work we have an action in terms of $\sigma$ and $\rho$, which take values in
the internal space, 
\begin{equation}
{\cal L}\,=\,\left( -\,\sigma ^{\prime }\,\dot{\sigma}\,-\,{\sigma ^{\prime }%
}^{2}\right) \,+\,\left( \rho ^{\prime }\,\dot{\rho}\,-\,\eta _{1}\,{\rho
^{\prime }}^{2}\right)  \label{44}
\end{equation}
where 
\[
\eta _{1}\,=\,\frac{1+(u^{0}+u^{1})^{2}}{1-(u^{0}+u^{1})^{2}}\;\;. 
\]
As we can easily see the $\sigma $ field describes a FJ chiral boson that now, at first sight, carries the dynamics of the system and the $Z_{2}$ duality group. However, in \cite
{baw2} it was shown that in all even dimensions, both $SO(2)$ and $Z_{2}$
duality symmetric groups could co-exist. As we have analyzed in the anterior
section, the $\rho $ field is responsible for the algebra \cite{pst2} of the
system. Hence, $\rho $ is the PST Hull's noton formulation.

The other possibility (see \cite{pst2}), which is more appropriate from the quantum point of view, is to construct a Lorentz covariant action like the following 
\begin{equation}
S\,=\,\int d^{2}x\,\left( \partial _{+}\phi \,\partial _{-}\phi \,+\,{%
\frac{1}{u^{2}}}\,u^{m}{\cal F}_{m}u^{n}{\cal F}_{n}\,-\,\epsilon
^{mn}\,u_{m}\partial _{n}B\right)  \label{pst3}
\end{equation}
where $B(x)$ is an auxiliary scalar field. After eliminating $B$ through the field equations and taking the values for $u_{m}$, the Lagrangian density of the above action can be
written as 
\begin{equation}
{\cal L}\,=\,\partial _{+}\phi \,\partial _{-}\phi \,-\,{\frac{\partial
_{+}\hat{\varphi}}{\partial _{-}\hat{\varphi}}}\,(\partial _{-}\phi
)^{2}\;\;,  \label{pst4}
\end{equation}
where $\hat{\varphi}$ is an another auxiliary field that helps in the
solution of the equation for $u_{m}$, i.e., $u_{m}(x)=\p_{m}\hat{\varphi}(x)$.  Hence, $u_{m}(x)$ can be seen as the gradient of a scalar $\hat{\varphi}(x)$.

In this case it is easy to see that the equation (\ref{pst4}) has the same
form as the Siegel chiral boson, where 
\[
\lambda \quad \rightarrow \quad {\frac{\partial_{+} \hat{\varphi} }{%
\partial_{-} \hat{\varphi}}} 
\]

\noindent and with a straightforward association we already have the
solution given in \cite{aw}

\begin{equation}
{\cal L}\,=\,\rho ^{\prime }\,\dot{\rho}\,-\,{\rho ^{\prime }}%
^{2}\,-\,\sigma ^{\prime }\,\dot{\sigma}\,-\,\eta _{2}\,{\sigma ^{\prime }}%
^{2}\;\;,
\end{equation}
where we have again $\sigma $ as the noton and 
\[
\eta _{2}\,=\,\frac{1\,-\,{\frac{\partial _{+}\,\hat{\varphi}}{\partial
_{-}\,\hat{\varphi}}}}{1\,+\,{\frac{\partial _{+}\,\hat{\varphi}}{\partial
_{-}\,\hat{\varphi}}}}\;\;. 
\]
In the face of this result we can make an analogous analysis for this second
possibility of Lorentz invariance restoration as we made it for the first one
and stress that the dynamics/$(Z_{2}+SO(2))$ duality groups and algebra
group are carried by $\rho$ and $\sigma $ respectively. We see clearly that
Hull's noton is present whatever the way we formulate the Siegel chiral
boson.

\subsubsection{The supersymmetric formulation}

The superfield generalization of (\ref{pst3}) is 
\begin{equation}  \label{super}
S\,=\,\int d^2 x d\theta^+ \,\left( D_+ \Phi\,\partial_{-}\Phi \,-\,{\frac{%
D_{+} \Lambda }{\partial_{-} \Lambda}}\,(\partial_{-}\,\Phi)^2 \right)
\end{equation}
where we are considering the bosonic superfields 
\[
\Phi\,(x^{-},x^{+},\theta^+)\,=\,\phi(x)\,+\,i\,\theta^+\,\psi_+ (x) 
\]
and 
\[
\Lambda\,(x^{-},x^{+},\theta^+)\,=\,\hat{\varphi}(x)\,+\,i\,\theta^+\,%
\chi_+ (x) \;\;, 
\]
which obey the conventional transformation laws under global shifts 
\[
\delta \theta^+\,=\,\epsilon^+ \;,\qquad \delta
x^{+}\,=\,i\theta^+\,\epsilon^+ \qquad \mbox{and} \qquad \delta
x^{-}\,=\,0 
\]
in $n=(1,0)$ flat superspace and where 
\begin{eqnarray}
D_+\,&=&\,{\frac{\partial }{\partial\theta^+}}\,+\,i\,\theta^+\,\partial_{+}
\nonumber \\
D^2_+\,&=&\,i\,\partial_{+}
\end{eqnarray}
is the supercovariant derivative.

After making the relevant substitutions described above in (\ref{super}) we
can write 
\begin{equation}
S\,=\,\int d^2 x d\theta^+ \left\{ \,{\frac{\partial \Phi }{\partial \theta^+%
}}\,\partial_{-}\,\Phi \,-\,\,{\frac{\partial \Lambda }{\partial \theta^+}}%
\,\,{\frac{(\partial_{-} \Phi)^2 }{\partial_{-} \Lambda}} \,+\,
\,i\,\theta^+\,\left[ \partial_{+}\Phi\,\partial_{-}\Phi \,-\,\,{\frac{%
\partial_{+} \Lambda }{\partial_{-} \Lambda}}\,(\partial_{-}\Phi)^2 %
\right] \right\}\;\;.
\end{equation}
Using the well known properties of the supersymmetric integrals we finally
have that, 
\begin{equation}  \label{super2}
{\cal L}\,=\, \partial_{+}\Phi\,\partial_{-}\Phi \,-\,{\frac{\partial_{+}
\Lambda }{\partial_{-} \Lambda}}\,(\partial_{-}\Phi)^2 \;\;,
\end{equation}
where a hidden global complex number is of no consequence here since it does
not affect the equations of motion and may be ignored. With this form, we
can make in (\ref{super2}) the same association that we made in (\ref
{pst4}). So, we can write, considering that now we are in a superspace that, 
\begin{equation}
{\cal L}\,=\, \Sigma^\prime\,\dot{\Sigma}\,-\,{\Sigma^\prime}%
^2\,-\,\Gamma^\prime\,\dot{\Gamma} \,-\,\eta_3\,{\Gamma^\prime}^2\;\;,
\end{equation}
where 
\[
\eta_3\,= \,\frac{1\,-\,{\frac{\partial_{+}\,\Lambda }{\partial_{-}\,%
\Lambda}}} {1\,+\,{\frac{\partial_{+}\,\Lambda }{\partial_{-}\,\Lambda}}}
\;\;. 
\]
The superfield $\Sigma$ represents the super-FJ chiral boson in the superspace
formulation and $\Gamma$ is the supersymmetric Hull's noton, the supernoton%
\footnote{%
A very interesting study of the duality groups in supersymmetric models can
be seen in \cite{rv}.}.

%\section{Conclusion}

To sum up, we have diagonalized the non-Abelian chiral field, disclosing a two
dimension internal structure for some compact Lie group $G$. This result was
expected based on the results found in \cite{aw} where it was shown that
notons, being non-dynamicals, couples to the gravitational backgrounds but
not to the electromagnetic field. Therefore, if a
gauge coupling is introduced before dual projection, it will be completely
decoupled by the dual projection procedure. Since the analysis was always
effected for first-order systems, an equivalence between the Lagrangian and
Hamiltonian approaches permitted us to use the concept of canonical
transformations. In other words we can say that the dual projection demanded
a change of variables which was, in the phase space, a canonical
transformation.

In \cite{pst2} it was proposed two ways to restore the manifest Lorentz
invariance of the FJ theory. The first was introducing an unit-norm time-like auxiliary
vector field in the FJ action and constructing an action
equivalent to the Siegel model. The other is constructing an action in the
light of the Maxwell manifestly Lorentz invariant duality symmetric action
proposed in \cite{pst2}, which is shown to be also Siegel's equivalent on
the mass shell. We show that in both cases the noton is present. Our last
result concerns the superfield generalization of the above action and a
supersymmetric formulation of the noton, the supernoton has been constructed.

%%%%%%%%%%%%%%%%%%%%%%%%%%%%%%%%%%%%%%%%%%%%%%%%%%%%%%%%%%%%%%%%%%%%%%%%%%%%%%%%%%%%%%%%%%%%%%%%%%%%%%%%%%%%%%%%%%%%%%%%%%%%%%%%%%%%%%%%%%%%%%%%

\section{Acknowledgments}

This work is partially supported by Funda\c{c}\~ao de Amparo \`a Pesquisa do Estado do Rio de Janeiro (FAPERJ) and Conselho Nacional de Densenvolvimento Cient\'{\i}fico e Tecnol\'ogico (CNPq) and PRONEX/CNPq/FAPESQ, Brazilian research agencies.

%%%%%%%%%%%%%%%%%%%%%%%%%%%%%%%%%%%%%%%%%%%%%%%%%%%%%%%%%%%%%%%%%%%%%%%%%%%%%%%%%%%%%%%%%%%%%%%%%%%%%%%%%%%%%%%%%%%%%%%%%%%%%%%%%%%%%%%%%%%%%%%%

%\begin{references}

\end{document}